\documentclass[acmsmall]{acmart}

\setcopyright{rightsretained}
\acmPrice{}
\acmDOI{10.1145/3622797}
\acmYear{2023}
\copyrightyear{2023}
\acmSubmissionID{oopslab23main-p50-p}
\acmJournal{PACMPL}
\acmVolume{7}
\acmNumber{OOPSLA2}
\acmArticle{222}
\acmMonth{10}
\received{2023-04-14}
\received[accepted]{2023-08-27}

\startPage{1}

\usepackage{booktabs}   %% For formal tables:
\usepackage{subcaption} %% For complex figures with subfigures/subcaptions
\usepackage{draftwatermark}
\usepackage{graphicx}
\usepackage{caption}
\usepackage{enumitem}
\usepackage{fancyvrb}
\usepackage{algorithmicx}
\usepackage{algorithm}
\usepackage{multirow}
\usepackage{verbatimbox}
\usepackage[noend]{algpseudocode}
\usepackage{xcolor,colortbl}
\usepackage{algpseudocode}% http://ctan.org/pkg/algorithmicx
\usepackage[subtle]{savetrees}

\begin{document}

%% Title informationhttps://www.overleaf.com/project/643888409537d55450220818
\title{Hardware-Aware Static Optimization of Hyperdimensional Computations}         %% [Short Title] is optional;
                                        %% when present, will be used in
                                        %% header instead of Full Title.
%\titlenote{with title note}             %% \titlenote is optional;
                                        %% can be repeated if necessary;
                                        %% contents suppressed with 'anonymous'
%\subtitle{Subtitle}                     %% \subtitle is optional
%\subtitlenote{with subtitle note}       %% \subtitlenote is optional;
                                        %% can be repeated if necessary;
                                        %% contents suppressed with 'anonymous'

%% Author information
%% Contents and number of authors suppressed with 'anonymous'.
%% Each author should be introduced by \author, followed by
%% \authornote (optional), \orcid (optional), \affiliation, and
%% \email.
%% An author may have multiple affiliations and/or emails; repeat the
%% appropriate command.
%% Many elements are not rendered, but should be provided for metadata
%% extraction tools.

%% Author with single affiliation.
\author{Pu (Luke) Yi}
%\authornote{with author1 note}          %% \authornote is optional;
                                        %% can be repeated if necessary
\orcid{0000-0001-6669-6520}             %% \orcid is optional
\affiliation{
  \position{PhD Student}
  \department{Computer Science Department}              %% \department is recommended
  \institution{Stanford University}            %% \institution is required
  \streetaddress{450 Jane Stanford Way}
  \city{Stanford}
  \state{California}
  \postcode{94305}
  \country{USA}                    %% \country is recommended
}
\email{lukeyi@stanford.edu}          %% \email is recommended

%% Author with two affiliations and emails.
\author{Sara Achour}
%\authornote{with author2 note}          %% \authornote is optional;
                                        %% can be repeated if necessary
\orcid{0000-0003-3444-1544}             %% \orcid is optional
\affiliation{
  \position{Assitant Professor}
  \department{Computer Science and Electrical Engineering Departments}             %% \department is recommended
  \institution{Stanford University}           %% \institution is required
  \streetaddress{450 Jane Stanford Way}
  \city{Stanford}
  \state{California}
  \postcode{94305}
  \country{USA}                    %% \country is recommended
}
\email{sachour@stanford.edu}         %% \email is recommended
% \affiliation{
%   \position{Position2b}
%   \department{Department2b}             %% \department is recommended
%   \institution{Institution2b}           %% \institution is required
%   \streetaddress{Street3b Address2b}
%   \city{City2b}
%   \state{State2b}
%   \postcode{Post-Code2b}
%   \country{Country2b}                   %% \country is recommended
% }
% \email{first2.last2@inst2b.org}         %% \email is recommended

%% Abstract
%% Note: \begin{abstract}...\end{abstract} environment must come
%% before \maketitle command

\algnewcommand\algorithmicmatch{\textbf{match}}
\algnewcommand\algorithmiccase{\textbf{case}}
\algnewcommand\algorithmicassert{\texttt{assert}}
\algnewcommand\Assert[1]{\State \algorithmicassert(#1)}%

\algdef{SE}[MATCH]{Match}{EndMatch}[1]{\algorithmicmatch\ #1\ \algorithmicdo}{\algorithmicend\ \algorithmicswitch}%
\algdef{SE}[CASE]{Case}{EndCase}[1]{\algorithmiccase\ #1}{\algorithmicend\ \algorithmiccase}%
\algtext*{EndMatch}%
\algtext*{EndCase}%

% https://flatuicolors.com/palette/ca
\definecolor{darkblue}{HTML}{341f97}
\definecolor{deepred}{HTML}{EA2027}
\definecolor{violet}{HTML}{8a2be2}
\definecolor{hotpink}{HTML}{ff69ff}
\definecolor{brickred}{HTML}{c23616}
\definecolor{lightred}{HTML}{ff5e57}
\definecolor{pixelgreen}{HTML}{009432}
\definecolor{skyblue}{HTML}{2e86de}
\definecolor{tangerine}{HTML}{ff9f43}
\definecolor{bluegrey}{HTML}{0a3d62}

\newcommand{\equationmacros}[0]{\footnotesize}
\newcommand{\luke}[1]{\textcolor{blue}{\textbf{LUKE:} #1}}
\newcommand{\todo}[1]{\textcolor{red}{\textbf{TODO:} #1}}

\newcommand{\syn}[1]{\texttt{\textcolor{deepred}{#1}}}
\newcommand{\tool}[0]{\textsc{Heim}}
\newcommand{\algname}[1]{\textsc{#1}}
\newcommand{\fillin}[0]{XXX}

\newcommand{\dtall}{\textbf{\texttt{dt-all}}}
\newcommand{\dtpar}{\textbf{\texttt{dt-par}}}

\newcommand{\code}[0]{c}
\newcommand{\codes}[0]{C}
\newcommand{\codebook}[0]{\mathcal{CB}}
\newcommand{\codebookspace}[0]{Codebooks}
\newcommand{\set}[1]{\{#1\}}
\newcommand{\codetuple}[0]{t}
\newcommand{\codetupleset}[0]{T}
\newcommand{\hwerror}[1]{err(\texttt{#1})}
\newcommand{\hyphen}[0]{\ensuremath{\_}}
\newcommand{\hdplus}{+}
\newcommand{\hdperm}[2]{\rho_{#2}(#1)}
\newcommand{\hdtimes}{\ensuremath{\odot}}
\newcommand{\expressions}[0]{expr}
\newcommand{\expr}{\expressions}
\newcommand{\hvsize}{n}
\newcommand{\hvsetsize}{m}
\newcommand{\meandist}[2]{M(#1,#2)}
\newcommand{\bitflipmeandist}[2]{M'(#1,#2)}
\newcommand{\threshold}[0]{thr}
\newcommand{\thresholdset}[0]{THR}
\newcommand{\accuracy}[0]{acc}
\newcommand{\ev}[1]{E[#1]}
\newcommand{\distance}[2]{dis(#1,#2)}
\newcommand{\tuple}[1]{\langle #1\rangle}
\newcommand{\realnumbers}[0]{Reals}
\newcommand{\size}[1]{SZ(#1)}
\newcommand{\product}[0]{\prod}
\newcommand{\powerset}[1]{2^{#1}}

\newcommand{\textsbf}[1]{{\fontseries{sb}\selectfont{#1}}}
\newcommand{\proseheading}[1]{\vspace{0.2em}\noindent\textsl{\textsbf{#1}}}
\newcommand{\subsubsecheading}[1]{\textbf{#1}}
\newcommand{\bulletheading}[1]{\textit{\textsbf{#1}}}
\newcommand{\codein}[1]{{\small\textbf{\texttt{#1}}}}
\newcommand{\codecap}[1]{\textbf{\texttt{#1}}}

\newcommand{\lit}[1]{\textcolor{darkblue}{\textbf{\texttt{#1}}}}
\newcommand{\swatch}[1]{{\textcolor{#1}{$\blacksquare$}}}
\newcommand{\datapoint}[1]{{\textcolor{#1}{\textbullet}}}
\newcommand{\circled}[1]{{\textcolor{#1}{$\circ$}}}

\newcommand{\tinybullet}[0]{{\tiny\raisebox{1ex}{$\blacktriangleright$}}}
\newcommand{\codecomment}[1]{\textcolor{pixelgreen}{#1}}
\newcommand{\magnitude}[1]{|#1|}

\newcommand{\analyticalmodel}{\textsc{M}}

\newcommand{\myheading}[1]{\textbf{#1}}
\newcommand{\mycodeheading}[2]{\textbf{#1 \codein{[\syn{#2}]}}}

\newcommand{\formulasize}{\footnotesize}

\begin{abstract}
  % background
Binary spatter code (BSC)-based hyperdimensional computing (HDC) is a highly error-resilient approximate computational paradigm suited for error-prone, emerging hardware platforms. In BSC HDC, the basic datatype is a \textit{hypervector}, a typically large binary vector, where the size of the hypervector has a significant impact on the fidelity and resource usage of the computation. Typically, the hypervector size is dynamically tuned to deliver the desired accuracy; this process is time-consuming and often produces hypervector sizes that lack accuracy guarantees and produce poor results when reused for very similar workloads. We present \tool{}, a hardware-aware static analysis and optimization framework for BSC HD computations. \tool{} analytically derives the minimum hypervector size that minimizes resource usage and meets the target accuracy requirement. \tool{} \textit{guarantees} the optimized computation converges to the user-provided accuracy target on expectation, even in the presence of hardware error. \tool{} deploys a novel static analysis procedure that unifies theoretical results from the neuroscience community to systematically optimize HD computations.

We evaluate \tool{} against dynamic tuning-based optimization on 25 benchmark data structures. Given a 99\% accuracy requirement, \tool{}-optimized computations achieve a 99.2\%-100.0\% median accuracy, up to 49.5\% higher than dynamic tuning-based optimization, while achieving  1.15x-7.14x reductions in hypervector size compared to HD computations that achieve comparable query accuracy and finding parametrizations 30.0x-100167.4x faster than dynamic tuning-based approaches.  We also use \tool{} to systematically evaluate the performance benefits of using analog CAMs and multiple-bit-per-cell ReRAM over conventional hardware, while maintaining iso-accuracy -- for both emerging technologies, we find usages where the emerging hardware imparts significant benefits.

% \todo{stopped here, finalize abstract}

% We use \tool{} to optimize 25 parametrized HDC data structures to attain 99\% query accuracy. \tool{} generated HD parametrizations achieve 99.2\%-100.0\% median accuracy, up to 39.3\% higher than dynamic tuning, and also deliver 1.15x-7.27x reductions in hypervector size compared to unoptimized and dynamically tuned hypervectors that achieve comparable query accuracy. Moreover, \tool{} computes parameters 303.0x-100167.4x faster than dynamic tuning-based approaches, and finds parametrizations for all data structures. We also use \tool{} to optimize HDC data structures for multiple-bit-per-cell resistive memory

% \luke{update when the results section are finalized} 

% \tool{}-optimized data structures deliver 1.31x-14.51x reductions in hypervector size and 2.191x-27.27x reductions in memory usage over unoptimized hypervectors that deliver the same accuracy, while attaining 98.96-99.75\% accuracy. \tool{}-optimized data structures deliver up to 41.40\% accuracy improvements over dynamically tuned parameters. \tool{} computes parameters significantly faster than dynamic approaches.
% \luke{\tool{} generated HD parametrization achieves 99.2\%-100.0\% median accuracy, up to 39.3\% higher than dynamic tuning, while delivering 303.0x-100167.4x speed-up over dynamic tuning.}

\end{abstract}

%% 2012 ACM Computing Classification System (CSS) concepts
%% Generate at 'http://dl.acm.org/ccs/ccs.cfm'.


%% End of generated code

%% \maketitle
%% Note: \maketitle command must come after title commands, author
%% commands, abstract environment, Computing Classification System
%% environment and commands, and keywords command.
\maketitle

\SetWatermarkText{Accepted in OOPSLA 23} % Text to be printed across the page
\SetWatermarkScale{0.5} % Size of the watermark text

\section{Introduction}

%\cite{imani2019fach,kim2020geniehd,imani2019quanthd,imani2018hierarchical}
%\cite{jones2007representing,karunaratne2020memory,eggimann20215,schlegel2022hdc,plate2000analogy,bloom1970space,gayler1998multiplicative,kanerva2010we,heddes2022hyperdimensional,pashchenko2020search,schlegel2021multivariate,rahimi2016hyperdimensional,imani2017voicehd,rahimi2018efficient}
%\cite{halawani2021fused,imani2017exploring,imani2019sparsehd,karunaratne2020memory,poduval2021stochd,kleyko2022vector}

Over the years, researchers have developed many emerging memory technologies (e.g., FeRAM, ReRAM, STT-MRAM) that offer non-volatility, better write endurance, and faster write speeds, and support integration into monolithic 3D integrated circuits due to low annealing temperatures.~\cite{halawani2021fused,imani2017exploring,imani2019sparsehd,karunaratne2020memory,poduval2021stochd,wu2018hyperdimensional,rahimi2017high} Moreover, resistive memories, such as ReRAM, have also been used to build analog in-memory computing fabrics that eliminate data movement by performing computation directly within memory cells, and as memory units in monolithic systems that employ other emerging device technologies (CNFETs) to realize extremely communication dense, next-generation hardware platforms. These new technologies offer unprecedented benefits but have not seen broad adoption, as they have much higher bit corruption rates than conventional hardware. These hardware errors often arise due to intrinsic properties in the involved materials, and therefore remain a significant problem despite investments from the devices community.~\cite{shulaker2014monolithic, imani2017exploring} 

\proseheading{Challenges with Approximate Classical Computation.}  Practitioners from the hardware and software communities have developed a range of techniques for statically and dynamically optimizing classical computations to execute reliably on error-prone hardware.~\cite{misailovic2014chisel,achour2015approximate,sharif2021approxtuner} All of these methods grapple with two truths of classical approximate computing: (1) certain bits are essential to the computation and must be retained accurately (e.g., exponent vs. mantissa bits), (2) some compute operations (e.g., branching) need to execute accurately to obtain a usable result. As a result, these techniques typically require computations and data to be partitioned into precise/approximation-amenable regions, where precise data and compute are either run separately on reliable hardware or run with a number of protection mechanisms (e.g., redundancy, ECC) that guard against bit corruptions. These requirements complicate the architectural designs of these platforms and introduce overheads into the computation. These inefficiencies and the relative difficulty associated with statically propagating error through programs without over-approximation make it exceedingly difficult to soundly and efficiently perform computation on emerging hardware.

\subsection{Hyperdimensional Computing / Vector Symbolic Architectures} 

Hyperdimensional Computing (HDC), or alternatively Vector Symbolic Architectures (VSA), is an approximate computing paradigm well-matched to error-prone, emerging computing platforms. The basic unit of data is a hypervector -- a large numeric or binary vector -- that distributes program information evenly across bits/values. There are many variants or dialects of HDC; this work focuses on the binary spatter code (BSC) variant of HDC that works with dense binary hypervectors.~\cite{kanerva1997fully} BSC HDC offers three key computational characteristics which together enable sound and robust approximate computing on emerging technologies: 

\begin{itemize}[leftmargin=*,label={$\tinybullet$}]
\item\bulletheading{Distributed Data Representation.} All hypervector bits are \textit{equally important}, and all bit errors have the same effect on the hypervector result. Therefore, a single-bit flip has the same effect on the computed result, regardless of where it occurs in memory, or within the computational pipeline.

\item\bulletheading{Distance-Based Computation.} Information is encoded over the relative similarity/dissimilarity of hypervectors, where the similarity of two hypervectors is computed with the Hamming distance metric. The Hamming distance is highly resilient to bit errors, as many bit corruptions are required to substantially influence the calculated distance. 

\item\bulletheading{Simple Operators.} The basic HD operators are implemented with circular shift and bit-wise XOR and majority operations. These operators are both hardware-efficient and amenable to analysis.
\end{itemize}

 In contrast, in classical computation, a single bit error can have an outsized effect on a computational result, the sensitivity of the final result to error is workload-dependent, and it is typically difficult to statically analyze the propagation of bit errors through the program without over-approximation.

\proseheading{Applications.} HD computing has enjoyed increased attention in the hardware and software research communities.~\cite{imani2019fach,kim2020geniehd,imani2019quanthd,imani2018hierarchical}  Practitioners have devised HD computations to build a range of data structures, including database records, graphs, trees, and finite-state automata,~\cite{yerxa2018hyperdimensional,osipov2017associative} and to perform a number of processing tasks, including signal and language classification, information retrieval, workload balancing, and analogical reasoning.~\cite{jones2007representing,karunaratne2020memory,eggimann20215,plate2000analogy,kanerva2010we,heddes2022hyperdimensional,pashchenko2020search,kleyko2022vector,kleyko2020autoscaling,rachkovskij2012similarity,simpkin2019constructing,gayler2009distributed} HD computation has also been successfully used in recent years to improve the accuracy and efficiency of edge ML models, and to embed intuition about problem structure into ML training tasks.~\cite{schlegel2022hdc,schlegel2021multivariate,rahimi2016hyperdimensional,imani2017voicehd,rahimi2018efficient,theiss2022unpaired}

% Hypervectors owe their error resilience to the robustness of this distance metric -- if the vector is sufficiently large, it will take many bit corruptions to appreciably change the distance. 
\proseheading{Challenges with VSA/HDC.} One drawback to HD computing is that large hypervectors are usually required to encode information reliably. The hypervector \textit{size}  strongly affects the accuracy of the implemented HD computation and determines the amount of information that can be reliably encoded with the hypervector. Lower dimensional bit-vectors consume less space but potentially reduce one's ability to retrieve information reliably. Typically, practitioners either leave the hypervector size unoptimized or dynamically tune the hypervector size by running Monte Carlo simulations for each parametrization of the target computation.~\cite{kanerva2014computing,kanerva2018computing,montagna2018pulp,kanerva2009hyperdimensional,rahimi2017high} Dynamic tuning is time-consuming and has a tendency to overfit -- the chosen hypervector sizes do not generalize well when minor adjustments are made to the computation.

\subsection{The \tool{} Optimizer}

% ~\footnote{for the remainder of this paper, references to HD computing specifically refer to the BSC variant of HD computing}
We present \tool{}, the first (to our knowledge) static analysis and optimization tool for BSC HD computations. To summarize, the HDC paradigm enables robust computation on error-prone hardware, and \tool{} delivers accuracy guarantees even in the presence of hardware error. Given a hardware error model and an accuracy specification, \tool{} derives the smallest hypervector size that meets the specified accuracy requirements on the target hardware platform:

\begin{itemize}[leftmargin=*,label={$\tinybullet{}$}]

 \item\bulletheading{Analysis.} \tool{} deploys a precise and sound static analysis that guarantees the convergence of the accuracy of the HD computation to the desired accuracy on expectation. The analysis uses several novel theoretical results to soundly derive the expected accuracy for HD computations (see Table~\ref{table:formulas}).
 
 \item\bulletheading{Hardware-Aware.}  \tool{} optimizes HD computations to execute accurately on hardware platforms that use error-prone and emerging device technologies. \tool{}'s analysis procedure works with a hardware error model and delivers accuracy guarantees in the presence of hardware error.
 
\item\bulletheading{Robust Optimization.} The \tool{}-derived parametrization is guaranteed to deliver the desired accuracy for all HD computations captured by the accuracy specification.  \tool{} analytically derives important HDC program parameters, including distance thresholds and HD operation-specific hypervector sizes, which together are used to optimize the computation.
 
\end{itemize}

% \todo{update}

 %We develop four analysis-amenable data structures (sets, graphs, finite-state automata, and analogical databases) and then use \tool{} to optimize these data structures to deliver a target accuracy of 99\% on both conventional and multiple-bit-per-cell memories while minimizing resource usage.\luke{Update} \tool{}-optimized data structures deliver 1.31x-14.51x reductions in hypervector size and 2.191x-27.27 reductions in memory usage while attaining 98.96-99.75\% accuracy. We also compare \tool{} to dynamic parameter tuning and find \tool{}-optimized data structures deliver up to 41.40\% accuracy improvements over dynamic tuning approaches. \tool{}'s analysis is highly efficient, taking just 0.036-9.344 milliseconds to complete depending on the benchmark.

\subsection{Contributions} 

%new theoretical formulations to enable the optimization of winner take all queries, the optimization of queries and data structures containing binding operations, and the optimization of queries in the presence of hardware error. 

\begin{itemize}[leftmargin=*,label={$\tinybullet$}]
\item \bulletheading{\tool{} Accuracy Analysis.} We present a novel accuracy analysis that employs novel theoretical results to derive the expected accuracy for a set of BSC computations on an emerging hardware technology. The accuracy analysis works with an accuracy specification that supports the description of HD computations and their associated accuracy constraints.
\item\bulletheading{\tool{} Optimizer.} We present an algorithm that uses the above accuracy analysis to statically derive thresholds and hypervector sizes that minimize resource usage for a given HD computation while satisfying the accuracy constraints provided in the \tool{} accuracy specification. 
\item \bulletheading{Evaluation.} We evaluate \tool{} against dynamic tuning-based optimization on 25 data structures. Given a 99\% accuracy requirement, \tool{}-optimized computations achieve a 99.2\%-100.0\% median accuracy, up to 49.5\% higher than dynamic tuning, and achieve 1.15x-7.14x reductions in hypervector size compared to iso-accuracy dynamically tuned executions. \tool{} also finds parametrizations 30.0x-100167.4x faster than dynamic tuning-based approaches. We also use \tool{} to find optimized computations at iso-accuracy for two emerging hardware technologies and find usages where the emerging hardware imparts significant benefits over the classical HDC implementation.
\end{itemize}
\section{Hyperdimensional Computing}

Hyperdimensional computing (HDC) is a highly error-resilient brain-inspired computational paradigm. In HDC, information is encoded by computing over randomly generated vectors corresponding to symbols (e.g., letters, colors, objects) in the application domain using binding, bundling, and permutation operations. A hypervector is a numerical vector which may contain binary, modular integer, complex, or real values depending on the HDC. Information is retrieved from HD-encoded data by computing the \textit{distance} ($\distance{hv}{hv'}$) between hypervectors, where hypervectors with small distances are similar. The \textit{distance threshold} ($\threshold$) determines the cutoff point that separates a "small" and a "large" distance.  Because information is evenly distributed across hypervector bits, and distance calculations are resilient to error, HD queries can complete successfully even when bit corruptions occur.

\proseheading{BSC HDC.} This work targets the binary spatter code (BSC) HDC variant, which works with dense binary hypervectors. Random hypervectors are generated by sampling bits from a $p=0.5$ Bernoulli distribution, and permutation, binding, and bundling operations are implemented with circular shifts, bit-wise exclusive-OR (XOR), and bit-wise majority operations, respectively. The Hamming distance  $\distance{hv}{hv'} = \frac{1}{\hvsize} \sum_i hv_i \wedge hv_i'$ is the BSC distance measure, where $\hvsize$ is the  hypervector size. The bit-wise majority operation computes whether there are a majority of "1" or "0" bits at each bit position, and can be alternatively interpreted as an element-wise sum, followed by a thresholding operation.
 
\proseheading{Operators.} The binding and bundling operations ($hv \hdtimes hv'$ and $hv + hv'$) respectively produce hypervectors that are \textit{dissimilar} and \textit{similar} to the input $hv$, $hv'$ hypervectors. The permutation operation $hv' = \hdperm{hv}{i}$ produces a hypervector $hv'$ that is dissimilar to the input hypervector $hv$, where the original hypervector can be recovered by inverting the permutation ($hv = \hdperm{hv'}{-i}$), where $i$ is an integer value. Generally, binding and permutation operations distribute over bundling, and for BSC HDC, bundling/bundling are commutative and associative, and binding is invertible.

\proseheading{Codebooks.} These operations are performed over a "codebook" of \textit{basis hypervectors}, which represent atomics in the problem domain. Examples of codebooks include letters of the alphabet (\textit{K}=26), numerical digits (\textit{K}=10), graph nodes (\textit{K}=\# nodes), and primary colors (\textit{K}=3). Each basis hypervector, or code, in the codebook is typically randomly generated; the associated atomics (e.g., letters) are, therefore, dissimilar from one another. Conceptually, this dissimilarity captures the idea that atomics are distinct -- the letter \codein{A} is distinct from the letter \codein{C}, for example. The HD permutation, binding, and bundling operations are then applied to encode information using these atomics. For example, bundling the "A" and "B" basis hypervectors  ($hv = hv_A + hv_B$) produces a basis hypervector similar to both \codein{A} and \codein{B}.

\subsection{Data Structure/Query Interpretation of HD Computing}

An HD computation can be thought of as a \textit{data structure encoding} operation that produces a \textit{data structure} hypervector \codein{ds} that can be queried by computing its distance from a \textit{query} hypervector \codein{q}.  Because hypervectors are lossy information encodings, a query against an HD data structure may occasionally return an incorrect result -- the \textit{accuracy} of a query is the probability that a query returns the correct result. The hypervector size and the distance thresholds together control the computation's accuracy.

\proseheading{Data Structures.} The bundling operation conceptually creates a set of elements where the membership of an element or subset can be tested with a distance calculation. For example, for the \codein{ds}=$hv_A + hv_B + hv_C \approx \set{A,B,C}$, The membership of a subset  \codein{q}=$hv_A + hv_B \approx \set{A,B}$ is tested by computing the distance $\distance{hv_A+hv_B}{hv_A+hv_B+hv_C}$. If the distance falls below a distance threshold $\threshold$, the set contains the subset; this is referred to as a \textit{match}. A false positive occurs when a data structure hypervector falsely \textit{matches} the query, and a false negative occurs when a data structure hypervector falsely fails to match the query. The binding operation conceptually creates a record of elements ($hv_A \hdtimes{} hv_B$ $\sim$ $\tuple{A, B}$), where each record is only similar to other matching records, and the permutation operation is used to encode positional information into the HD data structure.

Several complex data structures that compose sets, sequences, and records can be built from these basic operations. For example, $hv_A + \hdperm{hv_A}{1}$ builds the sequence $[A, A]$ that can be indexed at index $1$ by computing $\hdperm{hv_A + \hdperm{hv_A}{1}}{-1}$, the $hv_A \hdtimes hv_B + hv_C \hdtimes hv_D$ encoding builds the set of records $\set{\tuple{A, B}, \tuple{C, D}}$ that can be queried with record subsets, and the $hv_A \hdtimes \hdperm{hv_B}{1}$ encoding builds a tuple $\tuple{A, B}$ of ordered elements, such that $\tuple{A, B} \neq \tuple{B, A}$.

\proseheading{Item Memories.} Complex data structures, such as graphs and databases, can be encoded through the use of an \textit{item memory}, a key-hypervector data store that maps identifiers to hypervector item memory rows $hv_i$, which implements HD data structures. For example, an HD graph item memory maps nodes to hypervectors, and each hypervector "value" encodes the set of edges connected to the associated "key" node. Item memory-based data structures support two queries: (1) threshold-based queries and (2) $w$-winner winner-take-all queries. For both queries, the distance between the query hypervector and each item memory row $\distance{q}{hv_i}$ is computed. In threshold-based querying, all item memory rows that match a query hypervector $q$ are returned. In winner-take-all querying, the $w$ item memory rows closest to the query hypervector are returned. Threshold queries require a distance $\threshold$ to operate, which may be individually set for each item memory row, while winner-take-all queries take no additional parameters. See Section~\ref{sec:relatedwork} for more discussion on these two query types.

\section{Illustrative Example: Knowledge Graph}\label{sec:example}

\begin{figure}[t]
\small
\begin{subfigure}[b]{0.3\textwidth}
\centering
\includegraphics[width=0.7\textwidth]{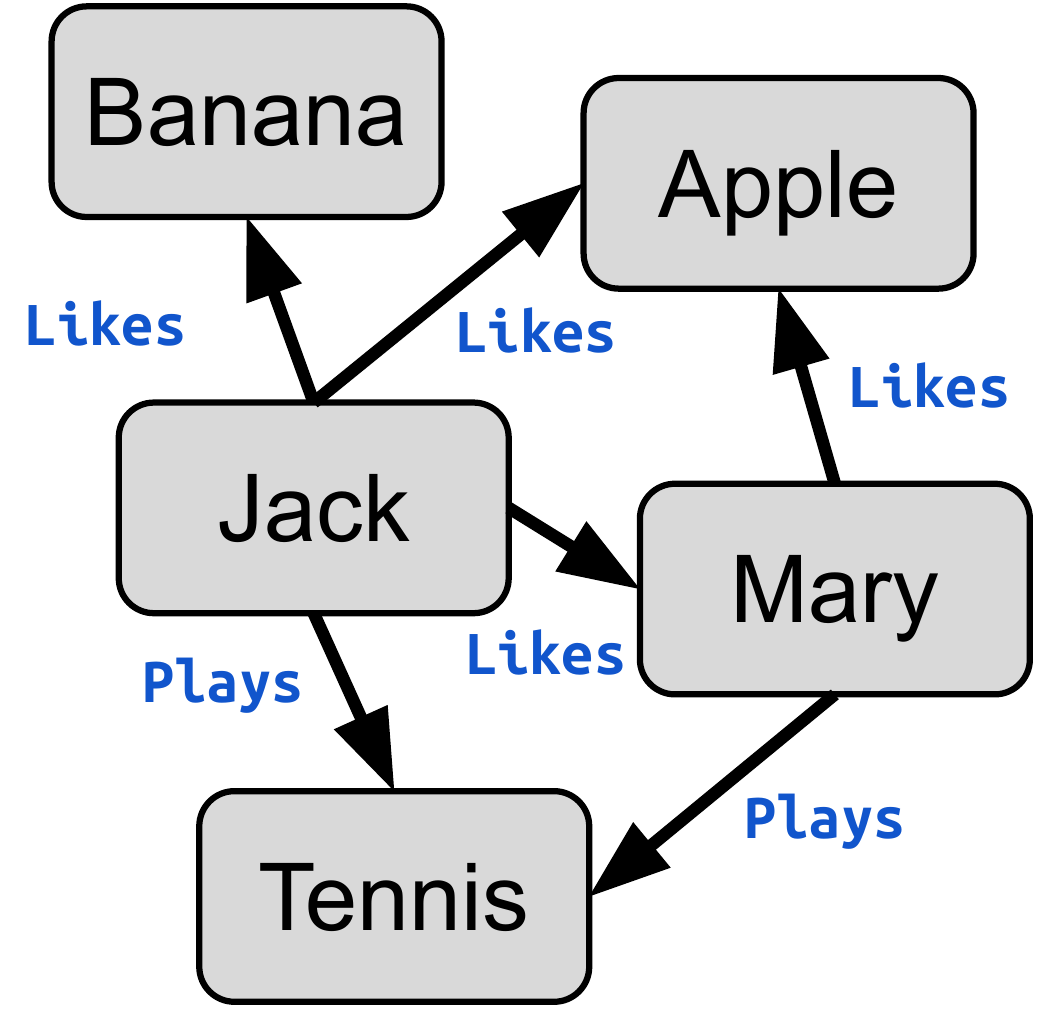}
\caption{Student knowledge graph.}
\label{fig:apple:graph}
\end{subfigure}%
\hfill
\begin{subfigure}[b]{0.67\textwidth}
\footnotesize
\begin{Verbatim}[formatcom=\sffamily,commandchars=\\\{\},numbers=left]
\codecomment{//data structure: one hypervector for each node in item memory}
ds[\syn{jack}] = \lit{act}\hdtimes{}(\lit{likes}\hdtimes{}(\lit{banana}+\lit{apple}+\lit{mary})+\lit{plays}\hdtimes{}\lit{tennis}) \codecomment{//jack}
ds[\syn{apple}] = \lit{target}\hdtimes{}\lit{likes}\hdtimes{}(\lit{jack}+\lit{mary}) \codecomment{//apple}
ds[\syn{mary}] = \lit{act}\hdtimes{}(\lit{likes}\hdtimes\lit{apple}+\lit{plays}\hdtimes\lit{tennis})+\lit{target}\hdtimes{}\lit{likes}\hdtimes{}\lit{jack} \codecomment{//mary}
ds[\syn{tennis}] = \lit{target}\hdtimes{}\lit{plays}\hdtimes{}(\lit{jack}+\lit{mary}) \codecomment{//tennis}
ds[\syn{banana}] = \lit{target}\hdtimes{}\lit{likes}\hdtimes{}\lit{jack} \codecomment{//banana}
query = \lit{act}\hdtimes{}\lit{likes}\hdtimes{}\lit{apples} \codecomment{//how many people like apples query}
\textbf{return} count(\distance{query}{ds[node]} <= \threshold{}[node]) \codecomment{//execute query on graph}
\end{Verbatim}
\caption{HDC encoding of student knowledge graph and \textit{apples} query. }
\label{code:apple:hd}
\end{subfigure}
\caption{Illustrative Example: Knowledge Graph}
\end{figure}

% <this link>
% https://docs.google.com/presentation/d/1-g7ujLwwIUSL0_bD8fCw6NL1goctNN36f-Z6zj56Z0E/edit#slide=id.g18d7b07dae5_0_789
Knowledge graphs capture networks of real-world entities (objects, people, situations), and model relationships between them. An outgoing edge indicates the originating node is \textit{acting} on a target node, and an incoming edge indicates the receiving node is the \textit{target} of another node. Nodes map to concepts (e.g., \lit{apple}, \lit{mary}, \lit{tennis}), and edges are labeled with relations (e.g., \lit{plays}, \lit{likes}, \lit{hates}).

\proseheading{HD Knowledge Graph.} Figure~\ref{code:apple:hd} presents the HD encoding of the student knowledge graph from Figure~\ref{fig:apple:graph} (Lines 2-6). This encoding works with codebooks that specify the relations \codein{\set{likes, plays}}, the interactions \codein{\set{act, target}}, and concepts \codein{\set{jack, mary, banana, apple, tennis}} that may appear in the student knowledge graph. Each node's edge information is then encoded as a hypervector \codein{ds[\syn{node}]} in item memory. A hypervector that encodes each incoming and outgoing edge is constructed by binding together interaction, relation, and concept tuples (e.g., $\codein{\lit{act} \hdtimes{} \lit{likes} \hdtimes{} \lit{apples}}$). The hypervector that encodes the set of edges connected to a given node is constructed by bundling (\codein{+}) all of the edge hypervectors containing the target node together. In BSC, binding distributes over bundling.  

Each edge set hypervector is then stored in an \textit{item memory}. The keys in the knowledge graph's item memory map to concepts, and the hypervectors are the constructed edge list. In this example, the item memory hypervectors are stored in 2-bit-per-cell resistive RAM (ReRAM). This storage medium is 2x denser than conventional binary storage but has a 2.15\% chance of corrupting a bit in memory -- this error rate is collected from a real ReRAM array (Section~\ref{sec:results}). Therefore, the emerging memory may sporadically corrupt bits at random positions in the hypervector data structure.

\proseheading{Queries.} We now want to query the knowledge graph to answer the following question: \textit{"How many students like apples?"}. To answer this question, we would count how many nodes have outgoing edges with the \codein{\lit{likes}} relation label pointing to apple -- this will be referred to as the \textit{apples} query. We want the \textit{apples} query to complete with 99\% accuracy, \textit{even in the presence of hardware error}.

This query can be dispatched on the hypervector representation of the student knowledge graph. We construct the \textit{apples} query hypervector by binding together the relation, concept, and interaction hypervectors together (Line 8) -- this is the same encoding used for the edge in the knowledge graph. We then determine if a node hypervector contains the query tuple by computing the Hamming distance \codein{\distance{query}{ds[\syn{node}]}} between the query hypervector and the node hypervectors in item memory (line 10). Node hypervectors with a distance below some node-specific distance threshold $\threshold$ contain $\tuple{\codein{\lit{act},\lit{likes},\lit{apples}}}$ query tuple in its edge set, and the corresponding key is returned as a match. This query cannot be expressed as a winner-take-all query as the number of matches is unknown.

\subsection{Naive Query Optimization}\label{sec:example:dynamic:tuning}

\begin{figure}[t]
\begin{minipage}{.55\textwidth} %
\begin{subfigure}[m]{0.95\textwidth}
\footnotesize
\begin{algorithmic}[1]
\Procedure{testAcc}{n,accReq,tests,nCbs,nTraces}
\State{data = []}
\For{q,ds,label \textbf{in} tests}
\For{i \textbf{in} \lit{range}(nCbs$\times$nTraces)}
\State{itemMemDist = \lit{execMonteCarlo}(q,ds,n)}
\State{data.\lit{append}($\langle$itemMemDist,label$\rangle$)}
\EndFor{}
\EndFor{}
\State{dists=\lit{sort}(getDists(data))}
\State{thrs = \lit{set}((dist[i]+dist[i+1])/2 \textbf{for} 0...|dists|-1)}
\State{thr,acc = \lit{bruteSearch}(thrs,data)}
\State{\textbf{return} $\langle$thr, acc >= accReq $\rangle$}
\EndProcedure
\Procedure{dynTune}{nMax,accReq,tests,nCbs,nTraces}
\State{fn = \textbf{$\lambda.$}n: testAcc(n,accReq,tests,nCbs,nTraces)}
\State{\textbf{return} \lit{binSearch}(0,nMax,fn)}
\EndProcedure
\end{algorithmic}
\vspace{-0.1in}
\caption{Dynamic tuning algorithm}
\label{algo:naive:tuning}
\end{subfigure}
\end{minipage}%
\begin{minipage}{.39\textwidth} %

\begin{subfigure}[m]{0.95\textwidth}
\footnotesize
\begin{Verbatim}[formatcom=\sffamily,commandchars=\\\{\},numbers=left]
\lit{spec} \{
    \lit{abs-data} query = \lit{prod}(itypes,rels,concepts);
    \lit{abs-data} ds = \lit{sum}(4,\lit{prod}(itypes,rels,concepts));
    \lit{thr-query}(query, ds, 1, 0.99, 0.01, 0.01); \}
\end{Verbatim}
\caption{Knowledge graph specification}
\label{fig:knowspec}
\end{subfigure}

\vspace{3em}

\begin{subfigure}[m]{0.95\textwidth}
\footnotesize
\begin{Verbatim}[formatcom=\sffamily,commandchars=\\\{\}, numbers=left]
\lit{hardware-model} \{
    \lit{mem codebook} = 0.00; \lit{mem item-mem} = 0.0215;
    \lit{op bind} = 0.00; \lit{op bundle} = 0.00; \}
\end{Verbatim}
\caption{Hardware error model}
\label{fig:hwerror}
\end{subfigure}
%\caption{\tool{} specifications.}
\end{minipage}
\caption{Dynamic tuning algorithm and \tool{} specifications.}
\end{figure}

We are interested in \textit{minimizing} the hypervector size to reduce the memory usage while still attaining this 99\% accuracy target. Typically, this is done by dynamically tuning the size and distance thresholds to execute the provided query with acceptable accuracy. Figure~\ref{algo:naive:tuning} presents a dynamic tuning algorithm that finds a minimum hypervector size between zero and \codein{nMax} and the associated distance threshold that achieves a query accuracy of \codein{accReq} over a test set of labelled matching/non-matching HD queries and data structures (\codein{tests}). The algorithm performs a binary search over hypervector sizes. For each size, the algorithm executes each test query and data structure for \codein{$nCbs \times{} nTraces$} Monte Carlo trials to build up the dataset (\codein{data}) of query-item memory distances. The algorithm then uses a brute-force search to find the distance threshold that maximizes the accuracy, or the fraction of correctly classified samples over the constructed dataset. In total, the dynamic tuning algorithm executes $log(\codein{nMax})\times\codein{|tests|}\times\codein{nCbs}\times\codein{nTraces}$ Monte Carlo trials of the computation.

\proseheading{Accuracy.} We parameterize the dynamic tuner with a test set containing the "apples" query and knowledge graph data structures, over 30 random codebooks, and 30 error traces -- in total, 900 trials. The dynamic tuner completes in 35.4 seconds (averaged over ten runs) and finds a hypervector size of \codein{117} bits and threshold of \codein{0.402} that empirically delivers a query accuracy of \codein{99.044\%} on the test set. With these dynamically tuned parameters, we attain an 85.47x reduction in the hypervector size, compared to the unoptimized 10,000-bit hypervector size used in previous literature -- dynamic tuning therefore significantly reduces the memory footprint of the knowledge graph item memory.~\cite{kanerva2009hyperdimensional,kanerva2014computing,kanerva2018computing,montagna2018pulp,rahimi2017high} 
%\footnote{We note the algorithm is unstable and will return different hypervector sizes when each size is evaluated over a few Monte Carlo simulations.}

While this parameterization delivers the desired accuracy for the \textit{apples} query, it does not generalize well to other knowledge graphs or queries of comparable complexity. To demonstrate this, we randomly construct 1000 knowledge graphs containing five nodes with a maximum degree of 4, generate five random edge queries for each graph, and then evaluate the accuracy of the edge queries over 900 trials. Over these 5000 randomly generated data structure-query combinations, we find that the accuracy for 0-degree nodes, 1-degree nodes, 2-degree nodes, 3-degree nodes, and 4-degree nodes are $100.0\%$, $98.8\%$, $98.8\%$, $98.9\%$, $97.6\%$ respectively. Notably, queries over 1-4 degree nodes fail to meet the $99\%$ accuracy target when the dynamically tuned parametrization is used. This issue can be addressed by dynamically tuning over randomly generated queries and data structures; however, doing so will drastically increase parameter tuning time.

\subsection{Optimizing the Apples Query with \tool{}}

With \tool{}, we can \textit{statically} optimize the size and distance thresholds for a given HD computation without performing any simulation. \tool{} delivers precise accuracy guarantees for data structure queries that generally hold, on expectation, over different query and data structure instantiations and can deliver these guarantees even in the presence of hardware error. 

 \proseheading{HD Specifications.}  \tool{} works with an \textit{specification} of the HD computation that describes the accuracy requirements for a set of data structure queries. Figure~\ref{fig:knowspec} presents a \tool{} specification that verifies that all edge queries over 5-node knowledge graphs with a maximum degree of 4 achieve a query accuracy of at least 99\%. The \textit{apples} query on the student knowledge graph presented in Figure~\ref{fig:apple:graph} is an example of a concrete data structure query that adheres to this specification. Line 2 defines a query as a product of interactions (\codein{itypes}), relations (\codein{rels}), and concepts (\codein{concepts}), and Line 3 defines a node hypervector in item memory as a bundle (\codein{\lit{sum}}) of up to 4 edge hypervectors, where each edge tuple is binding of an interaction, relation, and concept. The accuracy assertion on Line 4 requires all tuple queries made to a node hypervector to return the correct result at least 99\% of the time, with maximum false positive (incorrect match) and false negative (incorrect not match) rates of 1\%. See supplementary materials for more information on the analysis-amenable knowledge graph data structure.
 
\tool{} considers the effect of hardware error during optimization and works with a hardware error model that captures the error rates in the target device. Figure~\ref{fig:hwerror} presents the hardware error model for the 2 BPC ReRAM-based accelerator we are targeting in this example; this model defines the per-bit corruption probability for data in item memory as 0.0215. All other operations are error-free.

\proseheading{\tool{}-Optimized Parameters.} We use \tool{} to identify an optimal threshold and hypervector size for the specification in Figure~\ref{fig:knowspec} and the hardware error model in Figure~\ref{fig:hwerror}. \tool{} completes its analysis in 13.58 milliseconds (2606.8x faster than dynamic tuning) and returns a hypervector size of \codein{173} and distance thresholds of \codein{0.4116}, \codein{0.3795}, \codein{0.3795}, and \codein{0.1744} for nodes with degrees 4, 3, 2, and 1. The \tool{}-optimized \textit{apples} query attains an accuracy of 99.944\%. \tool{}, therefore, meets the 99\% accuracy target and delivers a 57.80x reduction in hypervector size over unoptimized 10,000 element vectors, reducing the number of 2 BPC ReRAM cells required to store each node hypervector from 5,000 cells to just 87 cells. The \tool{} hypervectors are \codein{1.48x} larger than the dynamically tuned hypervector size but much more reliably deliver the desired accuracy across data structures.  In fact, \tool{} guarantees that the derived threshold and hypervector size will classify edge queries on node hypervectors (with node degree $\leq 4$) with at least 99\% accuracy on expectation. We evaluate the \tool{}-optimized HD computation over the random knowledge graphs and queries described in Section~\ref{sec:example:dynamic:tuning} and find \tool{} empirically attains an accuracy of $100.0\%$, $100.0\%$, $99.9\%$, $99.9\%$, $98.9\%$ for 0-degree, 1-degree, 2-degree, 4-degree and 5-degree nodes respectively, all close to or higher than the target accuracy of $99\%$. Therefore, while \tool{}'s hypervector size is larger than the dynamically tuned hypervector size, the \tool{}-tuned computation more reliably meets the 99\% accuracy constraint.

\section{\tool{} Specification Languages}

\begin{figure}[t]
\begin{minipage}{0.58\linewidth}
\footnotesize
%\small
\setlength{\tabcolsep}{1pt}
\begin{tabular}{lll}
\multicolumn{3}{c}{$x,x',x'' \in \realnumbers$, $v \in Literals$, $i,k,w,m,t \in Integers$}\\
%$SExpr$ & ::= & $v$ | \syn{perm(}$i$,$SExpr$\syn{)}\\
$SExp$ & ::= & $v$ | \syn{perm(}$i$,$v$\syn{)} | $\syn{prod(}SExp^*\syn{)}$\\
$CExp$ & ::= & $\syn{sum(}i,SExp^*\syn{)}$ | $\syn{prod(}\syn{sum(}i,SExp^*\syn{)}^*\syn{)}$\\
$Expr$ & ::= & $SExp$ | $CExp$\\
%$Expr$ & ::= & 
%$SExpr$ | \syn{iproduct(}$SetExpr*$\syn{)} 
%| \syn{iset(}$Expr*$\syn{)}\\
$Stmt$ & ::= & \syn{thr-query(}$Expr$\syn{,}$Expr'$\syn{,}$k$\syn{,}$x$\syn{,}$x'$\syn{,}$x''$\syn{)}\\
%\syn{require-subset-accuracy(}$Expr$\syn{,}$Expr'$,$i$\syn{,}\syn{,}$x$\syn{,}$x'$\syn{,}$x''$\syn{)} |
&&| \syn{wta-query(}$Expr$\syn{,}$Expr'$\syn{,}$k$\syn{,}$w$\syn{,}$m$\syn{,}$x$\syn{,}$t$\syn{,}$x'$\syn{)} \\&&| \syn{abs-data} v \syn{=} $Expr$\\
$Spec$ & ::= & \syn{spec} $\syn{\{} Stmt^*$\syn{\}} \\
\end{tabular}
\end{minipage}
\hfill
\begin{minipage}{0.41\linewidth}
\footnotesize
%\small
\setlength{\tabcolsep}{1pt}
\begin{tabular}{lll}
\multicolumn{3}{c}{$x \in \realnumbers$}\\
$HOp$ & ::= & \syn{bundle} |  \syn{bind} |  \syn{perm}\\
$MLoc$ & ::= &   \syn{codebook} | \syn{item-mem} | \syn{query}\\
$Stmt$ & ::= & $HOp$ \syn{=} $x$ | \syn{mem} $MemLoc$ \syn{=} x\\
$Mdl$ & ::= & \syn{hardware-model} \syn{\{} $Stmt$* \syn{\}} \\
\end{tabular}
\end{minipage}
\caption{Program grammars - \tool{} accuracy specification language ($Spec$) and hardware model ($Mdl$).}
\label{fig:grammar:speclang}
\end{figure}

The \tool{} accuracy specification language $Spec$ enables practitioners to specify the structure of HD programs to optimize. The specification language supports the specification of \textit{abstract programs} that capture a set of HD computations. The language supports describing HD computations with the following statements:

%\item\textbf{Codebooks.} The \codein{\syn{codebook} v\syn{(}i\syn{)}} defines a codebook with name \codein{v} that contains \codein{i} codes. 

\proseheading{HD Expression.} Each \codein{\syn{abs-data} v \syn{=} Expr} statement maps an HD expression $Expr$ to a variable $v$. The HD expression $Expr$ statically describes the structure of the HD computation to analyze.  We break up HD expressions into simple ($SExp$) and complex ($CExp$) HD expressions. A simple HD expression can be, a code, a permutation of a code, where the basic permutation operator \codein{\syn{perm}} specifies the number of times to apply or unapply the permutation ($i$), or a tuple of (permutation of) codes.
A complex HD expression is either bundle of simple expressions \codein{$\syn{sum(}i,SExp*\syn{)}$}, or binding of several bundles of simple expressions \codein{$\syn{prod(}\syn{sum(}i,SExp*\syn{)}*\syn{)}$}.
All \codein{\syn{sum}} expressions specify the maximum number of hypervectors that will be summed together ($i$) -- this is necessary for \tool{} to complete its analysis.
% \end{itemize}

% \noindent\textbf{Accuracy Constraints.} \tool{} works with the following two types of accuracy constraints.

% \begin{itemize}[leftmargin=*,label=$\tinybullet$]

\proseheading{Thresholding Query Accuracy Constraint.} The \codein{\syn{thr-query(}$Expr$\syn{,}$Expr'$\syn{,}$k$\syn{,}$x$\syn{,}$x'$\syn{,}$x''$\syn{)}} statement imposes the requirement that the thresholding query $|Expr \cap Expr'| \geq k$ produces an accurate result with a probability of at least $x$. Intuitively, this formulation checks that thresholding on $\distance{Expr}{Expr'}$ can determine whether at least \codein{k} elements in the query $Expr$ are contained within the data structure $Expr'$ correctly with a probability of at least $x$. The statement also defines the maximum probability of a false positive ($x'$) and false negative ($x''$) occurrence. 

\proseheading{Winner-take-all Query Accuracy Constraint.} The \codein{\syn{wta-query(}$Expr$\syn{,}$Expr'$\syn{,}$k$\syn{,}$w$\syn{,}$m$\syn{,}$x$\syn{,}$t$\syn{,}$x'$\syn{)}} statement imposes the requirement that a WTA query produces an accurate result with a probability of at least $x$.
Specifically, the WTA query has a query in the form of $Expr$ and $m$ data structures in the form of $Expr'$ in item memory, out of which only $w$ match the query, satisfying that $|Expr \cap Expr'| \geq k$.
The WTA query returns the $w$ in $m$ data structures with the smallest distances to the query.
We define the result as \emph{accurate} when the returned $w$ ones are exactly the $w$ positives.
The statement also specifies a softer constraint that the $w$ true matches are contained within the top $t$ lowest distances to the query ($t\ge w$), with probability $x'$.

\subsection{Hardware Error Model}

\tool{} works with a hardware error model ($Mdl$) that describes the error rates for the basic HD computational operators. The \codein{$HOp$ \syn{=} x} statements define the per-bit error rates for the bundling (\codein{\syn{bundle}}), binding (\codein{\syn{bind}}),  and permutation (\codein{\syn{perm}}) operators. The  \codein{$MLoc$ \syn{=} x} defines the per-bit error rate associated with storing data in item memory, in codebook memory, or the query buffer. The item memory data storage location supports in-memory distance calculations, the query buffer stores the query to apply to item memory, and the codebook memory stores the basis vectors for the codebooks.

%\input{language-example}

%formalization

\section{\tool{} Accuracy Analysis}\label{sec:accuracyanalysis}

\begin{table}
\centering
\footnotesize
\begin{tabular}{|c|c|c|}
\hline
\textbf{Formulation} & \textbf{reference} & \textbf{description}\\ \hline
WTA-$acc$, $w=1$ (\ref{eq:frady:wta:1}) & \cite{frady2018theory} &  WTA accuracy for exactly one winner $w=1$\\
WTA-$acc$ (\ref{eq:wta:k}) & Section~\ref{sec:wta:analysis} & WTA accuracy for more than one winner $w>1$\\
WTA-$prob$ (\ref{eq:wta:prob}) & Section~\ref{sec:wta:analysis} & probability of the $w$ positives being in top $t$.\\
QDS I (\ref{eq:set}) & \cite{kanerva1997fully} & single-element sum-of-product set membership\\
QDS II (\ref{eq:partialsubset}) & \cite{kleyko2016holographic} & subset sum-of-product set membership\\
QDS III (\ref{eq:nwayprod}) & Section~\ref{sec:analysis:qds:3} & single-element product-of-sum set membership\\
Hardware Error (\ref{eq:distr}) & Section~\ref{sec:tonormal} & incorporation of hardware error\\
\hline
\end{tabular}
\caption{Summary of Theoretical Formulations}
\vspace{-0.16in}
\label{table:formulas}
\end{table}

At the heart of \tool{} is a novel static accuracy analysis that derives the query accuracy for threshold-based and winner-take-all data structure queries. The analytically derived accuracy is both precise and sound -- \tool{}'s analysis guarantees that the query under study will converge to the computed accuracy on expectation. This accuracy analysis works with analytical models of the query-data structure distance distributions that are parametrized over hypervector size, the hardware error model, and the query and data structure expressions -- these models are used to analytically derive the accuracy of each type of query. For the \tool{} analysis to be sound, \tool{} requires certain \textit{mutual independence} constraints to hold over the query and dataset. Section ~\ref{sec:model:threshold:intuition}-\ref{sec:model:wta:intuition} overviews the relationship between distance distributions and query accuracy, Sections~\ref{sec:thr:analysis}-\ref{sec:wta:analysis} present the accuracy analysis, and Section~\ref{sec:analyticalmodel} presents the analytical distance distribution models. Table~\ref{table:formulas} summarizes the novel and previously published theoretical results used in this analysis.

\proseheading{\tool{} Optimizer.} The \tool{} accuracy analysis is used by the \tool{} optimizer to find the minimum hypervector size for a given \tool{} specification. The \tool{} optimizer returns a set of query-specific thresholds that can be used to more accurately query from the data structure, and a query-specific hypervector size that can be used to soundly do partial computation. \tool{} also returns a set of query-specific mutual independence constraints that must hold for the above analysis to be fully sound; these constraints can be optionally checked at runtime with the \tool{} mutual independence checking algorithm. Section~\ref{sec:optimizer} presents the \tool{} optimization algorithm, and Section~\ref{sec:dynamic:checker} presents the algorithm for checking that the mutual independence constraints hold over concrete data structures and queries.

\begin{figure}
\begin{subfigure}[b]{0.45\textwidth}
\includegraphics[width=\linewidth]{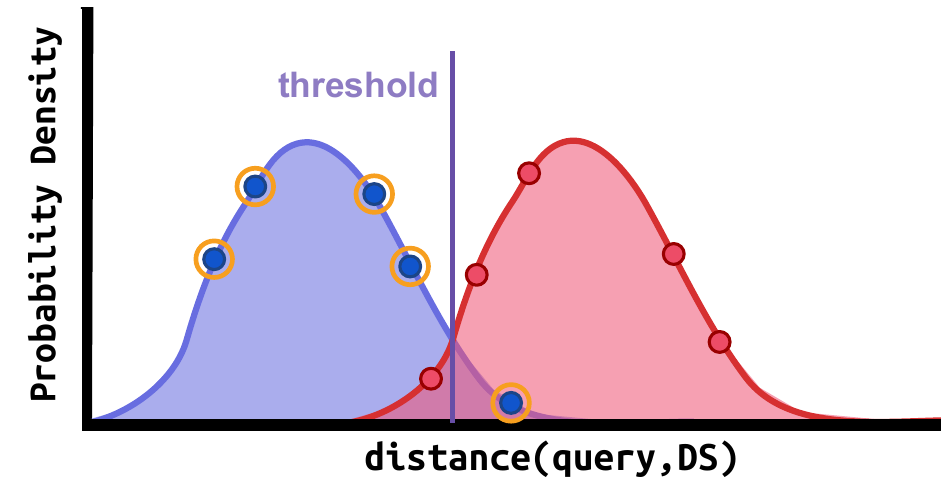}
\caption{Threshold Query}
\label{fig:distribution:thr}
\end{subfigure}
\begin{subfigure}[b]{0.45\textwidth}
\includegraphics[width=\linewidth]{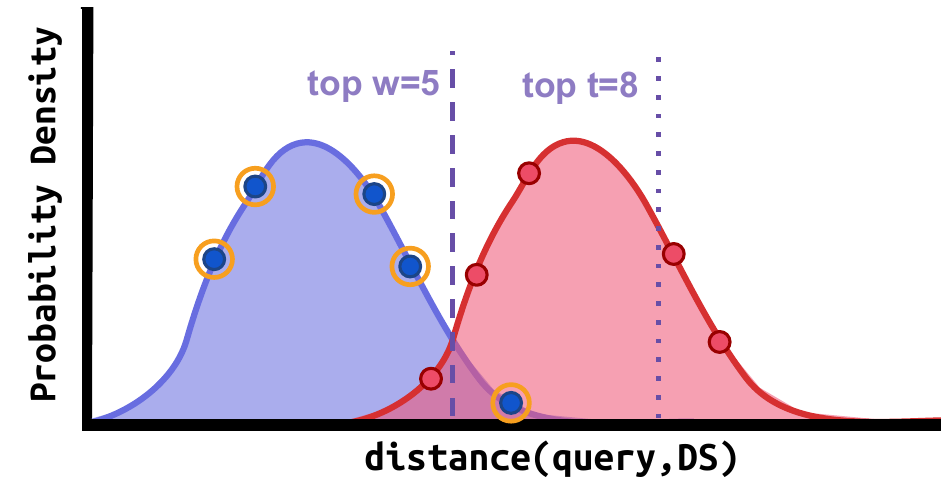}
\caption{WTA Query}
\label{fig:distribution:wta}
\end{subfigure}
\caption{Visualization of WTA/threshold query over match/no-match distance distributions. Points map to sampled match and no-match query-item memory row distances (\datapoint{skyblue} and \datapoint{lightred}) for a 10-element item memory, where match/no-match distances are sampled from match/no-match distance distributions (\swatch{skyblue} and \swatch{lightred}). Circled points (\circled{orange}) map to correct row matches for query.}
\label{fig:distr}
\end{figure}

% \todo{make a pass over this}
% $N(\mu(q,ds),\sigma(q,ds,n,p))$
% $N(\mu(q,ds),\sigma(q,ds,n,p))$

\subsection{Intuition: Accuracy of Threshold-Based Queries}\label{sec:model:threshold:intuition}

Consider the following \tool{} accuracy constraint over a threshold query \codein{thr-query(q,ds,k,x,x',x'')}, where \codein{q} is the query HD expression, \codein{ds} is the HD expression of the rows in a data structure's item memory. Each item memory row is a \textit{match} if it contains at least \codein{k} elements in the query expression, and a \textit{not match} otherwise. Figure~\ref{fig:distr} presents the match (blue) and not-match (red) query-data structure distance distributions. Each distance distribution is normally distributed, with a mean $\mu(q,ds,k,hw)$ that depends on the query and data structure expressions, the number of matching elements, the hardware error model, and a standard deviation $\sigma(\mu(q,ds,k),n)$ is a function of the mean and hypervector size. 

\proseheading{Example.} In Figure~\ref{fig:distr}, the points on the match and not-match distributions correspond to the query-data structure distances for a ten-element item memory with five match rows (blue, circled) and five not-match rows (red), given a sampled set of codebooks and error trace. Each point on the distance distribution corresponds to a distance between the query hypervector and a row hypervector in item memory. In threshold-based querying, points to the left of the distance threshold (grey line) are returned as a match, and points to the right are returned as not a match. The above example correctly returns four of five matches and incorrectly returns one not-matching distance as a match -- corresponding to 1 false positive and one false negative. The \textit{accuracy} of the threshold query corresponds to the probability of correctly classifying each row in the data structure's item memory. In the above example, 8/10 item memory rows are correctly classified, yielding an 80\% accuracy. 

\proseheading{Expected Accuracy.} The associated match and not-match distributions can be analyzed to compute the expected accuracy. The overlapping area between the match and not-match distributions is the probability that a query is erroneously misclassified. Because the distance distributions overlap, there is ambiguity on whether the query matches a given data structure row. The accuracy is, therefore, the probability that a distance is sampled from the non-overlapping regions of the match and not-match distributions. The false positive rate is the portion of the overlapping area left of the chosen threshold, and the false negative rate is the portion of the overlapping area to the right of the selected threshold. The degree of separation between the two distributions depends on how the distributions' mean and standard deviations are parameterized.

\subsection{Intuition: Accuracy of Winner-Take-All Queries}\label{sec:model:wta:intuition}

Consider a winner-take-all (WTA) query accuracy constraint \codein{wta-query(q,ds,k,w,m,x,t,x')} with \codein{w} winners for an \codein{m}-row item memory. The  \codein{q}, \codein{ds}, and \codein{k} arguments correspond to the query and data structure expressions and the number of query elements that must be contained in an item memory row to be considered a match. In \codein{w}-winner WTA queries, there are precisely \codein{w} true matches sampled from the match distribution. We note (1) that within matches sampled from the match distribution, there is no order to sampled matches, and (2) WTA queries are not used to query not-matching elements.

\proseheading{Example.} Figure~\ref{fig:distribution:wta} shows a 5-winner WTA query over a ten-element item memory. The above accuracy constraint requires all true winners $\omega$ (circled) to be contained within the top five lowest distances (left of \codein{top w=5} line) with probability \codein{x}, and all true winners $\omega$ to be contained within the \codein{t=8} lowest distances (left of \codein{top t=8} line) with probability \codein{x'}. Intuitively, the former requirement requires only rows that are true matches to be returned for a WTA query with probability \codein{x}. The latter requirement is a soft requirement that ensures all true matches are contained within the top \codein{t} lowest distances with probability \codein{x'}, where $t \geq k$. The above figure violates the hard constraint since the top 5 lowest distances contain one not-matching row, and satisfies the soft constraint since all true matches are in the top 8 lowest distances.

\proseheading{Expected Accuracy.}  Next, we provide an intuitive explanation of the expected accuracy. First, we draw $w$ distances from the match distribution and $m-w$ distances from the not-match distribution. Intuitively, the expected accuracy of a \codein{w}-winner WTA query corresponds to the probability that all \codein{w} matching distances are to the left of all \codein{m-w} not-matching distances. The probability the soft accuracy requirement is satisfied corresponds to the probability that the top \codein{t} distances contain all \codein{w} matching distances. The WTA distance distributions have the same $\mu$ and $\sigma$ parameters as the threshold-based distance distributions.

\subsection{Threshold-Based Query Accuracy Analysis (\lit{thrAccAnalysis})}\label{sec:thr:analysis}

% \todo{hacky.. we will present endpoints}
% We present how $acc$ and $prob$ are computed in this section.
% We start with $acc$. We denote the WTA query accuracy as $acc(w,m,\Phi_M,\Phi_{NM})$.
Given a hypervector size $n$, a hardware specification $hw$, and a threshold query constraint  \codein{\lit{thr-query}(q, ds, k, reqAcc, reqFp, reqFn)}, the analysis queries the analytical distance model $\analyticalmodel(q,ds,k,n,hw)$ to retrieve the corresponding match and not-match distance distributions $\Phi_M,\Phi_{NM}$ and the associated independence constraint $indepCstr$. \tool{} finds the optimal distance threshold $\threshold_{opt}$ maximizing $\accuracy$ for the given match ($\Phi_M$) and not-match distributions ($\Phi_{NM}$) and upper bounds on false negative $reqFn$ and false positive $reqFp$ rates. The analysis returns the computed threshold, the independence constraint, and whether the derived threshold satisfies the provided accuracy requirements.

\proseheading{Optimal Threshold Derivation.} For a $\threshold$, denoting $fp$ and $fn$ as the false positive and false negative rates, we have ($CDF(f)(x)$ is the cumulative distribution function of distribution $f$ evaluated on $x$)
\begin{equation}
\formulasize{}
fn=1-CDF(\Phi_{M})(\threshold),fp=CDF(\Phi_{NM})(\threshold)
\end{equation}
and with the increase of $\threshold$, $fn$ decreases and $fp$ increases.
Therefore, requirements $fp\le recFp$ and $fn\le recFn$ can be translated into bounds on $\threshold$.
\begin{equation}
\formulasize{}
    CDF(\Phi_M)^{-1}(1-reqFn)=\threshold_l\le \threshold\le \threshold_h=CDF(\Phi_{NM})^{-1}(reqFp)
\end{equation}
If $\threshold_l > \threshold_h$, no $\threshold$ satisfies the requirements and we set $success$ to \codein{False}.
Otherwise, we aim to maximize the accuracy $\accuracy=1-\frac12(fp+fn)$ (assuming balanced positive and negative queries).
\begin{equation}
\formulasize{}
    \max_{\threshold_l\le \threshold\le \threshold_h}\accuracy=\frac12(1+CDF(\Phi_{M})(\threshold)-CDF(\Phi_{NM})(\threshold))
\end{equation}
We take the derivative of $\accuracy$ as follows ($PDF(f)(x)$ is the probability density function of distribution $f$ evaluated on $x$)
\begin{equation}
\formulasize{}
\frac{\partial \accuracy}{\partial \threshold} = \frac12(PDF(\Phi_{M})(\threshold)-PDF(\Phi_{NM})(\threshold))
\end{equation}
We have $\frac{\partial \accuracy}{\partial \threshold}>0$ when $\mu_M<\threshold<x$, and $\frac{\partial \accuracy}{\partial \threshold}<0$ when $x<\threshold<\mu_{NM}$, where $x$ is the intersection of two PDF curves in range $\mu_M<x<\mu_{NM}$ (Figure~\ref{fig:distr}).
Solving $x$ is trivial as $PDF(\Phi_M)(x)=PDF(\Phi_{NM})(x)$ is a quadratic equation of $x$.
Therefore, the optimal $\threshold$ is the point closest to $x$ in range $[\threshold_l,\threshold_h]$, i.e., 
\begin{equation}
\formulasize{}
\threshold_{opt}=\max(\threshold_l,\min(\threshold_h,x))
\end{equation}

The analysis reports a satisfying threshold was found iff $\threshold_{opt}$ achieves the accuracy requirement and returns the optimal threshold and the independence constraint $indepCstr$ on success.

\subsection{Winner-Take-All Query Accuracy Analysis (\lit{wtaAccAnalysis})}\label{sec:wta:analysis}
%\label{sec:wta:acc:compute}

Given a hypervector size $n$, a hardware specification $hw$, and a WTA query constraint,  \codein{\lit{wta-query}(q, ds, k, w, m, x, t, x')}, the analysis queries the analytical distance model $\analyticalmodel(q,ds,k,n,hw)$ to retrieve the corresponding match and not-match distance distributions $\Phi_M,\Phi_{NM}$ and the associated independence constraint $indepCstr$. The algorithm computes the expected hard and soft accuracies ($acc$ and $prob$) from the  $\Phi_M,\Phi_{NM}$ distance distributions and the $w$, $m$, and $t$ WTA query parameters. The expected accuracy is then compared with the provided accuracy requirement ($acc \ge x$, $prob \ge x'$) to determine whether the hypervector size is sufficiently large. On success, the algorithm returns the independence constraint $indepCstr$.

\proseheading{Winner-Take-All Query Accuracy.} We present how the hard accuracy ($acc$) and soft accuracy ($prob$) are computed in this section.
We start with $acc$. We denote the WTA query accuracy as $acc(w,m,\Phi_M,\Phi_{NM})$. Frady et al.~\cite{frady2018theory} developed a perception theory that gives the expected WTA accuracy when $w=1$ as follows
\begin{equation}\label{eq:frady:wta:1}
\formulasize{}
    acc(1,m,\Phi_M,\Phi_{NM}) = \int_{-\infty}^{\infty}PDF(\Phi_M)(x)(1-CDF(\Phi_{NM})(x))^{m-1}dx
\end{equation}
An intuitive explanation of the above equation is that when the one positive vector has distance $x$ to $ds$ (with probability density $PDF(\Phi_M)(x)$), the result is accurate iff the other $m-1$ distractor vectors all have a distance greater than $x$ (each independently with probability $1-CDF(\Phi_{NM})(x)$), and the result is the integral of it for all possible $x$ over distribution $\Phi_M$.

\proseheading{Accuracy with Multiple Winners. }We extend the theory to handle the general cases when $w>1$.
Following the intuition of (\ref{eq:frady:wta:1}), if we have $PDF(MAX_{w,\Phi_M})$, the probability density function of the maximum of distances of $w$ positive vectors, then
\begin{equation}
\formulasize{}
acc(w,m,\Phi_M,\Phi_{NM}) = \int_{-\infty}^{\infty}PDF(MAX_{w,\Phi_M})(x)(1-CDF(\Phi_{NM})(x))^{m-w}dx
\end{equation}
because the results are accurate (i.e., the returned $w$ vectors are all positives) iff the maximum distance of the $w$ positives is no greater than the minimum of the $m-w$ negatives. We then derive $PDF(MAX_{w,\Phi_M})$. First, we have 
\begin{equation}
    \formulasize{}
    CDF(MAX_{w,\Phi_M})(x) = CDF(\Phi_M)^w(x)
\end{equation}
because $MAX_{w,\Phi_M}\le x$ iff all $w$ positive distances are no greater than $x$, each independently with probability $CDF(\Phi_M)(x)$. Then, with the relation of $CDF$ and $PDF$ and the chain rule, we have
\begin{equation}\label{eq:wta:pdf:max}
\formulasize{}
    PDF(MAX_{w,\Phi_M})(x) = CDF(MAX_{w,\Phi_M})'(x) = (CDF(\Phi_M)^w(x))' = wCDF(\Phi_M)^{w-1}(x)\cdot PDF(\Phi_M)(x)
\end{equation}
With (\ref{eq:wta:pdf:max}), we now conclude that
\begin{equation}\label{eq:wta:k}
\formulasize{}
    acc(w,m,\Phi_M,\Phi_{NM}) = \int_{-\infty}^{\infty}wCDF(\Phi_M)^{w-1}(x)PDF(\Phi_M)(x)(1-CDF(\Phi_{NM})(x))^{m-w}dx
\end{equation}

\proseheading{Analysis of Soft Accuracy Constraint.} Besides specifying the desired accuracy, \tool{} also enables users to specify \emph{soft accuracy constraint} for WTA queries, with the following form: with probability $x'$ the distances of the $w$ positives are all in the top-$t$ smallest. In other words, there can be at most $t-w$ in the other $m-w$ vectors in the codebook that have a distance smaller than any positive, with probability $x'$. To formulate this new constraint, we denote $prob(w,m,t,\Phi_M,\Phi_{NM})$ as the probability that $w$ positives all have at least top-$t$ smallest distances. To calculate $prob(w,m,t,\Phi_M,\Phi_{NM})$, we enumerate the number of negatives that have a smaller distance than any positives ($i$), which should be no more than $t-w$.
For each $i$, and $MAX_{w,\Phi_M}=x$, the probability is that exactly $i$ negatives have smaller distances than $x$, each independently with probability $CDF(\Phi_{NM})(x)$, and the other $m-w-i$ negatives have greater distances than $x$, each independently with probability $1-CDF(\Phi_{NM})(x)$.
Therefore, we have
\begin{equation}
\formulasize{}
prob(w,m,t,\Phi_M,\Phi_{NM})=\sum_{i=0}^{t-w}\binom{m-w}{i}\cdot\int_{-\infty}^{\infty}PDF(MAX_{w,\Phi_M})CDF(\Phi_{NM})^{i}(x)(1-CDF(\Phi_{NM})(x))^{m-w-i}dx
% \begin{array}{rl}
%     &prob(w,m,t,\Phi_M,\Phi_{NM})=\\ &\sum_{i=0}^{t-w}\binom{m-w}{i}\cdot\int_{-\infty}^{\infty}PDF(MAX_{w,\Phi_M})CDF(\Phi_{NM})^{i}(x)(1-CDF(\Phi_{NM})(x))^{m-w-i}dx
% \end{array}
\end{equation}

\noindent{}Substituting $PDF(MAX_{w,\Phi_M})$ with (\ref{eq:wta:pdf:max}), we have
\begin{equation}\label{eq:wta:prob}
\formulasize{}
\begin{array}{rl}
    &prob(w,m,t,\Phi_M,\Phi_{NM})=\\ &\sum_{i=0}^{t-w}\binom{m-w}{i}\cdot\int_{-\infty}^{\infty}wCDF(\Phi_M)^{w-1}(x)PDF(\Phi_M)(x)CDF(\Phi_{NM})^{i}(x)(1-CDF(\Phi_{NM})(x))^{m-w-i}dx
\end{array}
\end{equation}

%\newcommand{\binomsum}[2]{S_{#1}^{#2}}
%\newpage

\section{Analytical Model ($M$)}\label{sec:analyticalmodel}

The \tool{} analytical model derives the match distribution parameters $\tuple{\mu_{M},\sigma_{M}}$, the not-match distribution parameters $\tuple{\mu_{NM}, \sigma_{NM}}$ for the provided query, and the mutual independence constraint $indepCstr$ that must hold for the analysis to be valid:

% not clear what input and output are.
\vspace{3pt}
\begin{tabular}{rcl}
$\tuple{MeanDist_{M},MeanDist_{NM},indepCstr}$ & $=$ & $QDS(q,ds,k)$\\
$\tuple{\mu_{M},\sigma_{M}}$ & $=$ & $ToNormal(HwErr(hw, MeanDist_{M}),n)$\\
$\tuple{\mu_{NM},\sigma_{NM}}$ & $=$ & $ToNormal(HwErr(hw, MeanDist_{NM}),n)$\\
\end{tabular}
\vspace{3pt}

Sections~\ref{sec:qds}-\ref{sec:analysis:qds:3} describes how the match and not-match mean distances are derived from the query and data structure ($QDS$), Section~\ref{sec:hwerror} describes how hardware error is incorporated into the error-free mean distance ($HwErr$), and Section~\ref{sec:tonormal} derives the standard deviation from the mean of the same distance distribution ($ToNormal$).

\subsubsection*{Formalization of HD Computation}

A code $\code$ is a randomly generated hypervector that maps to a distinct atomic symbol (e.g., a letter).%, and a codebook $\codebook$ is a collection of codes.
%$\codes \subseteq \codebook$
A code set $\codes$ is a vector that is the superposition (bundle, $\hdplus$) of a set of codes.
We denote $\code \in \codes$ if $\code$ is a code in the set, and $\codes=\set{\code_1,\code_2,\ldots,\code_m}$ if $\codes$ is a superposition of codes $\code_1 \hdplus{} \code_2 \hdplus{} \ldots \hdplus{} \code_m$.
%The $\size{\codes}$ function retrieves the maximum size of the code set $m$.
We denote a code tuple $\codetuple$ as a product ($\hdtimes$) of two or more codes and create code tuples by binding codes, e.g., $\code \hdtimes \code' = \codetuple$. A code tuple set $\codetupleset$ is a set of code tuples.
The analysis works with codes, code tuples, code sets, and code tuple sets. All permutation operators over codes $\hdperm{\code}{k}$ are represented as distinct codes $\code'$ in our formalization. This transformation can be applied because the dependency of $\code$ and $\hdperm{\code}{k}$ does not affect distance computation.

%All permutation operators over codes $\hdperm{\codebook}{i}$ are represented as distinct codebooks $\codebook_{perm,i}$ in our formalization, where $\magnitude{\codebook} = \magnitude{\codebook_{perm,i}}$. This transformation can be applied because each bit of a code $\code$ is independently randomly generated, and thus $\code$ and $\hdperm{\code}{k}$ can be viewed as two independent codes. In other words, the dependency of $\code$ and $\hdperm{\code}{k}$ is negligible for computation and is elided in our formalization -- refer to the Section~\ref{sec:discussion} for more discussion on this simplification.

\subsection{Mutual Independence}\label{sec:mutual:independence}

The \tool{} analytical model assumes that both the data structure and query contain \textit{mutually independent} codes or tuples. In the analytical model, \tool{} identifies the mutual independence constraints that must hold. These mutual independence constraints are dynamically checked when constructing the data structure and query. The specific mutual independence constraint depends on the type of analytical procedure used to perform the analysis.
We next present two types of mutual independence constraints.
We discuss the implications of the mutual independence constraint in Section~\ref{sec:discussion:mutual:independence}, present an efficient dynamic independence checker and prove the equivalence of mutual independence constraints to statistical independence in Section~\ref{sec:dynamic:checker}.

\proseheading{Independent Set} We define independent sets as a set of codes or tuples that are \emph{mutually independent}.
Formally, given any HD expression, it can be flattened into the superposition (set) of code tuples $\codetupleset=\{t_1,t_2,\ldots,t_l\}$ ($\expr=\sum_{i=1}^lt_i$).
%This $\expr$ is an independent set if and only if for any $t_i,1\le i\le l$, its vector cannot be inferred from the vectors of other tuples in the set.
$\codetupleset$ is an independent set if and only if for any $t_i,1\le i\le l$, there exists no subset $\codetupleset'\subset \codetupleset$ other than the subset $\{t_i\}$ that contains only $t_i$, where $t_i$ is the binding of tuples in $\codetupleset'$ (i.e., $t_i=\odot_{t\in \codetupleset'}t$).
For example, $\expr=(a\hdplus b)\hdtimes c\hdplus a\hdtimes b=a\hdtimes c\hdplus b\hdtimes c\hdplus a\hdtimes b$ is not an independent set, because $a\hdtimes b=(a\hdtimes c)\hdtimes (b\hdtimes c)$.
Recalling a tuple in an independent tuple set is similar to recalling a code from a simple code set (they fall into the same QDS type, see Section~\ref{sec:analysis:qds:1}).

\proseheading{Independent Product.} A product of sets $\expr=\hdtimes_{i=1}^ls_i$ is called an independent product if and only if the multiplicant sets are disjoint and the sum of all the multiplicant sets $\sum_{i=1}^ls_i$ is an independent set.
For example, $(a\hdplus b)\hdtimes (c\hdplus d)$ is an independent product because these two sets are disjoint and $a,b,c,d$ are mutually independent.
$(a\hdtimes b\hdplus a\hdtimes c)\hdtimes (b\hdtimes c\hdplus b\hdtimes d)$ is not an independent product because $a\hdtimes b, a\hdtimes c, b\hdtimes c,b\hdtimes d$ are not mutually independent, even though the two multiplicant sets are both independent sets. Note that a product of sets can also be flattened into a set of tuples, and the flattened set is usually not independent, e.g., $(a\hdplus b)\hdtimes (c\hdplus d)=a\hdtimes c\hdplus a\hdtimes d\hdplus b\hdtimes c\hdplus b\hdtimes d$ is not an independent set because $(b\hdtimes c)\hdtimes (b\hdtimes d)\hdtimes (a\hdtimes d) = a\hdtimes c$.
%Therefore, product independence is a weaker constraint than set independence.

\subsection{Query-Data Structure (QDS) Predicates}\label{sec:qds}

 We introduce the concept of a query-data structure (QDS) predicate, a unifying formalization that enables \tool{} to implement an analysis that both leverages theoretical results from previous literature~\cite{kleyko2021survey1,kleyko2022vector}, and the novel derivations (Section~\ref{sec:analysis:qds:3}). A query-data structure predicate is a set membership expression with the formulation $|q \cap ds|\ge k$ for which the symbolic match and not-match mean distances have been analytically derived. The \tool{} analysis supports three forms of QDS predicates (tuple $\codetuple$ and tuple set $\codetupleset$ can also be simply code $\code$ and code set $\codes$):

\vspace{3pt}
\begin{tabular}{ll}
    $|\set{\codetuple} \cap \codetupleset| \ge 1$ & \textbf{Type I, Single Element, Independent Tuple-Set} [Section~\ref{sec:analysis:qds:1}]\\
    $|\codetupleset \cap \codetupleset'| \ge k$ & \textbf{Type II, Subset, Independent Tuple-Set} [Section~\ref{sec:analysis:qds:2}]\\
    $|\set{\codetuple} \cap \odot_i \codetupleset_i| \ge 1$ & \textbf{Type III, Single Element, Independent Product} [Section~\ref{sec:analysis:qds:3}]\\
\end{tabular}
\vspace{5pt}

Type I QDS predicates test if a code/tuple is in an independent code/tuple set, type II QDS predicates test if $k$ elements of a code/tuple set is a subset of an independent code/tuple set, and type III predicates test if a tuple is in an independent product.
Each QDS type is associated with an independence constraint on the data structures.
Type I QDS can be seen as a special case of type II QDS, but it is listed as a separate QDS as it is most frequently used. Note that these 3 QDS predicates systematically cover all cases when the data structure is independent, except for the subset query in the independent product.
Subsets of independent products may have dependent tuples, which are theoretically challenging~\cite{clarkson2023capacity}.
This kind of queries are rarely used in data structures~\cite{kleyko2021survey1,kleyko2022vector}.
We leave incorporating this kind of queries as future work.

\proseheading{QDS Classification.} Given a query and data structure, \tool{} classifies the query into 3 supported QDS predicates by inspecting the data structure and the query expression form.
If the data structure is in the form of product of sums (bound tuple sets), then it must be a type III query and the query must be a tuple.
Otherwise, the data structure must be an independent code/tuple set, in the form of sum or products (bundle of code tuples).
Then if the query is a code set or code tuple set, it must be a Type II query.
If the query is a code or code tuple, then it must be a Type I query.

\proseheading{Mutual Independence Constraints.}
Depending on the QDS Type, \tool{} returns a mutual independence constraint that must hold over the data structure for the corresponding analysis to be valid.
For sum-of-product formed data structure $\expr$, the independence constraints $indepCstr=iset($\expr$)$ requires that $\expr$ is an independent set.
For product-of-sum formed data structure $\expr$, the independence constraint $indepCstr=iproduct($\expr$)$ requires that $\expr$ is an independent product.
The definitions of independent set and product are in Section~\ref{sec:mutual:independence}.

\subsection{Most Common Not-Match Distribution - Independent Vectors}\label{sec:analysis:independent}
We start with the simplest and the most commonly used not-match distribution - the distance distribution between two independent/unrelated vectors $\expr_1$ and $\expr_2$.
We denote the mean distance between two vectors $\expr_1$ and $\expr_2$ as $\meandist{\expr_1}{\expr_2}=\ev{\distance{\expr_1}{\expr_2}}$.
In the following analysis, we assume all the code sets and code tuple sets are of odd size.
The reason is that when bundling even number of vectors, the common practice is to add one more randomly generated vector as an operator to prevent the potential ties for majority~\cite{kleyko2021survey1}.

Because Hamming distance is the average distance across all vector dimensions, the mean distance between two vectors is then the probability of them to differ in any one dimension. Since no correlation exists between two independent vectors, each dimension of one is equally likely (with probability $0.5$) to be 0/1 (same/different) from the perspective of the other. The mean distance is therefore:

\begin{equation}\label{eq:independent}
\formulasize{}
    \meandist{\expr_1}{\expr_2} = \frac12, \forall\ \mathrm{independent}\ \expr_1,\expr_2
\end{equation}

\subsection{Type I (Single Element - Independent Set) Analytical Model}\label{sec:analysis:qds:1}

In this QDS type, the query is a code $\code$/tuple $\codetuple$ and the data structure is a code set $\codes$/tuple set $\codetupleset$.
Since the distance distributions in this QDS are the same for $|\set{\code}\cap\codes|\ge 1$ or $|\set{\codetuple}\cap\codetupleset|\ge 1$, we describe the $|\set{\code}\cap\codes|\ge 1$ case.
The query asks whether $\code\in \codes$.
Suppose $|\codes|=m$.
In the not-match case $|\set{\code}\cap \codes|<1$, i.e., $\code\notin \codes$, the mean distance is (\ref{eq:independent}) as the two vectors have no correlation.
The mean distance of the match case $\code\in \codes$ has been derived by Kanerva~\cite{kanerva1997fully}:

{\formulasize{}
\begin{equation}\label{eq:set}
    \meandist{\set{\code_1}}{\set{\code_1,\code_2,\ldots,\code_m}} = \frac{1}{2} - \frac{\binom{m-1}{\frac{m-1}{2}}}{2^m}
\end{equation}
}

For Type I QDS queries, equations (\ref{eq:set}) and (\ref{eq:independent}) are the match mean distance $MeanDist_M$, and not-match mean distance ($MeanDist_{NM}$) respectively. The independence constraint ($indepCstr$) for Type I queries requires all codes be mutually independent. ($iset{\set{\code_1,\code_2,\ldots,\code_m}}$)

\subsection{Type II (Subset - Independent Set) Analytical Model}\label{sec:analysis:qds:2}

QDS Type II queries are a generalization of Type I QDS.
In this QDS type, the query is a code set $\codes$/tuple set $\codetupleset$, and the data structure is a code set $\codes'$/tuple set $\codetupleset'$.
Again, in this QDS type, the distance distributions are the same for $|\codes\cap \codes'|\ge k$ or $|\codetupleset\cap\codetupleset'|\ge k$, we describe the $|\codes\cap \codes'|\ge k$ case.
Assume $\codes'=\set{\code_1,\code_2,\ldots,\code_m}$, $\codes=\set{\code_1,\code_2,\ldots,\code_n,\code_1',\code_2',\ldots,\code_p'}$ has $l$ ($l\le m$) codes $\code_1,\code_2,\ldots,\code_l$ also in $\codes'$ and the $p$ other codes $\code_1',\code_2',\ldots,\code_p'$ not in $\codes'$.
The mean distance of $\codes$ and $\codes'$ in this case has been derived by Kleyko et al.~\cite{kleyko2016holographic}:

{\formulasize{}
\begin{equation}\label{eq:partialsubset}
    \meandist{\set{\code_1,\code_2,\ldots,\code_l,\code_1',\code_2',\ldots,\code_p'}}{\set{\code_1,\code_2,\ldots,\code_m}} = 1-\frac1{2^{p+m-1}}\sum_{i=0}^{\min(\frac{l+p-1}{2},\frac{m-1}{2})}\binom{l}{i}\sum_{j=0}^{\frac{l+p-1}{2}-i}\binom{p}{j}\sum_{k=0}^{\frac{m-1}{2}-i}\binom{m-l}{k}
\end{equation}}

For Type II QDS queries, equation (\ref{eq:partialsubset}) implements the match mean distance ($MeanDist_M$) when $p=k$, and the not-match mean distance ($MeanDist_{NM}$) when $p=k-1$. The Type II independence constraint ($indepCstr$) for Type II queries requires both subset and set codes be mutually independent ($iset{\set{\code_1,\code_2,\ldots,\code_m}} \wedge iset{\set{\code_1,\code_2,\ldots,\code_l,\code_1',\code_2',\ldots,\code_p'}}$).

\subsection{Type III (Tuple-Set) Analytical Model }\label{sec:analysis:qds:3}
In this QDS type, the query vector is a code tuple $\codetuple$ and the data-structure vector is a product-of-sum formed code tuple set $\codetupleset=\hdtimes_{i=1}^w \codetupleset_i$, i.e., $\codetupleset$ is a binding of several code tuple sets.
We derive for this QDS type because binding code sets is a common operation used in constructing data structures, e.g., analogical database~\cite{kanerva2010we}, finite-state-automata~\cite{pashchenko2020search}.
Although $\codetupleset$ is also a tuple set (it can be flattened to sum-of-product form), this QDS type differs from type I in that the tuples in $\codetupleset$ have dependencies, while QDS type I assumes independence of tuples in the set.
For example, consider $\codetupleset=(\code_1\hdplus \code_2)\hdtimes (\code_3\hdplus \code_4)=\code_1\hdtimes \code_3+\code_1\hdtimes \code_4+\code_2\hdtimes \code_3+\code_2\hdtimes \code_4$.
The first tuple $\code_1\hdtimes \code_3$ is the binding of the other three.
The dependencies make the distance distributions different.

Since this analysis requires that $\codetupleset=\hdtimes_{i=1}^w \codetupleset_i$ is an independent product (Section~\ref{sec:mutual:independence}), enabling us to view tuples in $\cup_{1\le i\le w}\codetupleset_i$ as independent codes, in the following analysis we assume $\codetupleset=\hdtimes_{i=1}^w \codes_i$, i.e. $\codetupleset$ is a product of code sets, and the results generalize to the $\codetupleset=\hdtimes_{i=1}^w \codetupleset_i$ case.
We first consider the simple case where $\codetupleset=\codes\hdtimes \codes'$ is the product of two code sets.
Assume $\codes=\set{\code_1,\code_2,\ldots,\code_l}$ is of size $l$ and $\codes'=\set{\code_1',\code_2',\ldots,\code_m'}$ is of size $m$.
The not-match case is $|\set{\codetuple}\cap \codetupleset|< 1$, i.e., $\codetuple\notin \codetupleset$, then the two independent vectors have mean distance (\ref{eq:independent}).
Otherwise, suppose $t=\code_1\hdtimes \code_1'$, we derive that:

{\formulasize{}
\begin{equation}\label{eq:twowayprod}
\begin{aligned}
    \meandist{\code_1\hdtimes \code_1'}{\set{\code_1,\code_2,\ldots,\code_l}\hdtimes \set{\code_1',\code_2',\ldots,\code_m'}} &= \frac{1}{2^{l+m-2}}\left[\left(\sum_{i=0}^{\frac{l-1}{2}}\binom{l-1}{i}\right)\left(\sum_{i=0}^{\frac{m-3}{2}}\binom{m-1}{i}\right)\right.\left.+\left(\sum_{i=0}^{\frac{l-3}{2}}\binom{l-1}{i}\right)\left(\sum_{i=0}^{\frac{m-1}{2}}\binom{m-1}{i}\right)\right]
\end{aligned}
\end{equation}
}

\noindent{}The derivation is as follows.
Since binding is commutative, we have:

{\formulasize{}
\[
\begin{aligned}
    \distance{\code_1\hdtimes \code_1'}{\set{\code_1,\code_2,\ldots,\code_l}\hdtimes \set{\code_1',\code_2',\ldots,\code_m'}}&=\frac1{\hvsize}|(\code_1\hdtimes \code_1')\hdtimes (\set{\code_1,\code_2,\ldots,\code_l}\hdtimes \set{\code_1',\code_2',\ldots,\code_m'})|\\&=\frac1{\hvsize}|(\code_1\hdtimes \set{\code_1,\code_2,\ldots,\code_l})\hdtimes (\code_1'\hdtimes \set{\code_1',\code_2',\ldots,\code_m'})|\\&=\distance{\code_1\hdtimes \codes_1}{\code_1'\hdtimes \codes_1'}
    %&=\distance{\code_1\hdtimes \set{\code_1,\code_2,\ldots,\code_n}}{\code_1'\hdtimes \set{\code_1',\code_2',\ldots,\code_m'}}
\end{aligned}
\]
}

Therefore, for one dimension of $\codetuple$ and $\codetupleset$ to differ, either in the dimension $\code_1$ and $\codes$ are the same while $\code_1'$ and $\codes'$ differ, or the $\code_1'$ and $\codes'$ are the same while $\code_1$ and $\codes$ differ.
The probability of $\code_1$ and $\codes$ to be the same in a dimension is $\frac1{2^{l-1}}\sum_{i=0}^{\frac{l-1}{2}}\binom{l-1}{i}$ because it requires less than half (at most $\frac{l-1}2$) of $\code_2,\code_3,\ldots,\code_l$ to differ from $\code_1$ in the dimension, and the number of possible choices satisfying this is $\sum_{i=0}^{\frac{l-1}{2}}\binom{l-1}{i}$ and there are $2^{l-1}$ choices for $\code_2,\code_3,\ldots,\code_l$ in total.
For $\code_1$ and $\codes$ to be different in a dimension, there has to be at most $\frac{l-3}{2}$ of $\code_2,\code_3,\ldots,\code_n$ to be the same as $\code_1$ in the dimension, with probability $\frac1{2^{l-1}}\sum_{i=0}^{\frac{l-3}{2}}\binom{l-1}{i}$.
By symmetry, we also get the probability for $\code_1'$ and $\codes'$ to be the same or differ in one dimension.
Combining them together gives (\ref{eq:twowayprod}). Note that the computation of (\ref{eq:twowayprod}) and (\ref{eq:partialsubset}) can be sped up by pre-computing the binomial coefficients and prefix sums of them with Pascal's triangle. More generally, $\codetupleset$ can be the binding of $w,\forall w\ge 2$ sets. Assume $\codetupleset = \hdtimes_{i=1}^w\codes_i$, and $\codes_i=\set{\code_{i1},\code_{i2},\ldots,\code_{il_i}}$ is a code set of size $l_i$.
In this general case, we have:

{\formulasize{}
\begin{equation}\label{eq:nwayprod}
\begin{aligned}
    \meandist{\hdtimes_{i=1}^w\code_{i1}}{\hdtimes_{i=1}^w\set{\code_{i1},\code_{i2},\ldots,\code_{il_i}}} &= \frac{1}{2^{\sum_{i=1}^wl_i-w}}\sum_{\sum_{i=1}^we_i\mathrm{\ is\ odd}}\left[\prod_{j=1}^w\left(\sum_{k=0}^{\frac{l_i-1}{2}-e_i}\binom{l_i-1}{k}\right)\right]
\end{aligned}
\end{equation}
}

\noindent{}The derivation is similar to (\ref{eq:twowayprod}). By commutativity of binding, we have:

{\formulasize{}
\[
\begin{aligned}
    \frac1{\hvsize}|(\hdtimes_{i=1}^w\code_{i1})\hdtimes (\hdtimes_{i=1}^w\set{\code_{i1},\code_{i2},\ldots,\code_{il_i}})|=\frac1{\hvsize}|\hdtimes_{i=1}^w(\code_{i1}\hdtimes \set{\code_{i1},\code_{i2},\ldots,\code_{il_i}})|\\
    %&=\distance{\code_1\hdtimes \set{\code_1,\code_2,\ldots,\code_n}}{\code_1'\hdtimes \set{\code_1',\code_2',\ldots,\code_m'}}
\end{aligned}
\]
}

Therefore, denoting $e_i$ as the value of one dimension of $\codes_{i1}\hdtimes \set{\code_{i1},\code_{i2},\ldots,\code_{il_i}})$ ($0$ or $1$), for $\codetuple\hdtimes \codetupleset$ to be $1$ in the dimension, odd number of $e_i$s should be $1$, i.e., $\sum_{i=1}^we_i$ is odd.
The probability for each $e_i$ to be $0$ or $1$ has been derived for (\ref{eq:twowayprod}).
Adding the probability of all the independent cases gives (\ref{eq:nwayprod}). Note that for large $w$, (\ref{eq:nwayprod}) is non-trivial to compute, as there are exponential number of cases where $\sum_{i=1}^we_i$ is odd.
However, common HDC computations do not involve binding of more than $2$ sets.
We leave the problem of computing (\ref{eq:nwayprod}) more efficiently to future work. For Type III QDS queries, equation (\ref{eq:nwayprod}), or (\ref{eq:twowayprod}) when $w=2$ computes the match mean distance ($MeanDist_M$). Equation (\ref{eq:independent}) computes the not-match mean distance ($MeanDist_{NM}$). The mutual independence constraint ($indepCstr$) for this QDS query is $iproduct(\hdtimes_{i=1}^w\set{\code_{i1},\code_{i2},\ldots,\code_{il_i}})$.

\subsection{Hardware Error-Aware Mean Distance Model ($HwErr(hw,MeanDist)$)}\label{sec:hwerror}

HDC is a suitable computing paradigm for emerging hardware platforms because it is highly resilient against noises in them~\cite{halawani2021fused,imani2017exploring,imani2019sparsehd,karunaratne2020memory,poduval2021stochd}.
\tool{} incorporates the noise present in hardware, which works simultaneously for all the distance distributions above.
We consider the bit-flip error model, where the probability of each bit in hyper-vectors to flip is $p$. Bit flips change the mean distance between two vectors -- we denote $\bitflipmeandist{\expr_1}{\expr_2}$ as the mean distance of two vectors considering possible bit flips:

{\formulasize{}
\begin{equation}\label{eq:bitflip}
    \bitflipmeandist{\expr_1}{\expr_2} = (p^2+(1-p)^2)\meandist{\expr_1}{\expr_2} + 2p(1-p)(1-\meandist{\expr_1}{\expr_2})
\end{equation}}

The derivation is as follows.
$\meandist{\expr_1}{\expr_2}$ is the probability of two vectors to be the same in one dimension, and $\bitflipmeandist{\expr_1}{\expr_2}$ is the probability considering bit flips.
There are two cases where they are the same in one dimension with possible bit flips.
First, they can be the same before possible bit flips with probability $\meandist{\expr_1}{\expr_2}$, and then two vectors either both have a bit flip, or both have no bit flip in the dimension, with probability $p^2+(1-p)^2$.
Second, they differ before possible bit flips with probability $1-\meandist{\expr_1}{\expr_2}$, and a bit flip occurs only to one of the two vectors in this dimension, with probability $2p(1-p)$.

\proseheading{Hardware Errors Increase the Expected Distance Between Vectors.} In all cases we consider, $\meandist{\expr_1}{\expr_2}\le \frac12$.
The maximum $\meandist{\expr_1}{\expr_2}$ is $\frac12$ when $\expr_1$ and $\expr_2$ are unrelated, as shown in (\ref{eq:independent}), and relatedness of vectors makes their mean distance smaller.
We show that possible bit flips increase the mean distances between vectors.
{\formulasize{}
\begin{equation}\label{eq:bitfliplargermean}
    \bitflipmeandist{\expr_1}{\expr_2}-\meandist{\expr_1}{\expr_2} = 2p(1-p)(1-2\meandist{\expr_1}{\expr_2}) \ge 0
\end{equation}
}
Larger mean distance means closer to distribution of unrelated vector, implying loss of relation information encoded in the hypervectors.
The implication is that hardware noise decreases the information resolution, which is intuitive.
We note that information loss increases with $p$ in a reasonable noise range, as in (\ref{eq:bitfliplargermean}) $2p(1-p)$ increases monotonically for $0<p<0.5$.
This enables us to use a upper bound of $p$ in our analysis and deliver a sound accuracy guarantee.

%\subsubsection{\subsubsecheading{Bit-Flip Probability}}\label{sec:inferbfp}
\proseheading{Bit-Flip Probability.}
We derive the bit flip error probability from the hardware specification $hw$.
As a standard practice, raw bit flip error rate is commonly used to characterize hardware noises.~\cite{le2021radar, grossi2019resistive, li2016hyperdimensional}
Given the hypervector operators and memory locations $op \in Op =$ \{\codein{bind}, \codein{bundle}, \codein{codebook}, \codein{item\hyphen{}mem}, \codein{query}\} 
from the hardware specification, we denote the error of an operator as $\hwerror{op}$. We can compute the $p$ as follows:

{\formulasize{}
\[
\begin{aligned}
p &=& 1-[\product_{op \in Op} (1-\hwerror{op})]
%(1-\hwerror{bundle})(1-\hwerror{bind})\times (1-\hwerror{distance})\times (1-\hwerror{permute})\ &&\times (1-\hwerror{codebook}) \times (1-\hwerror{item\hyphen{}mem}) \times (1-\hwerror{query})
\end{aligned}
\]}

$p$ captures the probability of at least one bit flip happens in certain operator or memory location.
Note that $p$ is a probability upper bound of a bit flip occurs in query or data structure during distance calculations, and using $p$ delivers sound accuracy analysis as the information loss increases with $p$ in a reasonable range $0<p<0.5$ (shown in (\ref{eq:bitfliplargermean})).

\subsection{Correspondance between Mean Distance and Distance Distributions ($ToNormal$)}\label{sec:tonormal}

Given a mean distance $\bitflipmeandist{\expr_1}{\expr_2}$ considering bit flips and hypervector size $\hvsize$, we can derive the standard deviation to get the corresponding normal distribution $N(\mu,\sigma)$. Denote $x$ as the value of one dimension in $\expr_1 \hdtimes \expr_2$.
Note that $\distance{\expr_1}{\expr_2}=\frac1{\hvsize}|\expr_1 \hdtimes \expr_2|$.
Since $x$ is either $0$ or $1$, the following holds:
{\formulasize{}
\[
\ev{x^2}=\ev{x}=\bitflipmeandist{\expr_1}{\expr_2},
Var[x]=\ev{x^2}-E^2[x]=\bitflipmeandist{\expr_1}{\expr_2}(1-\bitflipmeandist{\expr_1}{\expr_2})
\]
}
Since all $\hvsize$ dimensions in $\expr_1 \hdtimes \expr_2$ are independent and symmetric, we have
{\formulasize{}
\[
    Var[\distance{\expr_1}{\expr_2}]=\frac1{\hvsize} Var[x]=\frac1{\hvsize} \bitflipmeandist{\expr_1}{\expr_2}(1-\bitflipmeandist{\expr_1}{\expr_2})
\]
}
To sum up, the distance distribution is determined by the mean distance and hypervector size $\hvsize$.
{\formulasize{}
\begin{equation}\label{eq:distr}
    \distance{\expr_1}{\expr_2}\sim N\left(\bitflipmeandist{\expr_1}{\expr_2}, \sqrt{\frac1{\hvsize} \bitflipmeandist{\expr_1}{\expr_2}(1-\bitflipmeandist{\expr_1}{\expr_2}})\right)
\end{equation}
}

\subsection{Discussion on Model Simplifications}\label{sec:discussion:modelsimpl}

\proseheading{Use of Normal Distributions.} We next justify the use of a normal distributions to model match and not-match distances. Recall, the distance metric is the Hamming distance, essentially the average distance in all dimensions. Since all the dimensions are symmetric, the distance of each dimension follows the same distribution. Therefore, the distance is the average of many i.i.d. variables. Furthermore, the hypervectors are long in HD computation, meaning that the number of i.i.d. variables averaged is large. By the central limit theorem, the distance distributions can be well approximated by normal distributions.
In fact, the Kolmogorov–Smirnov difference (supremum of absolute distance) of the cumulative distribution functions (CDF) of the binomial distribution and its corresponding normal distribution is bounded by $\Omega(\hvsize^{-\frac12})$~\cite{nagaev2011bound}. Approximation with normal distributions is also a standard practice by theoreticians in this field~\cite{frady2018theory}, and we note that using binomial distributions to model hypervector distances is computationally expensive (costly to compute PDF and CDF, compared with normal distributions).
Therefore, we view the distance distributions as normally distributed with a standard deviation and mean that both depend on the hypervector dimension, query and data structure sizes, and bit error probability.

\proseheading{Elimination of Permutation Operations.} Because each bit of a code $\code$ is independently randomly generated, each bit of $\code$ and the corresponding bit of $\hdperm{\code}{k}$ are independent, and thus the distance distribution between $\code$ and permutations of it $\hdperm{\code}{k}$ are exactly the same as that of two independently generated codes, as in (\ref{eq:independent}), unless $k$ is equal to $\hvsize$ (number of dimensions) or multiple of $\hvsize$ times.
Therefore, we may treat permuted codes as an independent code of the original codes.

\subsection{Discussion on Mutual Independence}\label{sec:discussion:mutual:independence}

Heim requires the HDC data structure and query to satisfy mutual independence constraints for the analysis to hold.
We show that hypervectors that satisfy mutual independence can represent set, knowledge graph, analogical database, and NFA (see supplementary materials).
%(Section~\ref{sec:datastructures}).
Besides, a number of data structures, including stacks, sequences, and 2D images can be also be represented. These data structures are useful for signal and language classification, information retrieval, workload balancing, and analogical reasoning workloads.~\cite{kleyko2021survey1} In data structures with correlated information, independence can be induced by partitioning the data structure across multiple hypervectors, where each hypervector encodes mutually independent elements. Because hypervector size linearly increases with the number of stored elements (Figure~\ref{fig:storage:set}), the independent sub-hypervectors take up almost the same amount of space. We note this technique does not work well for applications where information loss induced by the bundling operation is a feature, such as feature encoding for machine learning applications, and cannot be used on queries with correlated elements.

%\subsubsection{\subsubsecheading{Correlated Data Structures}}
\proseheading{Correlated Data Structures.}
The analysis of models with correlation is known to be hard and is an open problem in the HDC community.~\cite{clarkson2023capacity} This work establishes a core HDC analysis that is precise and sound. In the future, the analysis can be extended to directly support data structures that have correlations -- these extensions would likely need to use overapproximations or use empirically derived information, and will not deliver the same guarantees as Heim’s core analysis.

\proseheading{Example.} It is possible to implement data structures with correlations and satisfy \tool{}'s mutual independence constraints. Consider the edge set $\{a\odot b, a\odot c, a\odot d, b\odot d, c\odot d\}$ for a 4-node graph. If we encode this set as one vector $G=a\odot b + a\odot c + a\odot d + b\odot d + c\odot d$, the expected distance of vectors of $G$ and $a\odot b$ is $\frac12$ (obtained by enumerating all $2^4$ value combinations), totally indistinguishable from the distance of two independent vectors, although $a\odot b$ is a member of the set. In this case, no threshold can have a $>50\%$ accuracy (random guessing). However, one can decompose a dependent set into a number of independent sets, each stored in one vector. Therefore, instead of storing the graph as a set of all edges $G$, we can store the set of incident edges of each node as one vector, similar to adjacency lists.
This way, each vector is a mutually independent set, and a query falls into the QDS I predicate (Section~\ref{sec:analysis:qds:1}).
%practical tools

\begin{figure}[t]
\footnotesize
\begin{algorithmic}[1]
\Procedure{optimize}{hwModel,heimSpec,maxN}
\State{optDim = 0, queryParams \textbf{=} \set{}}
\For{queryCstr \textbf{in} \lit{getConstraints}\textbf{(}heimSpec\textbf{)}}
\Match{queryCstr}
\Case{queryCstr \textbf{is} \syn{threshold-query}}
\State{success,indepCstr,minDim,thrs = \textbf{thrAccAnalysis}.\lit{binsearch}(hw,0,maxN,queryCstr)}
\State{queryParams = queryParams $\cup$ $\langle$queryCstr,indepCstr,minDim,thrs$\rangle$}
\EndCase
\Case{queryCstr \textbf{is} \syn{wta-query}}
\State{success,indepCstr,minDim = \textbf{wtaAccAnalysis}.\lit{binsearch}(hw,0,maxN,queryCstr)}
\State{queryParams = queryParams$\cup$ $\langle$queryCstr,indepCstr,minDim,[]$\rangle$}
\EndCase
\EndMatch
\State{optDim = \lit{max}\textbf{(}optDim,minDim)\textbf{)}}
\State{\textbf{assert}(success, \syn{"failed to find size that satisfies query accuracy requirements"})}
\EndFor
\State{\textbf{return} $\langle$optDim,queryParams$\rangle$}
\EndProcedure
\end{algorithmic}
\caption{\tool{} accuracy analysis}
\vspace{-0.16in}
\label{algo:opt}
\end{figure}

\section{\tool{} Optimization Framework}\label{sec:optimizer}

%\luke{I'm thinking, maybe we move this whole detailed section into supplementary materials, because the optimizer is just a binary search on top of the accuracy analysis, and the dynamic checker is optional. We can briefly introduce them here (e.g., one paragraph each) and then point to supplementary materials for more details.}
Figure~\ref{algo:opt} presents the \tool{} optimization algorithm. The optimizer takes a hardware error specification (\codein{hwModel}), a \tool{} specification (\codein{heimSpec}), and a maximum hypervector size (\codein{maxN}) as input, and returns both the smallest hypervector size that satisfies all accuracy constraints (\codein{optDim}), query-optimized collection of hypervector sizes, independence constraints, and distance thresholds (\codein{queryParams}) (line 13). \tool{} iterates over each query accuracy constraint in the \tool{} specification, and derives the minimum hypervector size \codein{minDim} required for the query, the mutual independence constraints \codein{indepCstr} that must hold for the analysis to be sound, and a set of distance thresholds to use for threshold-based queries \codein{thr} (lines 6-7, 9-10). The \codein{\textbf{wtaAccAnalysis}.\lit{binsearch}} and \codein{\textbf{thrAccAnalysis}.\lit{binsearch}} routines derive the minimum hypervector size \codein{minDim} and associated thresholds and independence constraints for the given query by performing a binary search over hypervector sizes \codein{0..maxN} and invoking the appropriate accuracy analysis for each candidate size. The final hypervector size returned by is the maximum required dimension across all queries. The \tool{} returns early with an error if the accuracy analysis fails to find an appropriate size for any one query.

\subsection{Dynamic Independence Checker}\label{sec:dynamic:checker}

% \subsection{Mutual Independence Checking Algorithm}\label{sec:dynanalysis:dyncheck}

\tool{} offers an optional dynamic checking algorithm that validates the concrete data structure and query hypervectors that meet all independence constraints associated with the selected query. The independence checker helps users build data structures that satisfy \tool{}'s independence requirements and can be disabled if the user is sure these requirements are met. The algorithm tests if a concrete HD expression contains \textit{mutually independent} tuples or codes. Intuitively, a set of elements is mutually independent if the existence of each set element does not depend on the existence of other set elements.

\proseheading{Algorithm.} We present an efficient dynamic checking algorithm for validating the mutual independence of products and sets of tuples. Since the definition of product independence is derived from set independence, an efficient set independence checker can also check product independence. To avoid checking independence with a time-intensive exhaustive search, we derive an equivalent condition that can be efficiently checked to ensure independence. Given a tuple set $\expr=\sum_{i=1}^nt_i$, denote $c_1,c_2,\cdots,c_m$ as the codes that are a factor of some $t_i$.
For each tuple $t_i,1\le i\le n$ in $\expr$, we create a binary vector $V_i$ of length $m$, where the $j$-th element is $1$ if $c_j$ is a factor of $t_i$, and $0$ otherwise.
The equivalent condition of independence of set $\expr=\sum_{i=1}^nt_i$ is that $V_1,V_2,\cdots,V_n$ are linearly independent in $GF(2)$.
In other words, if we make a $n\times m$ matrix $M$, where the $i$-th row is $V_i$, the equivalent condition is that the $M$ has rank $n$ in $GF(2)$.
For example, for $\expr=a\hdtimes c\hdplus b\hdtimes c\hdplus a\hdtimes b$, if $c_1,c_2,c_3$ are $a,b,c$ respectively, the vectors for $a\hdtimes c,b\hdtimes c,a\hdtimes b$ are $[1,0,1],[0,1,1],[1,1,0]$ respectively.
These $3$ vectors are not linearly independent because $[1,0,1]+[0,1,1]=[1,1,0]$ in $GF(2)$, so $\expr$ is not an independent set.
Intuitively, this is because the binary vector addition in $GF(2)$ represents the binding of tuples, e.g., $1+1=0$ in the previous example's last vector element is because binding $a\hdtimes c$ and $b\hdtimes c$ cancels out $c$.
%One can prove that when $V_i,1\le i\le n$ are linearly independent in $GF(2)$, the $t_i,1\le i\le n$ can be viewed as $n$ independently randomly generated hypervectors.
%We omit the proof here due to the space limit.
Calculating the rank of a $n\times m$ matrix can be done with a $O(n^2m)$ or $O(m^2n)$ Gaussian elimination algorithm.
%Appendix~\ref{sec:dynanalysis:dyncheck}

\proseheading{Correctness.} A binding of tuples corresponds to the sum of their $V$s in $GF(2)$. And for an index set $s\subset \{1,2,\ldots,n\}$, $t_i=\odot_{j\in s}t_j$ is equivalent to a linear equation $V_i+\sum_{j\in s}V_j=\mathbf 0$ in $GF(2)$. Therefore, when $C$ is an independent set,  there exist no such linear equations, which is equivalent to linear independence of $V$s.

%\subsubsection{\subsubsecheading{Connection to statistical independence}}\label{sec:connection:statistical:independence}
\proseheading{Connection to Statistical Independence.}
From the formalism of the independence checker, we can derive that $\codetupleset$ is an independent set that is equivalent to that the bit values of $t_i\in \codetupleset$ are statistically mutually independent. We consider the value of one dimension for all the vectors, and all the other dimensions follow the same arguments. Denote the values of the code vectors as $\mathbf x$, i.e., $\mathbf{x}_i$ is the value of $c_i$'s vector. $\mathbf{x}$ can be any of the $2^m$ values, each with probability $2^{-m}$, as codes are independently randomly generated. For an assignment of the tuple vectors $\mathbf y$, we have $\mathbf{y}=M\mathbf{x}$. There are $2^n$ possible $\mathbf y$s. For each given assignment $\mathbf y$, the probability mass of it is $2^{-m}$ times the number of solutions $\mathbf x$ of equation $M\mathbf{x}=\mathbf{y}$. Since $M$ is rank $n$, it is guaranteed that there is at least one solution, and after Gaussian elimination, the Echelon form has $m-n$ free variables in $\mathbf{x}$, each of which can be either $0$ or $1$. This means that there are exactly $2^{m-n}$ solutions of $\mathbf x$, and thus the probability mass of this assignment $\mathbf y$ is $2^{m-n}\cdot 2^{-m}=2^{-n}$. Thus, the joint distribution of the vector bit values of $t_i$ is a uniform distribution over all possible $2^{n}$ values (each bit is $0$ or $1$, with equal 50\% probability). This is exactly the same as the joint distribution of $n$ randomly generated codes, so we can analyze them as if they were independent codes.

\subsection{Discussion}

\proseheading{Use of Binary Search.} \tool{}'s optimization algorithm exploits the fact that the accuracy of the HD computation increases monotonically with hypervector size to parametrize the HD computation efficiently.
This has been shown theoretically~\cite{kleyko2022vector,frady2018theory,gallant2013representing}, and we also verify it in our evaluation (Figures~\ref{fig:storage:set}-\ref{fig:storage:nfa}).
Because the accuracy is monotonic with respect to hypervector size, we can perform a binary search over hypervector sizes to identify the smallest hypervector size that satisfies a minimum accuracy requirement.

\proseheading{Metadata for Independence Checker.} Many usage patterns involve building a data structure once and then querying the data structure hypervector. Once the data structure is checked for independence and built as hypervectors, no independence metadata about the data structure needs to be stored. Only the subset queries must be checked for independence when the data structure is queried. This can be done with the algorithm we described. We note it may be possible to directly embed this check in the encoding computation, which could be more lightweight.
% data structures
%\input{datastructures}

\section{Evaluation on Error-Free Hardware}\label{sec:results}

\begin{table}
\footnotesize
\centering
\caption{Summary of randomly generated data structure and query characteristics as a function of data structure size ($k$ or $(k,m)$). Non-standard queries are described in grey. WTA queries only support matches. }
\begin{tabular}{|c|c|ccccc|}
\hline
\textbf{benchmark} & \textbf{query type}&\multicolumn{5}{c|}{\textbf{data structure and query sizes}}\\
\hline
set  & threshold & \multicolumn{5}{c|}{$50k$-$100k$ element sets, 1 element/query} \\
db-match & threshold & \multicolumn{5}{c|}{$5k$-$10k$ fields/record, 50-100 records, $m$ fields/query} \\
kgraph & threshold & \multicolumn{5}{c|}{1-$100k$ edges/concept, $100k+10$ concepts, $800k$-$1000k$ edges,} \\
&&\multicolumn{5}{c|}{1 edge/query, 2 relations}\\
nfa & threshold & \multicolumn{5}{l|}{recognizes \textit{str} with length $k$, 1-$k$ character strings/query, 26 letters/alphabet} \\
& \multicolumn{6}{c|}{\cellcolor{lightgray}{\textit{query}: matches are substrings of \textit{str},  non-matches are partial substrings of \textit{str}}}\\
db-analogy & winner-take-all (WTA) & \multicolumn{5}{c|}{$k/2$-$k$ fields/record, $50m$-$100m$ records, 1 analogy query} \\
& \multicolumn{6}{c|}{\cellcolor{lightgray}{\textit{query}: select rows $a$,$b$ where $\tuple{k,v} \in a$, $\tuple{k,v'} \in b$, infer $v$ from item memory and $v'$}}\\
\hline
\textbf{benchmark} & \textbf{size parameters}&\multicolumn{5}{c|}{\textbf{benchmark sizes}}\\
\hline
set& $k$ &1&2&3&4&5\\
db-match& $(k,m)$& (1,2)&(2,4)&(3,6)&(4,8)&(5,10)\\
kgraph&$k$& 1&2&3&4&5\\
nfa& $k$ & 6&8&10&12&14\\
db-analogy&$(k,m)$&(4,1)&(8,2)&(12,3)&(16,4)&(20,5)\\
\hline
\end{tabular}
\vspace{-0.2in}
\label{tbl:benchmarks}
\end{table}

We evaluate \tool{} on five analysis-amenable HDC-based data structures over five different data structure complexities ($k$ or $(k,m)$) -- Table~\ref{tbl:benchmarks} summarizes the complexity of the randomly generated data structures and queries at each size.~\footnote{See supplementary materials for implementation details} For example, the \codein{set-100} benchmark has complexity $k=100$, and is evaluated over random sets containing 50-100 elements and single element queries over the set. Each \tool{} data structure parametrization is evaluated over 100 randomly generated data structures and 20 randomly generated match/not-match queries, where half the queries evaluate to "match". The accuracy of each data structure-query computation is evaluated over ten randomly sampled codebooks. All baseline and \tool{} executions are evaluated over the same randomly sampled data structures, queries, and codebooks to reduce the effect of variance on the evaluation. 

\proseheading{Query Accuracy Metric.}  Given \codein{P} matching query executions and \codein{N} not-matching query executions that produce \codein{TP} true positive and \codein{TN} true negative results, the accuracy of each benchmark is defined as one minus the average of the true positive and true negative rates ($\frac 1 2 (\frac{TP}{P} + \frac{TN}{N})$). We employ a balanced strategy where false positive and negative rates are equally important since the real distributions of positive and negative queries depend on the target applications. \tool{} supports unbalanced false positive and false negative rates, so unbalanced query distributions can also be handled. We also report the accuracy ratio (\codein{rat}) for benchmark applications, corresponding to the percentage of random data structure instantiations satisfying the target accuracy.

\proseheading{Baselines.} We evaluate \tool{}-optimized hypervector size and threshold parametrizations against dynamic tuning-based baselines. These comparisons isolate the accuracy and performance benefits delivered by \tool{} over traditional parameter tuning approaches. Each parameterization is optimized to deliver a target query accuracy of 99\%, all dynamic tuning baselines use the dynamic tuning algorithm from Figure~\ref{algo:naive:tuning}, and all described baselines have error injection disabled:

\begin{itemize}[leftmargin=*,label=\tinybullet]

\item\bulletheading{\codein{\tool{}}:} \tool{} is used to statically optimize the distance thresholds and hypervector size for each benchmark with \tool{} to get 99\% accuracy. For queries composed of sets of elements (\codein{db-match}, \codein{nfa}), \tool{} computes distance thresholds for different query set sizes. 

\item\bulletheading{\codein{dt-par}:} The hypervector size is fixed at 10,000 bits and a single distance threshold is dynamically tuned to attain 99\% query accuracy. WTA queries accept no settable query parameters and, therefore, do not have \codein{dt-par} executions.

\item\bulletheading{\codein{dt-all}:} The hypervector size is dynamically tuned to find the smallest size between 1-100,000 bits that attains a query accuracy of 99\%. For each candidate size, a single distance threshold is dynamically tuned to maximize accuracy. The search saturates at 100,000 bits.

\item\textbf{\codein{dt-hybrid}:} The hypervector size is dynamically tuned similarly to \codein{dt-all}, but \tool{} is used to find the theoretically optimal distance thresholds for each candidate hypervector size. This baseline isolates the effect of using \tool{}-derived thresholds for hypervector queries.

\end{itemize}

\subsection{Query Accuracy Comparison}\label{sec:results:accuracy}

\begin{figure}
\begin{subfigure}[b]{0.28\textwidth}
\includegraphics[width=\linewidth]{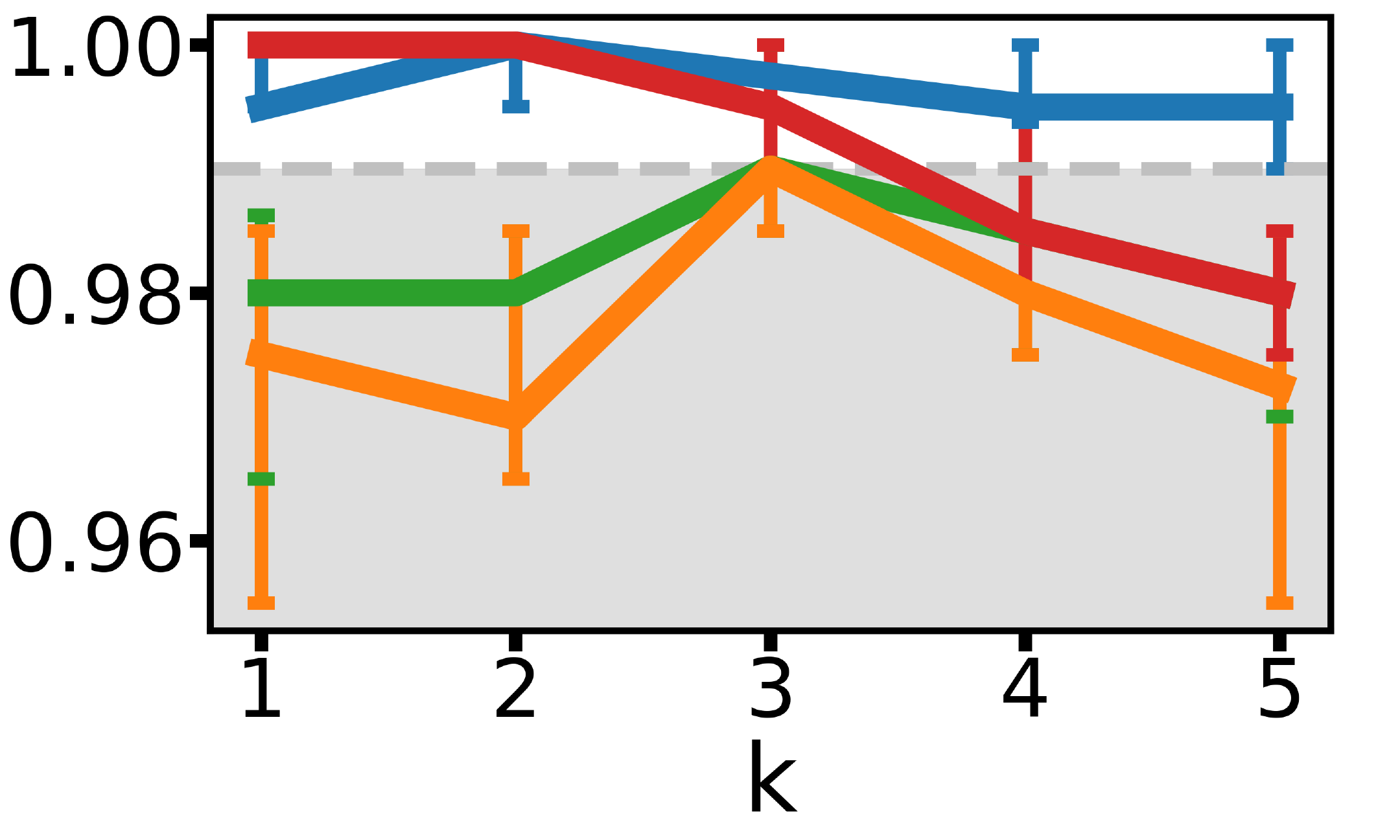}
\vspace{-0.25in}
\caption{\codein{set}}
\label{fig:acc:set}
\end{subfigure}
\begin{subfigure}[b]{0.28\textwidth}
\includegraphics[width=\linewidth]{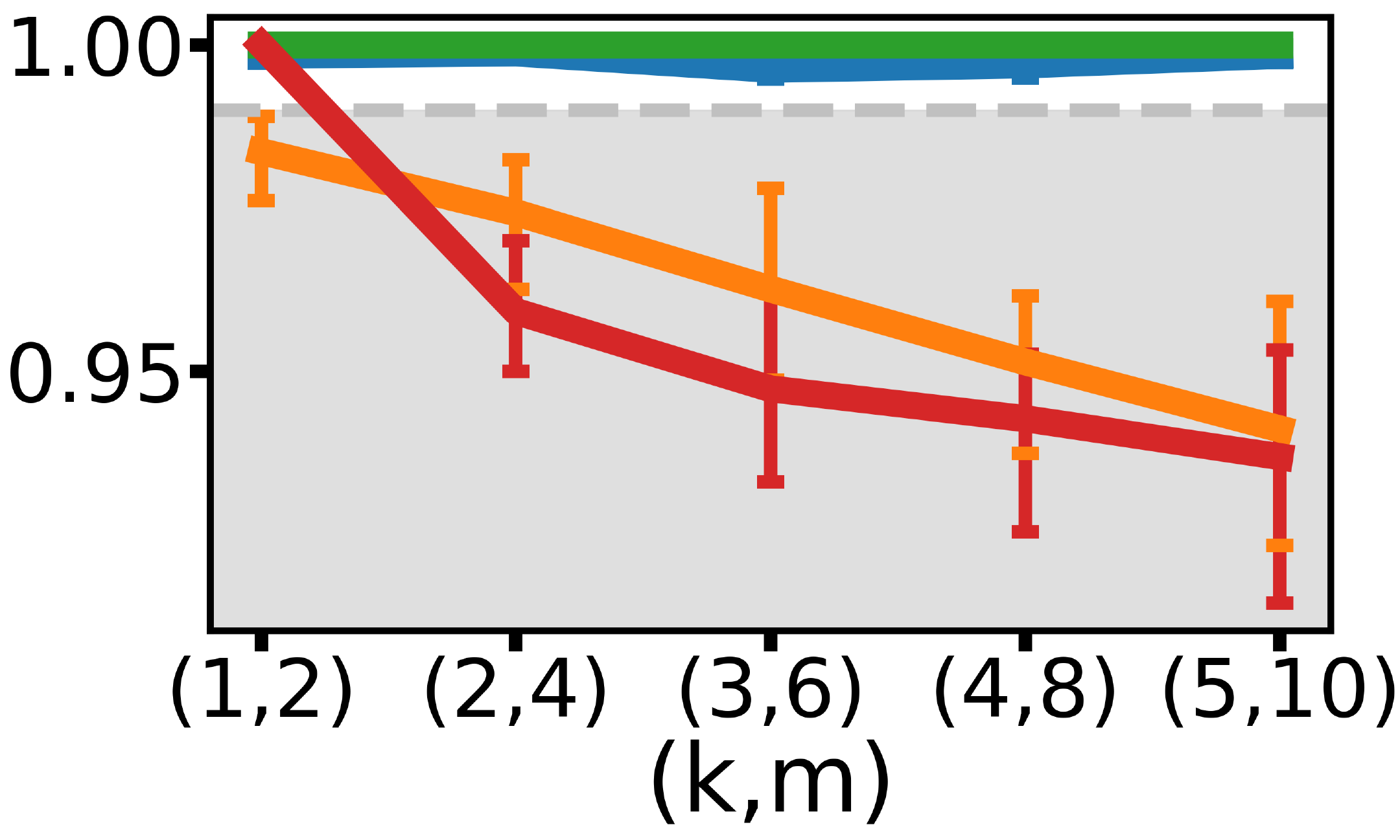}
\vspace{-0.25in}
\caption{\codein{db-match}}
\label{fig:acc:dbmatch}
\end{subfigure}
\begin{subfigure}[b]{0.28\textwidth}
\includegraphics[width=\linewidth]{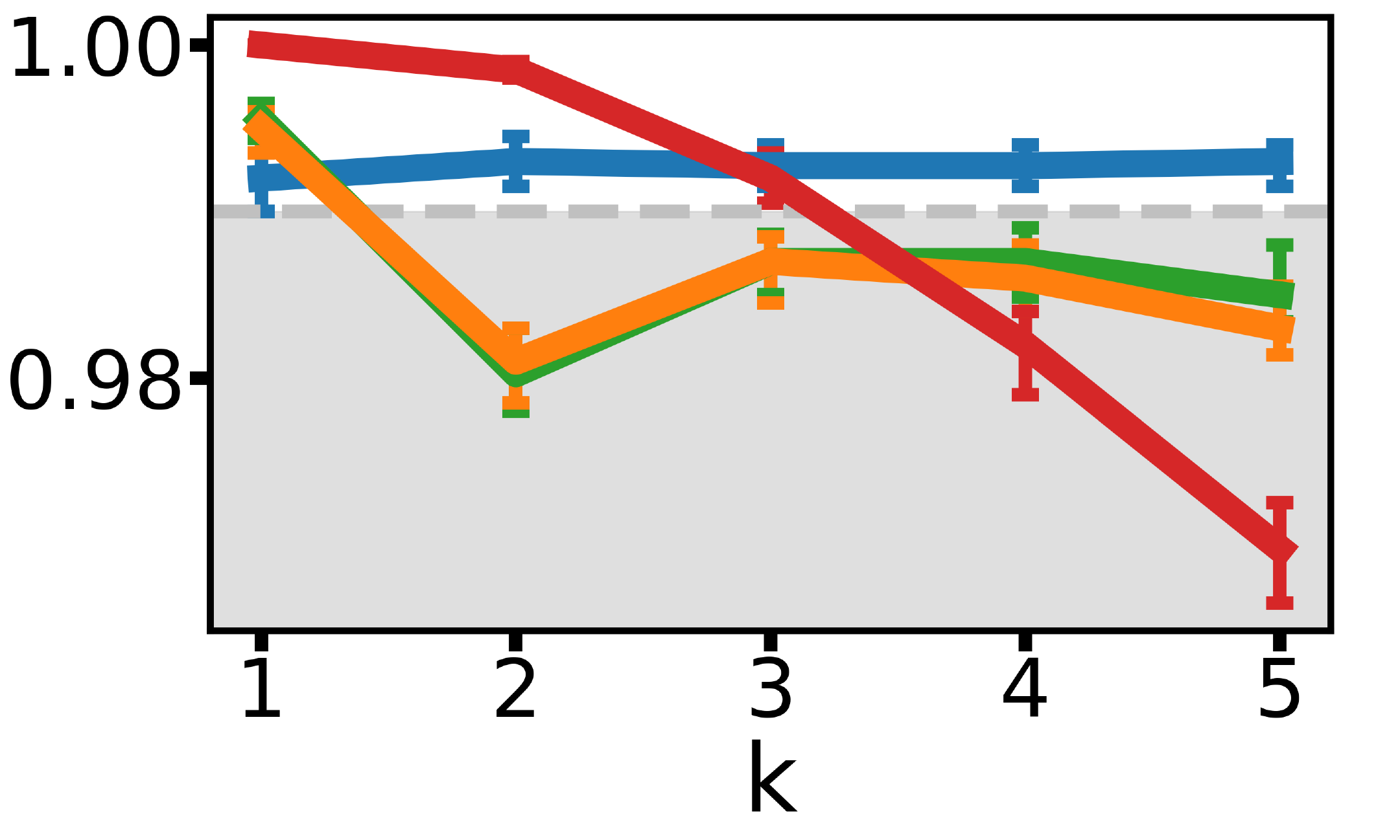}
\vspace{-0.25in}
\caption{\codein{kgraph}}
\label{fig:acc:graph}
\end{subfigure}
\begin{subfigure}[b]{0.28\textwidth}
\includegraphics[width=\linewidth]{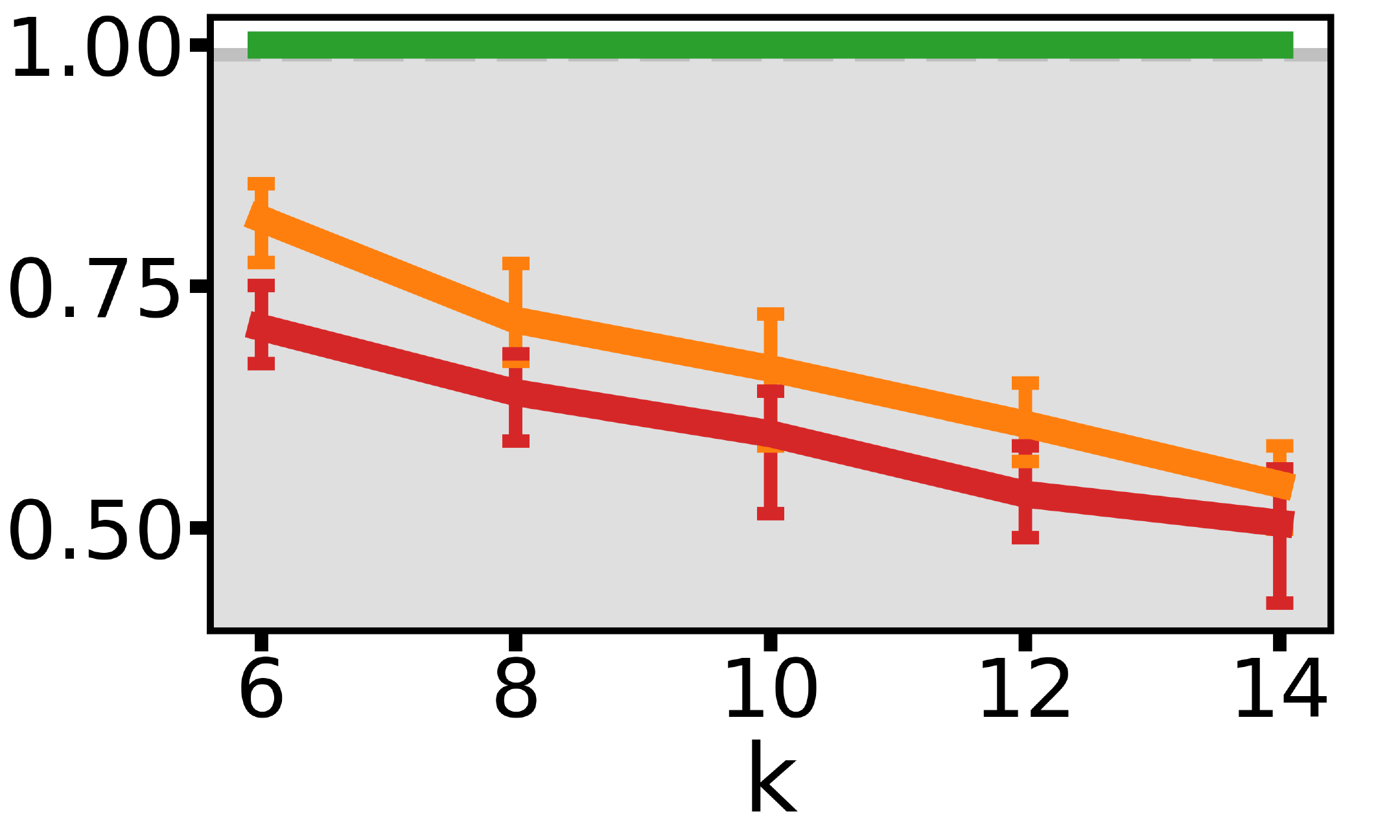}
\vspace{-0.25in}
\caption{\codein{nfa}}
\label{fig:acc:nfa}
\end{subfigure}
\begin{subfigure}[b]{0.28\textwidth}
\includegraphics[width=\linewidth]{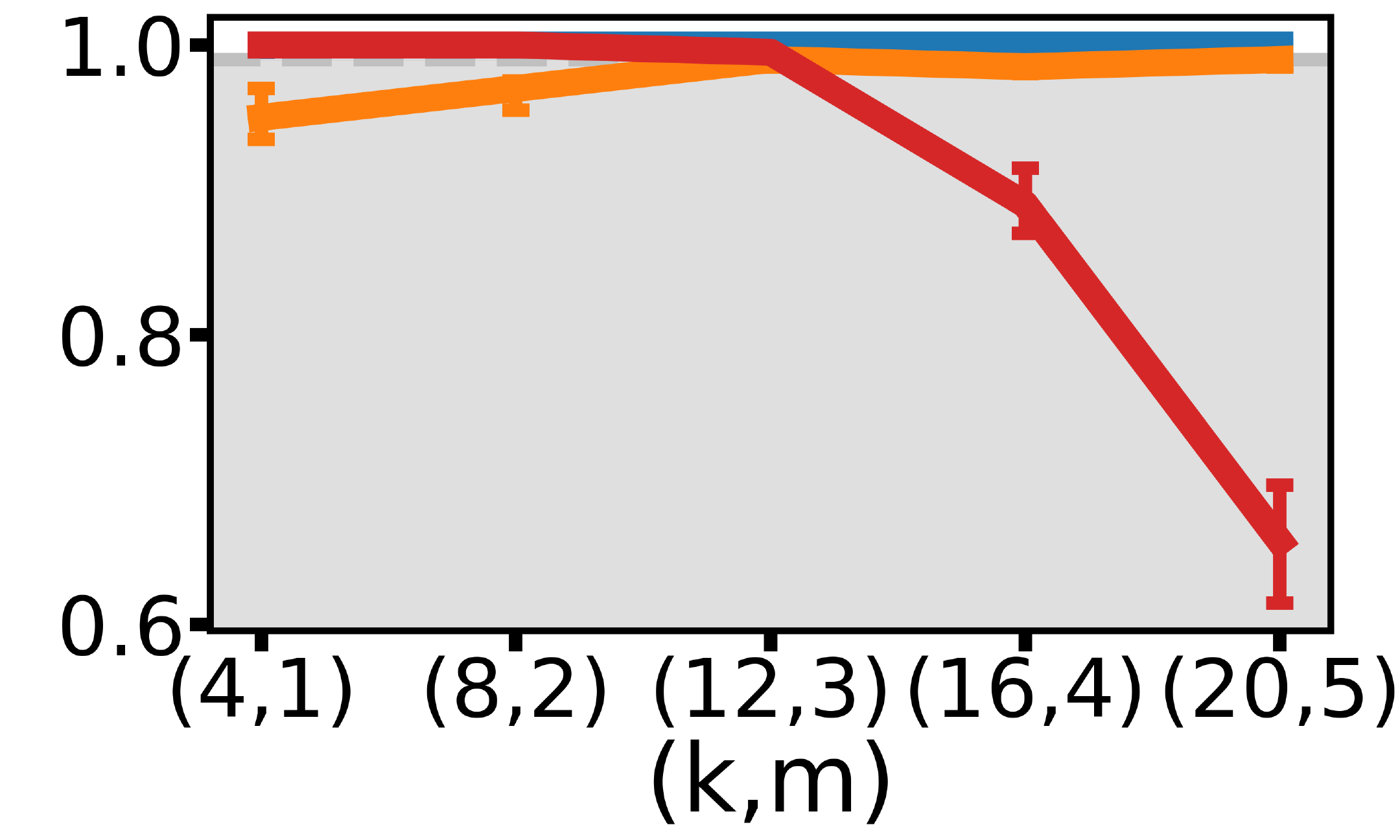}
\vspace{-0.25in}
\caption{\codein{db-analogy}}
\label{fig:acc:dbanalogy}
\end{subfigure}
% \begin{subfigure}[b]{0.73\textwidth}
% \centering
% \small
% \begin{tabular}{|c|ccccc|}
% \hline
%  \textbf{benchmark} & \textbf{set}  & \textbf{db-match} & \textbf{graph} & \textbf{db-analogy} & \textbf{nfa}\\
% \hline
% \codein{\tool{}-nom} & 0.0871 & 0.1891 & 0.0378 & 9.3440 & 0.0361 \\
% \codein{bl-thr} & 15851 & 5049.9 & 1065.7 & 1171.1 & 3464.9 \\
% \hline
% \end{tabular}
% \caption{Parameter optimization times (ms) for the largest problem size in each benchmark.}
% \label{table:runtimes}
% \end{subfigure}
% on 100 random data structures for each benchmark specification.
\caption{Y-axis is median of the reported accuracies (0.0-1.0), error bars are 25\% and 75\% percentiles. Shaded area is below the target 99\% accuracy. \swatch{skyblue} for \codecap{\tool{}}, \swatch{tangerine} for \codecap{dt-all}, \swatch{brickred} for \codecap{dt-par}, \swatch{pixelgreen} for \codecap{dt-hybrid}.}
\label{fig:acc}
\end{figure}

Figure~\ref{fig:acc} compares the query accuracy of \tool{}-optimized programs against the baseline executions. The plot charts the median (timeseries), Q1, and Q3 (vertical bar) for each execution, and the query accuracies that violate the 99\% accuracy requirement are shaded grey. \tool{} achieves 99.2\%-100.0\% median accuracy across all benchmarks -- the required accuracy target of 99\% is therefore met on expectation. Qualitatively, \tool{}-optimized executions have low variance in accuracy (vertical bars) across trials and generally deliver consistent accuracy across different benchmark sizes compared to the dynamically tuned and statically sized baselines. Therefore, \tool{}-optimized data structures reliably satisfy the desired query accuracy targets and consistently deliver the expected accuracy. We note that 80\% of the executed trials exceed the 99\% accuracy target (80\% \codein{rat}) across all benchmarks -- we do not observe adherence to the accuracy constraint 100\% of the time because \tool{}'s guarantees hold on expectation.

In contrast, the dynamically tuned (\codein{dt-all}) baseline delivers 54.5\%-99.5\% median query accuracy, where only four benchmark evaluations (4 points) have median accuracies that meet the accuracy target of 99\%. The \codein{dt-all} benchmark evaluations also experience more significant fluctuations in query accuracy (vertical bars) than \tool{}-optimized HD computations and experience degradations in accuracy as the benchmark size increases, likely because the empirically derived parametrizations do not generalize well, especially as the queries and data structures grow in complexity. Statically fixing the hypervector size to 10,000 bits and dynamically tuning only the hypervector threshold (\codein{dt-par}) attains higher median accuracies than full dynamic tuning and \tool{} when the data structure is small. For 3 of 5 benchmarks, \codein{dt-par} achieves at least 99\% median accuracy for the smaller 1-2 benchmark executions. The median accuracy of \codein{dt-par} evaluations substantially degrades as the size of the data structures increases -- over all executions, \codein{dt-par} optimized programs attain a 50.5\%-100.0\% median accuracy. This phenomenon occurs because threshold-only tuning cannot expand the hypervector size to accommodate hypervectors that implement larger data structures and encode more information.

\proseheading{Hybrid Optimization.} We also evaluate the query accuracy of a hybrid optimization approach (\codein{dt-hybrid}) that uses \tool{} to find thresholds, given dynamically tuned query size. For 24 of the 25 benchmark executions, \codein{dt-hybrid} executions attain median accuracies that are 0.1\%-45.5\% higher than \codein{dt-all}.\footnote{The dt-hybrid execution reports a 0.1\% lower median accuracy than \codein{dt-all} on the remaining execution due to sample randomness.}  Notably, the \codein{dt-hybrid} executions attain substantially better accuracy than \codein{dt-all} on the \codein{nfa} and \codein{db-match} benchmarks, which supports the claim that specializing the distance threshold to the query size is important for queries containing multiple elements. We observe the \codein{db-match} and \codein{nfa} benchmarks use queries containing multiple elements and, therefore, likely work best when the distance thresholds are selected based on the query size. Because the dynamic tuning baselines only tune one threshold, the threshold may not work well across different query sizes. While it is possible to tune multiple thresholds dynamically, this would be prohibitively expensive to do with dynamic tuning.

%In Figure~\ref{fig:acc}, both the \codein{dt-all} and \codein{dt-par} dynamic tuning baselines fail to find a parametrization that meets the accuracy target on the \codein{nfa} benchmark application, while \tool{}  finds a good parametrization with a <100,000 bit hypervector size. 
 
\subsection{Hypervector Size Comparison}\label{sec:results:size}

\definecolor{lightyellow}{HTML}{f6e58d}

\definecolor{salmon}{HTML}{ffb8b8}
\definecolor{fog}{HTML}{d1ccc0}
\newcommand{\ltfifty}[0]{\cellcolor{salmon}}
\newcommand{\lteighty}[0]{\cellcolor{fog}}

\begin{figure}
\footnotesize
\centering
\setlength{\tabcolsep}{1.5pt}
\begin{tabular}{|c|c|c|c|c|c|c|c|c|c|c|c|c|c|}
\hline

 & \textbf{s-100} & \textbf{s-200} & \textbf{s-300} & \textbf{s-400} & \textbf{s-500} & \textbf{m-10-2} & \textbf{m-20-4} & \textbf{m-30-6} & \textbf{m-40-8} & \textbf{m-50-10} & \textbf{k-100} & \textbf{k-200} & \textbf{k-300}\\\hline
\tool{} &  3407 &  6807 &  10208 &  13608 &  17009 &  1376 &  4224 &  8633 &  14601 &  22129 &  3407 &  6807 &  10208\\\hline
\dtpar{} &  0.34 &  0.68 & \lteighty{} 1.02 & \ltfifty{} 1.36 & \ltfifty{} 1.70 &  0.14 & \ltfifty{} 0.42 & \ltfifty{} 0.86 & \ltfifty{} 1.46 & \ltfifty{} 2.21 &  0.34 &  0.68 &  1.02\\\hline
\dtall{} & \ltfifty{} 1.69 & \ltfifty{} 1.76 & \lteighty{} 1.27 & \ltfifty{} 1.63 & \ltfifty{} 1.81 & \ltfifty{} 0.86 & \ltfifty{} 0.04 & \ltfifty{} 0.09 & \ltfifty{} 0.15 & \ltfifty{} 0.22 &  0.87 & \ltfifty{} 1.39 & \ltfifty{} 1.20\\
 & \ltfifty{} 2016 & \ltfifty{} 3865 & \lteighty{} 8049 & \ltfifty{} 8324 & \ltfifty{} 9394 & \ltfifty{} 1604 & \ltfifty{} 100000 & \ltfifty{} 100000 & \ltfifty{} 100000 & \ltfifty{} 100000 &  3909 & \ltfifty{} 4908 & \ltfifty{} 8496\\\hline
 & \textbf{k-400} & \textbf{k-500} & \textbf{n-6} & \textbf{n-8} & \textbf{n-10} & \textbf{n-12} & \textbf{n-14} & \textbf{a-4-100} & \textbf{a-8-200} & \textbf{a-12-300} & \textbf{a-16-400} & \textbf{a-20-500}&\\\hline
\tool{} &  13608 &  17009 &  3594 &  6517 &  10387 &  15235 &  21088 &  1509 &  6119 &  14157 &  25804 &  41186&\\\hline
\dtpar{}  & \ltfifty{} 1.36 & \ltfifty{} 1.70 & \ltfifty{} 0.36 & \ltfifty{} 0.65 & \ltfifty{} 1.04 & \ltfifty{} 1.52 & \ltfifty{} 2.11 &  0.15 &  0.61 &  1.42 & \ltfifty{} 2.58 & \ltfifty{} 4.12&\\\hline
\dtall{}  & \ltfifty{} 1.21 & \ltfifty{} 1.29 & \ltfifty{} 0.04 & \ltfifty{} 0.07 & \ltfifty{} 0.10 & \ltfifty{} 0.15 & \ltfifty{} 0.21 & \ltfifty{} 2.22 & \ltfifty{} 1.96 & \lteighty{} 1.52 & \ltfifty{} 1.62 & \lteighty{} 1.51&\\
 & \ltfifty{} 11254 & \ltfifty{} 13195 & \ltfifty{} 100000 & \ltfifty{} 100000 & \ltfifty{} 100000 & \ltfifty{} 100000 & \ltfifty{} 100000 & \ltfifty{} 681 & \ltfifty{} 3125 & \lteighty{} 9318 & \ltfifty{} 15897 & \lteighty{} 27241&\\\hline
\end{tabular}
\caption{Baselines' hypervector size in each benchmark. \codecap{s}, \codecap{m}, \codecap{k}, \codecap{n}, \codecap{a} abbreviate \codecap{set}, \codecap{db-match}, \codecap{kgraph}, \codecap{nfa}, \codecap{db-analogy} respectively. Row 1, 4 present  \codecap{dt-par}, \codecap{dt-all} hypervector sizes (in bits), and rows 2,3 present ratio of \tool{} size to \codecap{dt-all} and \codecap{dt-par} size. \codecap{dt-par} uses hypervector size $10000$ in all benchmarks. \swatch{salmon} and \swatch{fog} cells have a \codein{rat} of less than 50\% and 80\% respectively.}
\label{table:dimensions}
\end{figure}

Figure~\ref{table:dimensions} compares the hypervector sizes of \tool{}-optimized executions against dynamically tuned (\codein{dt-all}) and statically sized (\codein{dt-par}) executions. The red-shaded cells fail to meet the 99\% median accuracy requirement, and the grey shaded cells meet the 99\% accuracy requirement less than 80\% of the time. For the 4 of 25 \codein{dt-all} benchmark executions which achieve 99\% median accuracy, \codein{dt-all} finds 1.27-1.52x smaller hypervector sizes than \tool{} for three executions, and a 1.15x larger hypervector size than \tool{} for one execution. Of the benchmarks that do not meet the accuracy requirement, 9 of the 25 executions max out at the largest hypervector size. Therefore, dynamic tuning (\codein{dt-all}) typically produces smaller hypervectors than \tool{} when it can find a parametrization that meets the desired accuracy target, as \tool{} employs a conservative strategy, but \codein{dt-all} is rarely able to find a parametrization that reliably delivers a 99\% query accuracy on average. In contrast{}, \tool{} chooses larger hypervectors than dynamic tuning but reliably meets the accuracy target on expectation with low variance in all cases. In cases where dynamic tuning cannot meet the target query accuracy (size=100k), the dynamic tuning algorithm selects hypervector sizes 4.55-25x larger than \tool{}. \tool{} can analytically derive smaller hypervectors by searching over a larger parameter space that includes thresholds and sizes tailored to specific queries and data structures.

For the 10 of 25 \codein{dt-par} benchmark executions which achieve 99\% median accuracy, \tool{} finds hypervectors that are 1.47x-7.14x smaller than 10k bits for 7 of 10 executions, and finds hypervectors that are 1.02x-1.42x larger than 10k bits for 3 of 10 executions, where 10k is the statically configured hypervector size. We also find \codein{dt-par}'s accuracy degrades for larger benchmarks and selects unnecessarily large hypervectors for small benchmark executions. These issues arise because \code{dt-par} cannot flexibly adjust the size to accommodate larger or smaller benchmark executions. Therefore, \tool{} more consistently meets the target accuracy requirement than static allocation strategies while also delivering space savings for executions that can execute with smaller numbers of bits.

%Figures~\ref{fig:acc} and Figure~\ref{table:dimensions} compare the query accuracy and hypervector size of \codein{\tool{}}-optimized benchmarks against the \codein{dt-par}, \codein{dt-all}, \codein{dt-hybrid} dynamically tuned baselines, and Figure~\ref{fig:runtime} presents the runtimes of \tool{}-based and dynamic tuning optimization approaches. For each benchmark evaluation, we optimize a benchmark with a given size using either dynamic tuning or \tool{}, and then report the accuracy over 20 random queries (200 monte carlo trials). The median, Q1, and Q3 accuracy for each benchmark evaluation map to points and vertical bars in the accuracy figure, and the optimized hypervector size for each benchmark evaluation and maps to a cell in the hypervector size table. All benchmark executions work with a query accuracy target of 99\%. Any \codein{dt-all} executions that cannot satisfy the accuracy target terminate with the maximum allowed vector size of 100,000, and a single distance threshold is computed for all dynamic tuning baselines. 
 
\subsection{Optimization Time Comparison}\label{sec:results:runtime}

\begin{figure}
\begin{subfigure}[b]{0.28\textwidth}
\includegraphics[width=\linewidth]{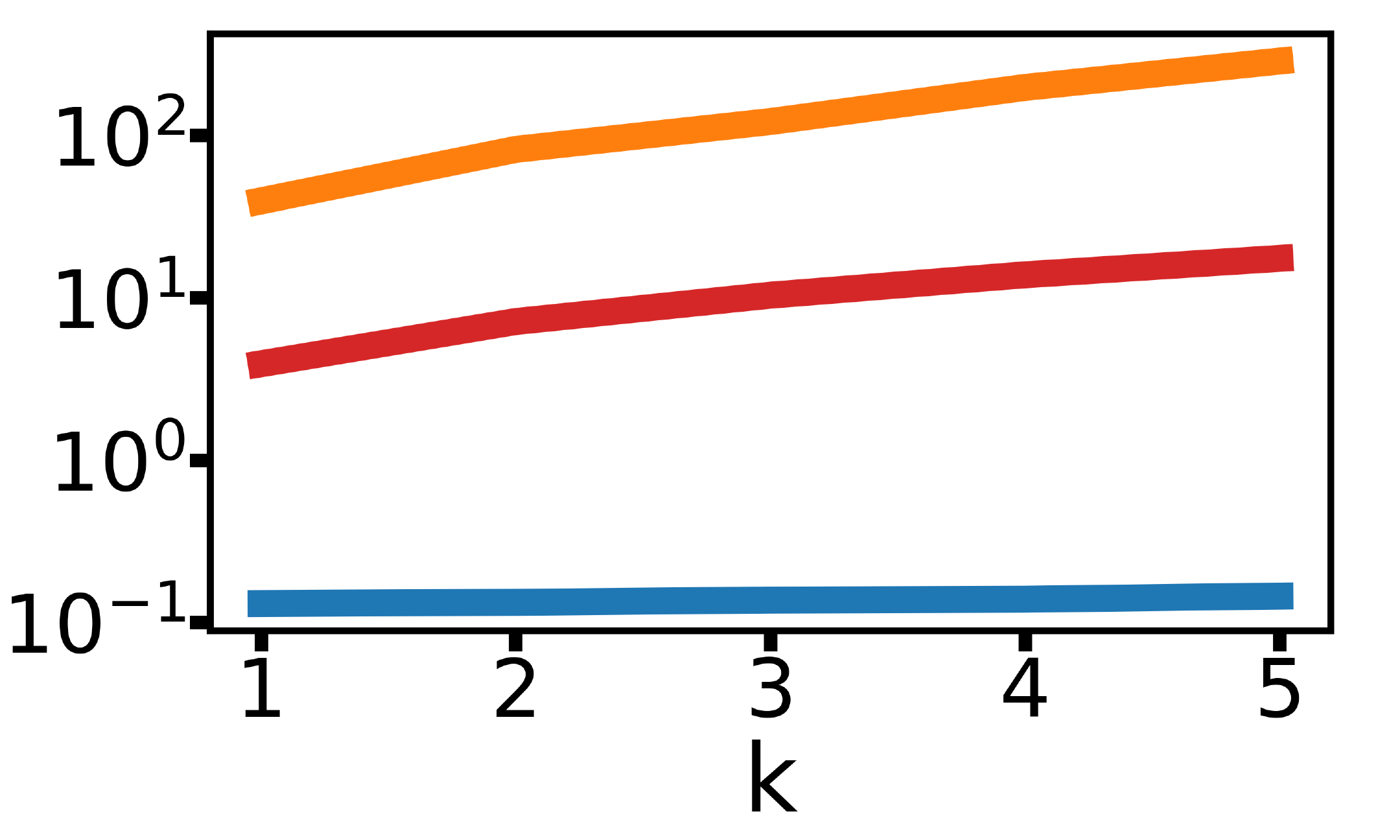}
\vspace{-0.25in}
\caption{\codein{set}}
\label{fig:runtime:set}
\end{subfigure}
\begin{subfigure}[b]{0.28\textwidth}
\includegraphics[width=\linewidth]{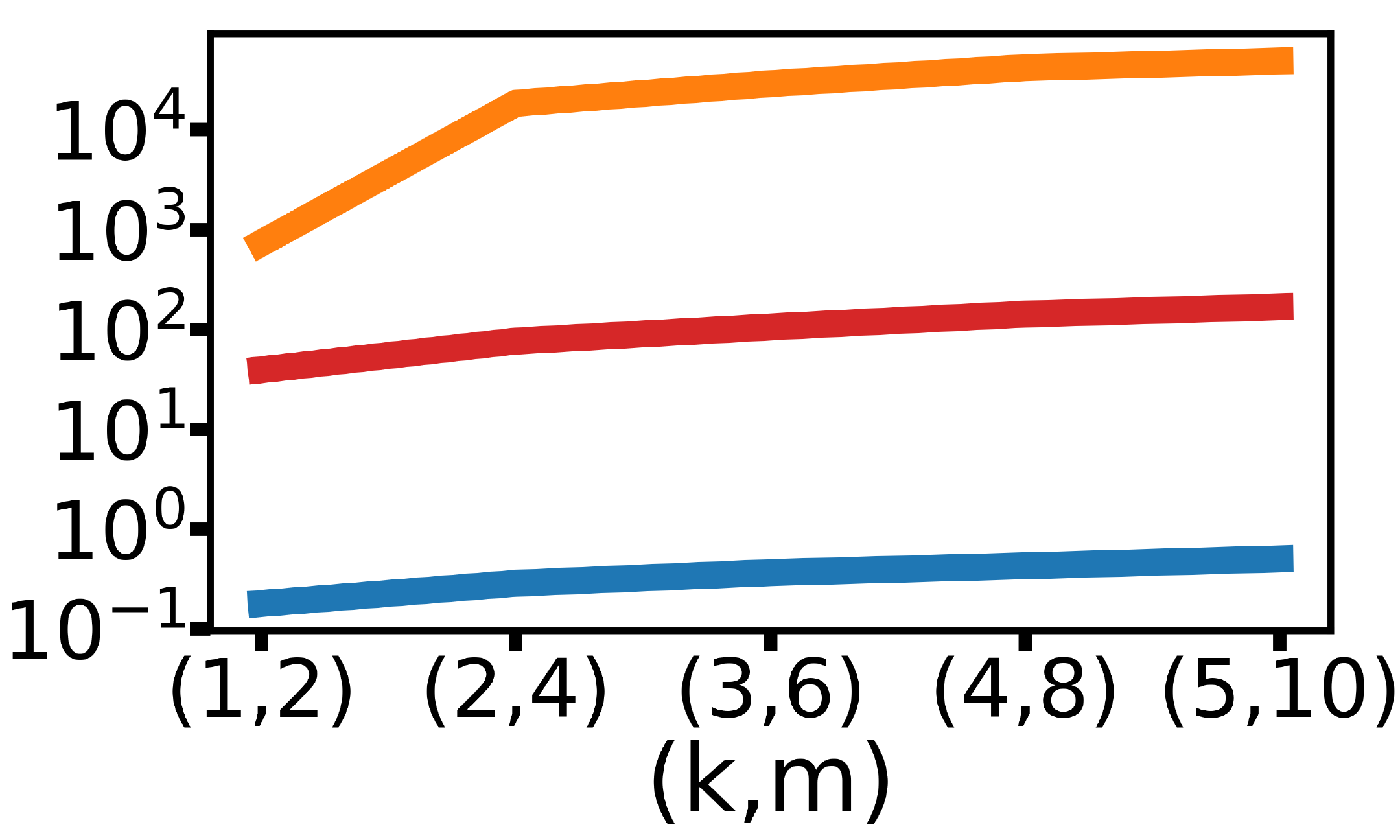}
\vspace{-0.25in}
\caption{\codein{db-match}}
\label{fig:runtime:match}
\end{subfigure}
\begin{subfigure}[b]{0.28\textwidth}
\includegraphics[width=\linewidth]{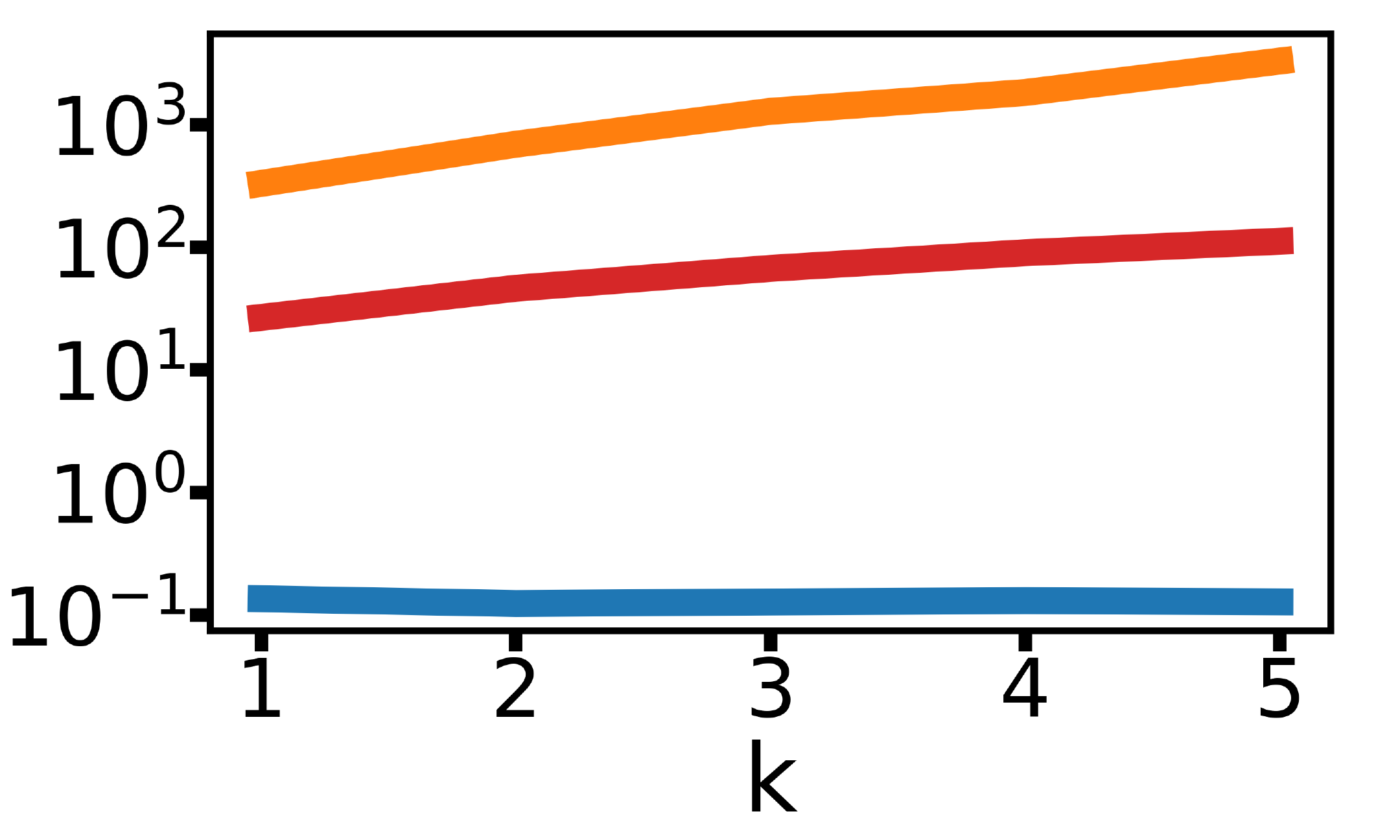}
\vspace{-0.25in}
\caption{\codein{kgraph}}
\label{fig:runtime:kg}
\end{subfigure}
\begin{subfigure}[b]{0.28\textwidth}
\includegraphics[width=\linewidth]{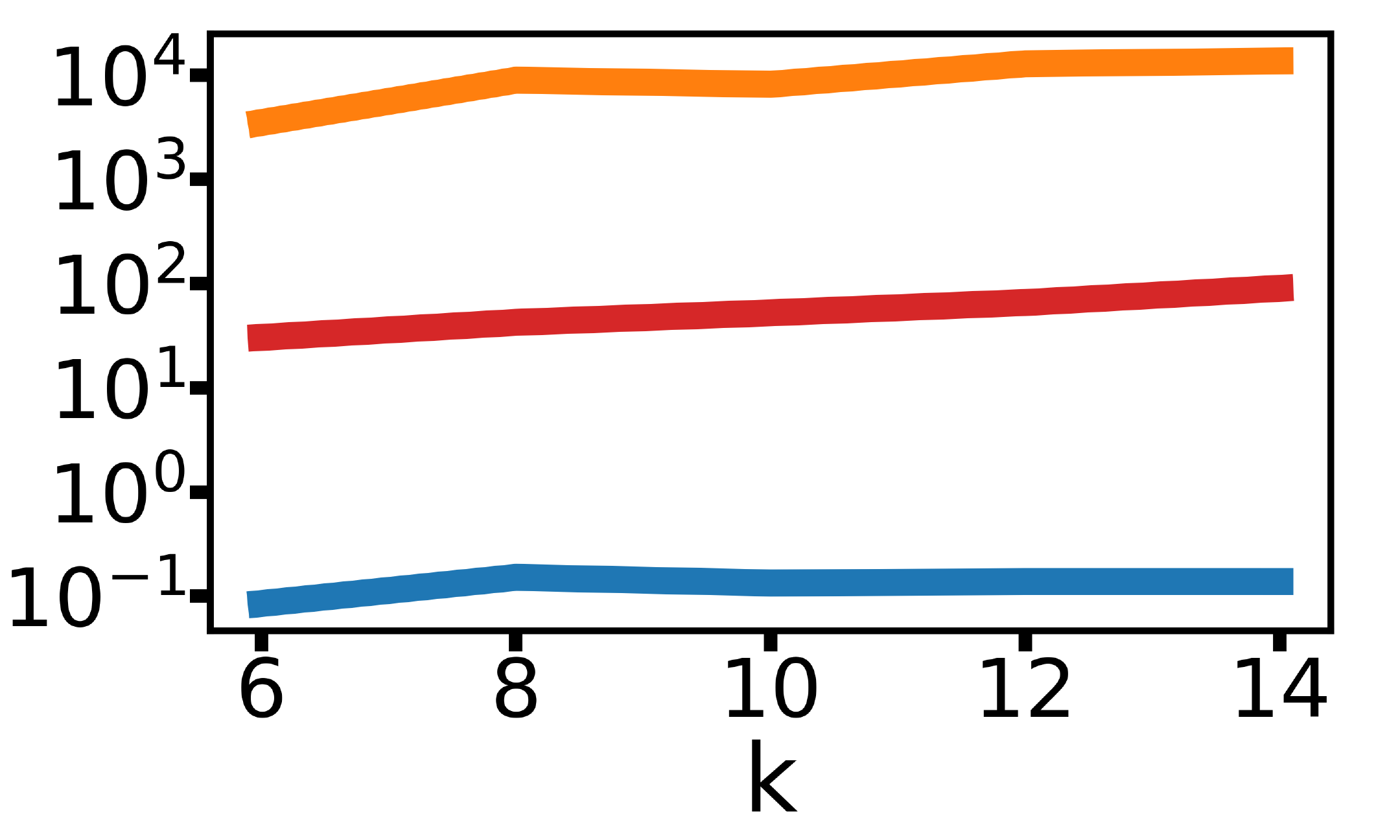}
\vspace{-0.25in}
\caption{\codein{nfa}}
\label{fig:runtime:nfa}
\end{subfigure}
% \begin{subfigure}[b]{0.30\textwidth}
% \includegraphics[width=\linewidth]{figures/thr_analogy_runtime.pdf}
% \caption{\codein{db-thr-analogy}}
% \label{fig:runtime:thr:analogy}
% \end{subfigure}
\begin{subfigure}[b]{0.28\textwidth}
\includegraphics[width=\linewidth]{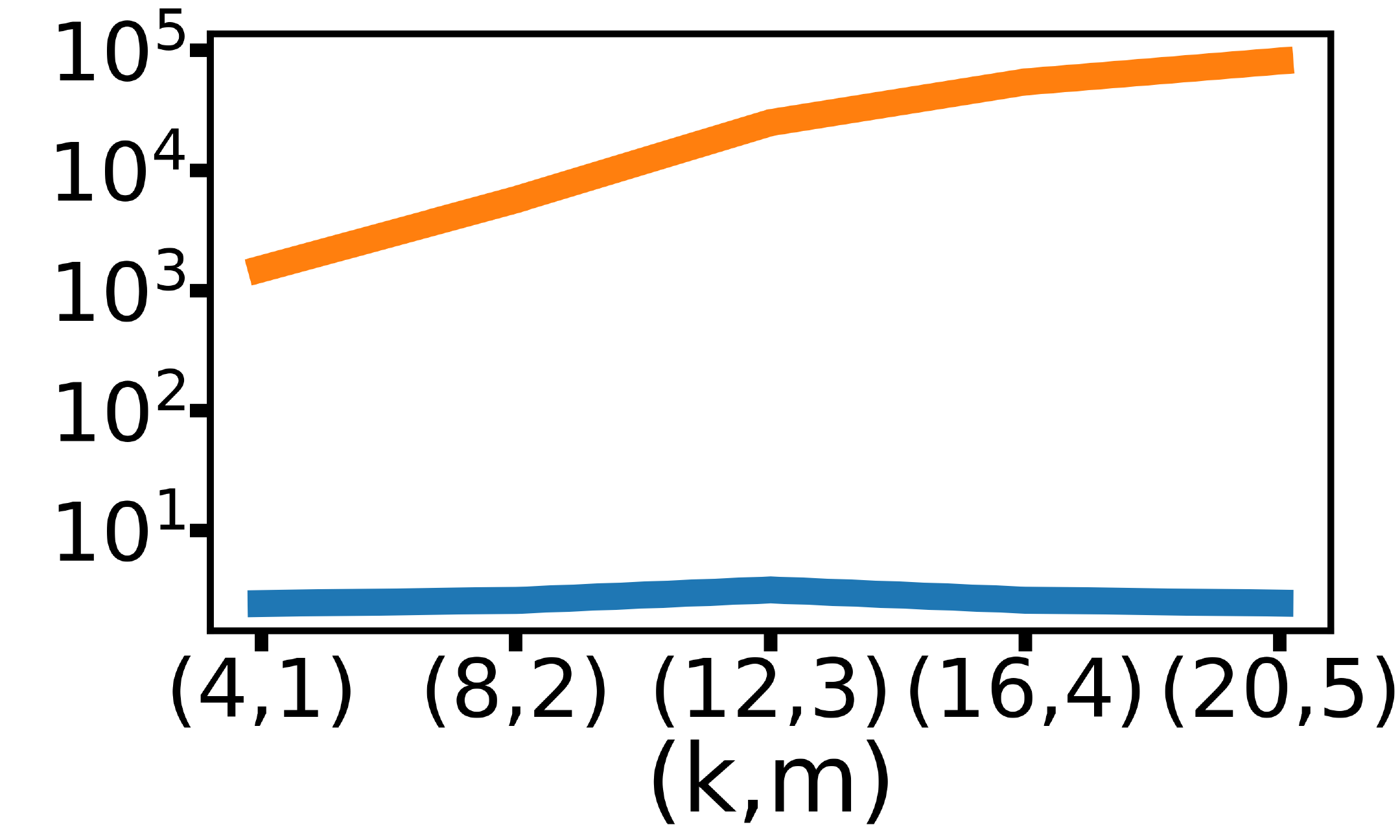}
\vspace{-0.25in}
\caption{\codein{db-analogy}}
\label{fig:runtime:wta:analogy}
\end{subfigure}
\caption{Y-axis is measured runtime (s) for tuning the hypervector size and threshold, averaged over 10 runs on a single-core machine. The \swatch{skyblue} is \codein{\tool{}}, \swatch{tangerine} is 
 \codein{dt-all}, \swatch{brickred} is \codein{dt-par}. The \codein{dt-par} trendline is omitted in the \codein{db-analogy} figure, because the benchmark uses a WTA query and does not require a threshold.  } 
 \label{fig:runtime}
\end{figure}

Figure~\ref{fig:runtime} compares the optimization runtimes for \tool{} against the dynamic tuning baselines. \tool{} completes its analysis in 85-3210 milliseconds, while \codein{dt-all} takes 40 seconds to 22.72 hours to optimize the hypervector size and threshold. Fixing the hypervector size and dynamically tuning only the threshold (\code{dt-par}) is substantially faster than full dynamic tuning, taking between 3.9 seconds and 168.8 seconds to compile. The \tool{} optimizer is 303.0x-100167.4x faster than full dynamic tuning (\codein{dt-all}) and 30.0x-874.4x faster than threshold-only dynamic tuning (\codein{dt-par}), and generally scales better as the benchmark size increases. Because both dynamic tuning approaches are simulation-based, execution time scales poorly as the number of parameters to tune and the complexity of the data structure increases. In contrast, \tool{}'s static analysis procedure is model-based and computes the optimal threshold and hypervector size in constant time. The performance of the parameter derivation algorithm is insensitive to the size of the optimized data structure, enabling scalable analysis.

\section{Evaluation of \tool{} on Emerging Hardware Technologies}\label{sec:emerging:results}
%  cima
\tool{}'s accuracy analysis enables sound optimization of HD computations to execute with acceptable accuracy in the presence of hardware error. This capability enables the optimization of HD computations for error-prone emerging hardware technologies. We use \tool{} to systematically study the benefits and drawbacks of using different emerging technologies for hyper-dimensional computation. We analyze the performance benefits offered by analog content-addressable memories (CAMs) (Section~\ref{sec:results:cam}), and the storage benefits offered by analog multi-bit storage arrays.~(Section~\ref{sec:results:mlc}).

\subsection{\tool{} Storage Density Analysis with MLC ReRAMs}\label{sec:results:mlc}

\begin{figure}
\begin{subfigure}[b]{0.28\textwidth}
\includegraphics[width=\linewidth]{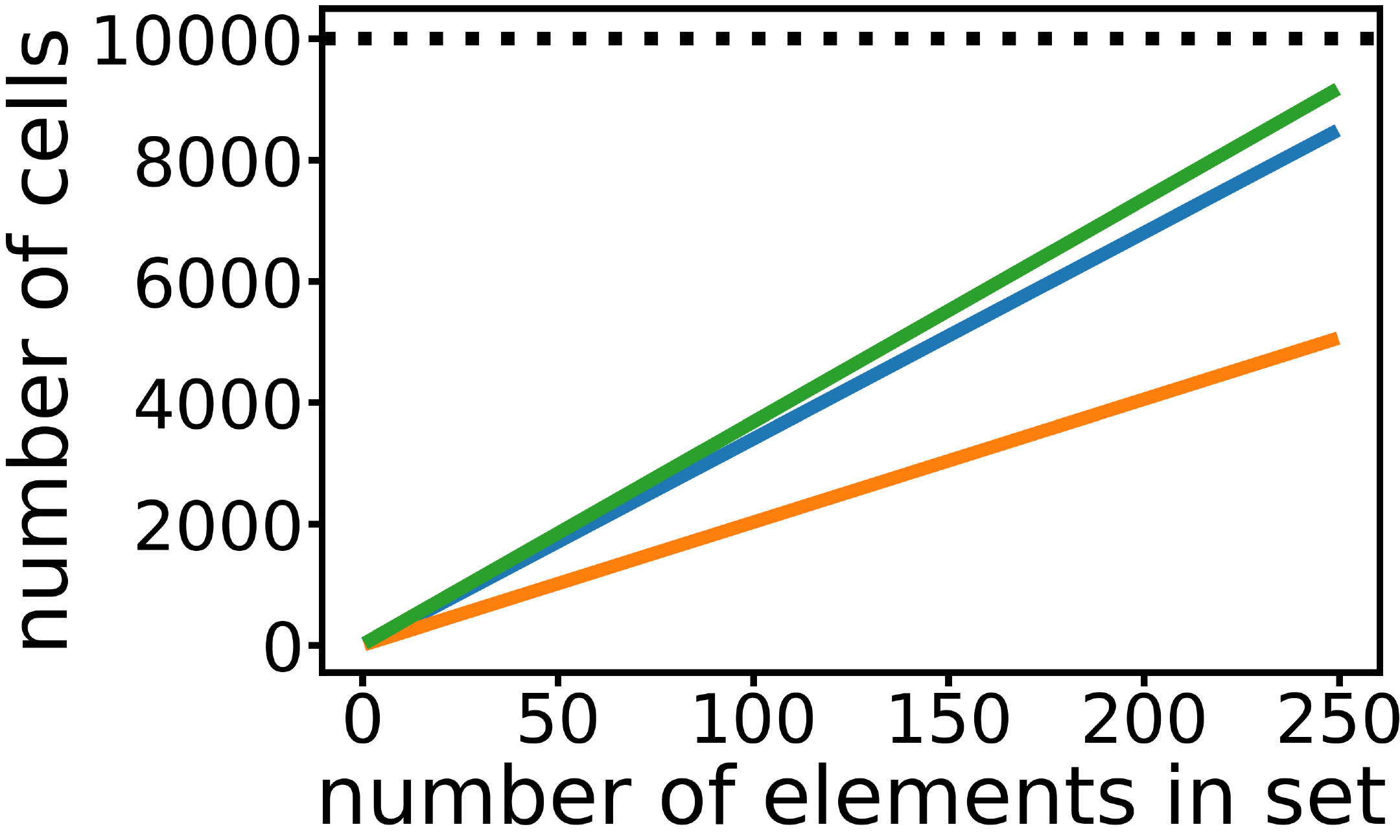}
\vspace{-0.25in}
\caption{\codein{set}}
\label{fig:storage:set}
\end{subfigure}
\begin{subfigure}[b]{0.28\textwidth}
\includegraphics[width=\linewidth]{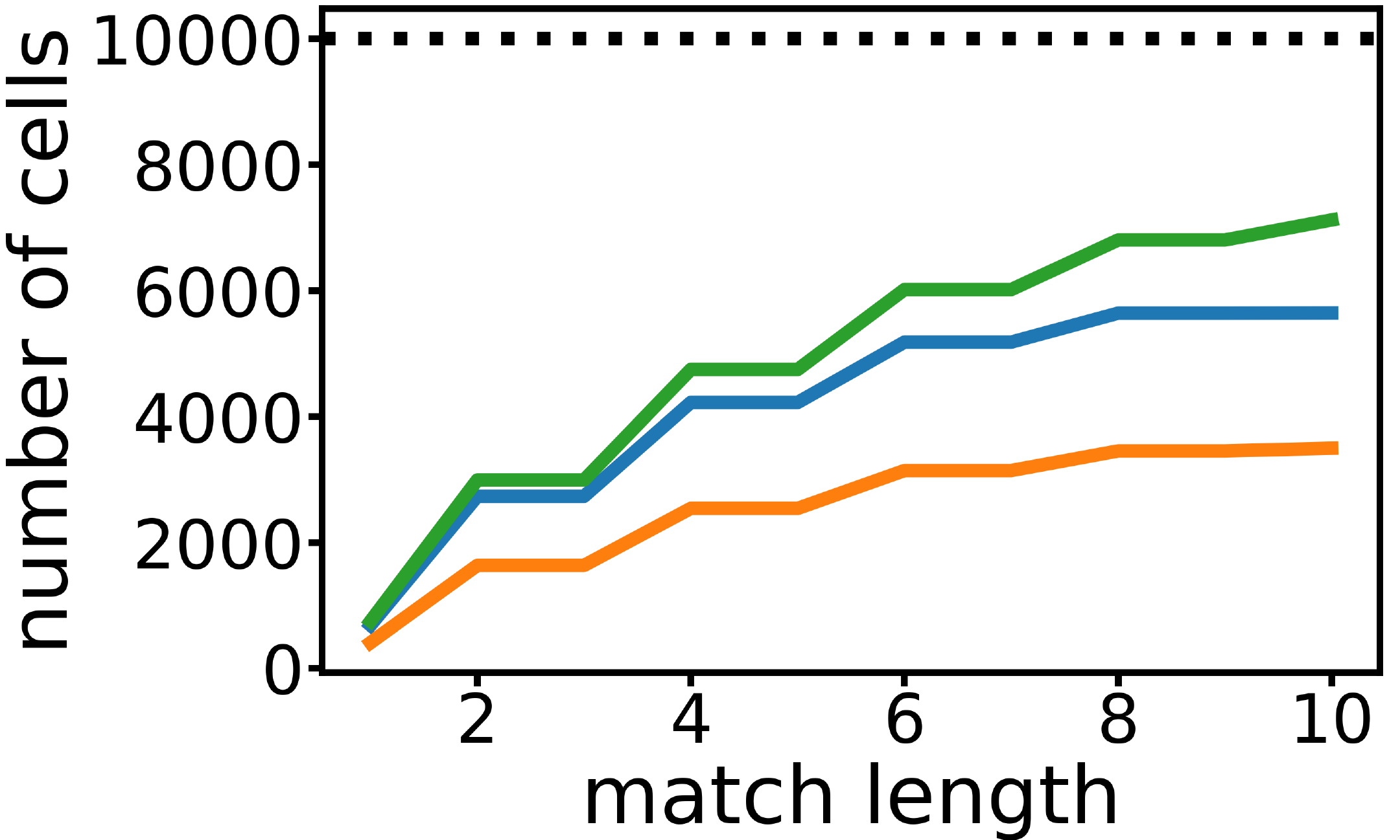}
\vspace{-0.25in}
\caption{\codein{db-match}}
\label{fig:storage:match}
\end{subfigure}
\begin{subfigure}[b]{0.28\textwidth}
\includegraphics[width=\linewidth]{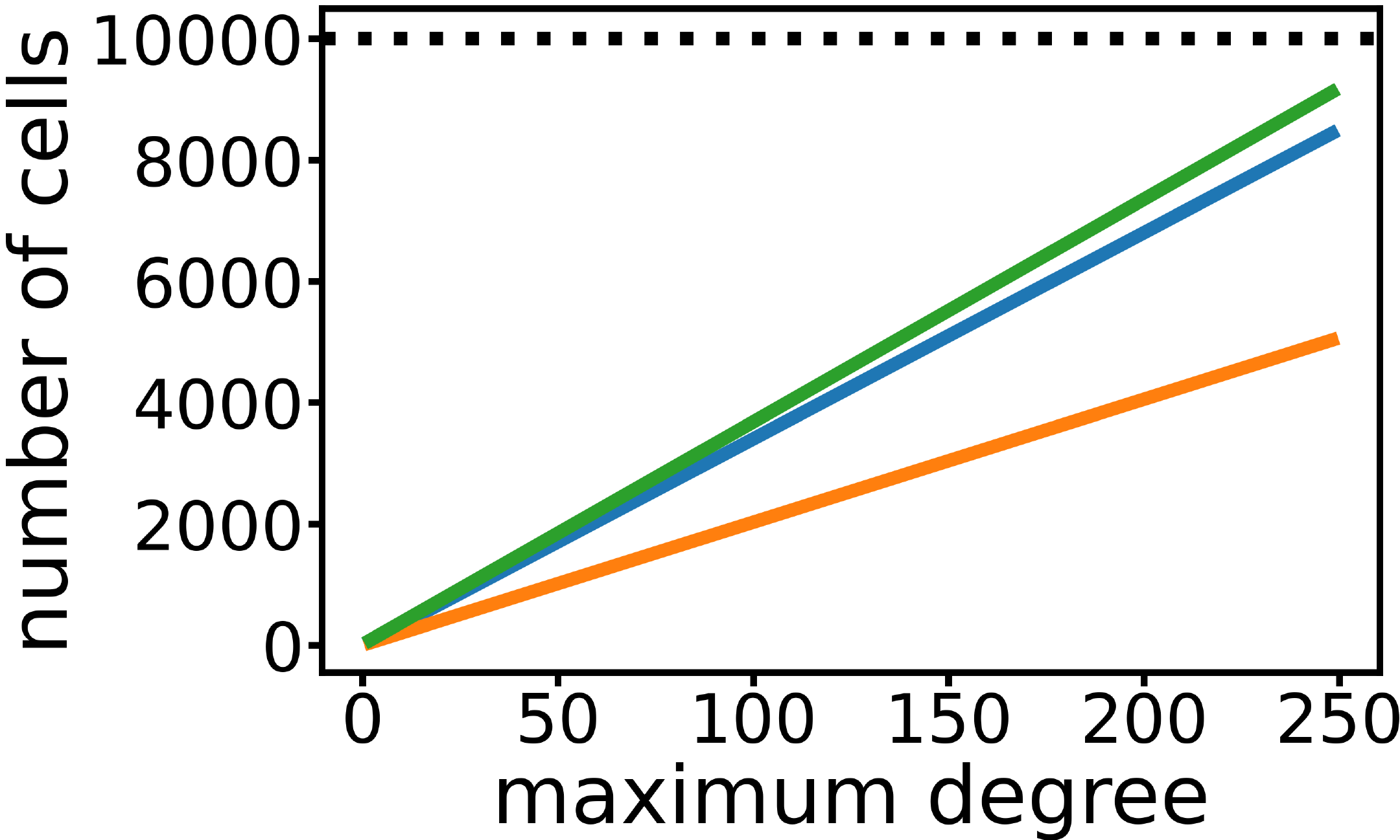}
\vspace{-0.25in}
\caption{\codein{kgraph}}
\label{fig:storage:kg}
\end{subfigure}
\begin{subfigure}[b]{0.28\textwidth}
\includegraphics[width=\linewidth]{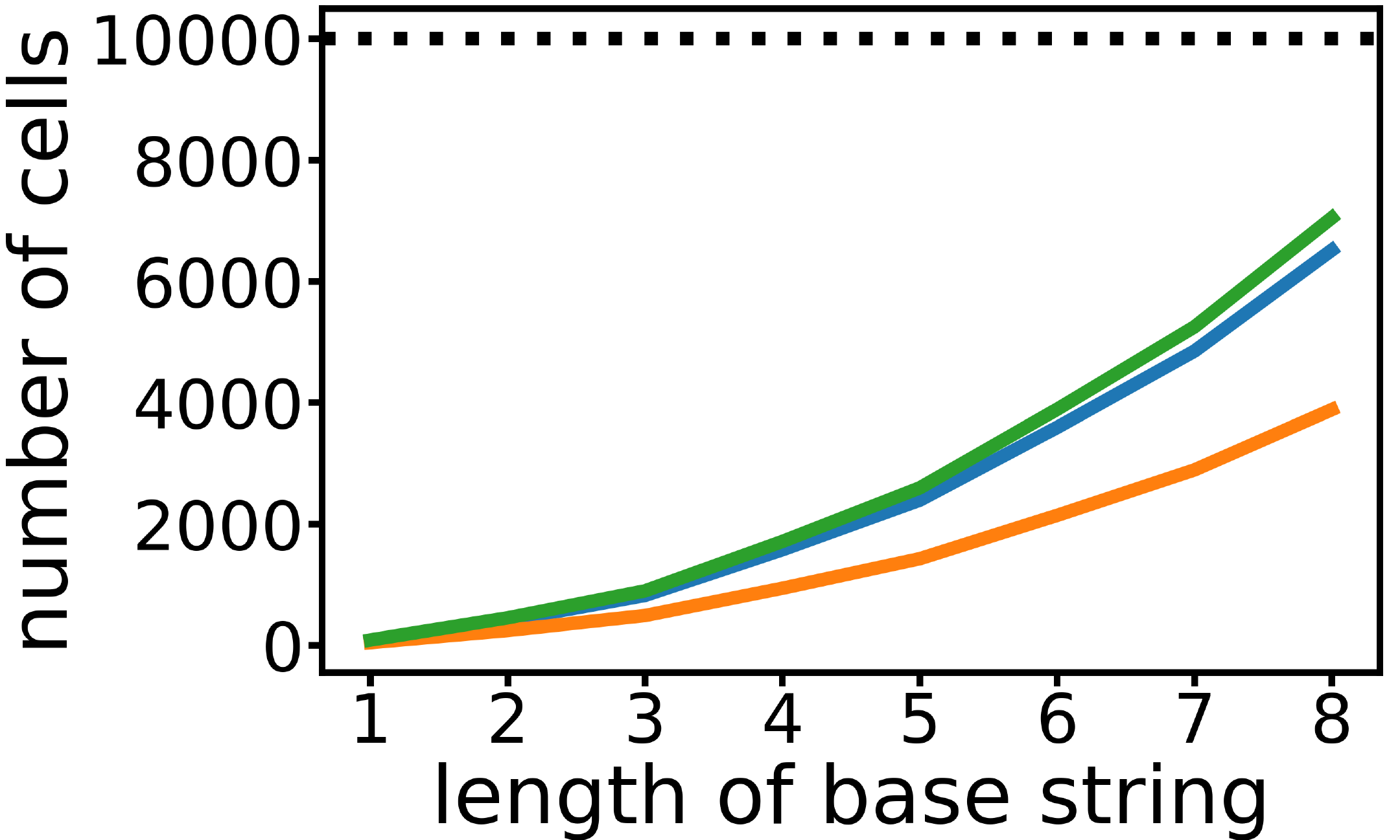}
\vspace{-0.25in}
\caption{\codein{nfa}}
\label{fig:storage:nfa}
\end{subfigure}
% \begin{subfigure}[b]{0.30\textwidth}
% \includegraphics[width=\linewidth]{figures/thr_a_Theory_Cells_BitFlip.pdf}
% \caption{\codein{db-thr-analogy}}
% \label{fig:storage:thr:analogy}
% \end{subfigure}
\begin{subfigure}[b]{0.28\textwidth}
\includegraphics[width=\linewidth]{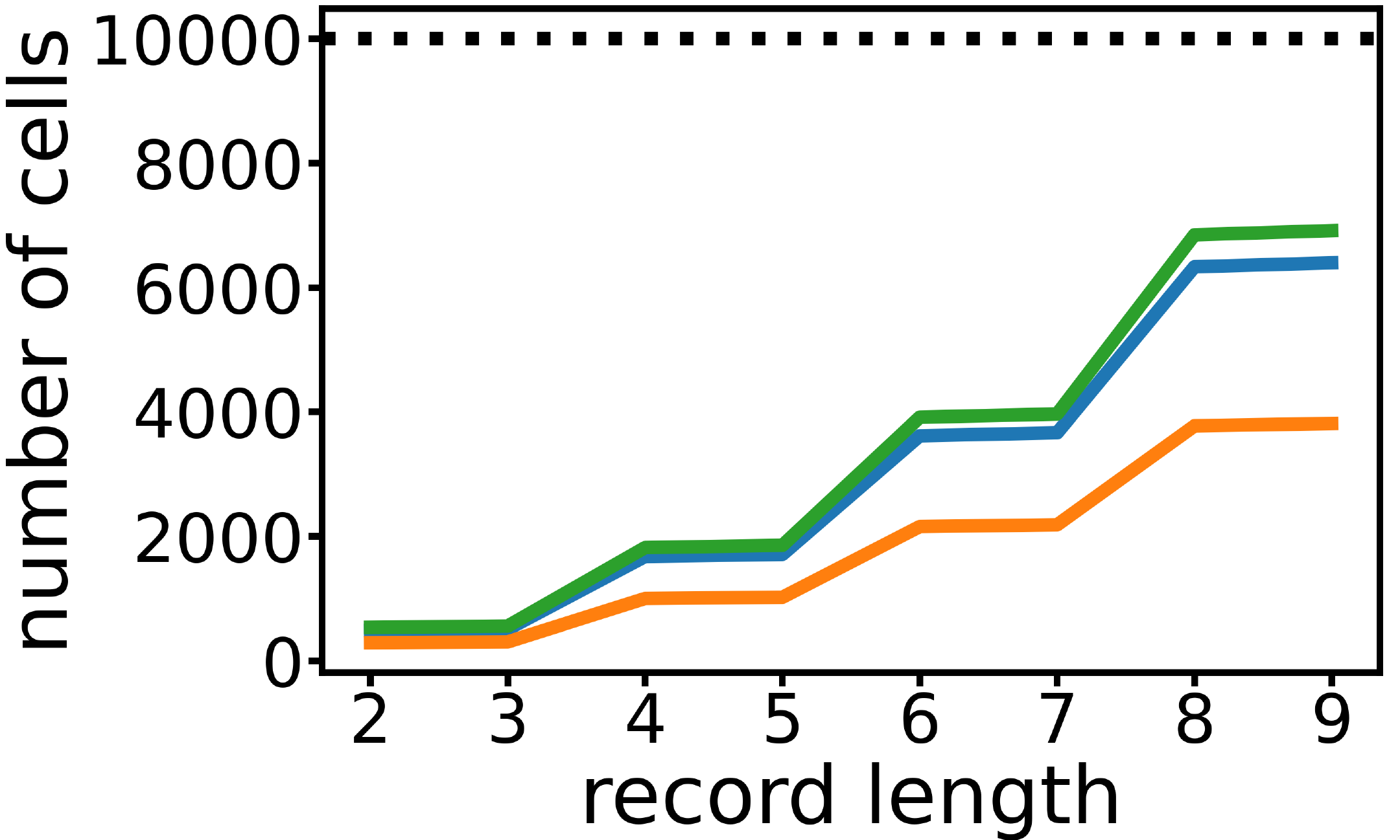}
\vspace{-0.25in}
\caption{\codein{db-analogy}}
\label{fig:storage:wta:analogy}
\end{subfigure}
% \begin{subfigure}[b]{0.74\textwidth}
% \small
% \centering
% \setlength{\tabcolsep}{1.5pt}
% \begin{tabular}{|c|cc|cc|cc|cc|cc|}
% \hline
%  & \multicolumn{2}{c|}{\textbf{set}}  & \multicolumn{2}{c|}{\textbf{db-match}} & \multicolumn{2}{c|}{\textbf{db-analogy}} & \multicolumn{2}{c|}{\textbf{graph}} & \multicolumn{2}{c|}{\textbf{nfa}}\\
%  \textbf{benchmark}&\textit{mean}&\textit{std}& \textit{mean}&\textit{std}& \textit{mean}&\textit{std}& \textit{mean}&\textit{std}& \textit{mean}&\textit{std}\\
% \hline
% \codein{\tool{}-nom} & 99.04 & 0.12 & 99.10 & 0.29 & 99.18 &0.52 & 99.75 & 0.34 & 99.50 & 0.85 \\
% \codein{\tool{}-2bpc} & 99.00 & 0.07 & 98.96 & 0.18 & 99.03 & 0.46 & 99.75 & 0.33 & 99.25 & 1.41 \\
% \codein{\tool{}-3bpc} & 99.00 & 0.05 & 99.05 & 0.20 & 99.00 & 0.32 & 99.75 & 0.34 & 99.31 & 1.19 \\
% \hline
% %\codein{\tool{}-nom} & (0.9904,0.0012) & (0.9910,0.0029) & (0.9918,0.0052) & (0.9975,0.0034) & (0.9944,0.0140) \\
% %\codein{\tool{}-2bpc} & (0.9900,0.0007) & (0.9896,0.0018) & (0.9903,0.0046) & (0.9975,0.0033) & (0.9950,0.0122) \\
% %\codein{\tool{}-3bpc} & (0.9900,0.0005) & (0.9905,0.0020) & (0.9900,0.0032) & (0.9975,0.0034) & (0.9381,0.0480) \\
% \end{tabular}
% \caption{Measured \% accuracy (\codein{accuracy $\times$ 100\%}), target accuracy is \codein{99.0}\%}
% \label{table:empirical:accuracy}
% \end{subfigure}

\caption{Memory cells required to store each hypervector with \codein{99\%} accuracy. The \swatch{black} is unoptimized hypervector size (10,000 bits, 1-bit-per-cell), \swatch{skyblue} is \codein{\tool{}}, \swatch{tangerine} is 
 \codein{\tool{}-2bpc}, \swatch{pixelgreen} is \codein{\tool{}-3bpc}. \codein{db-match} records use 20 fields, \codein{db-analogy} databases have 300 records.}
 \vspace{-0.16in}
 \label{fig:storage}
\end{figure}

We use \tool{} to analyze the storage benefits of using multiple-bit-per-cell (MLC, $k$ BPC) ReRAM-based item memories for HD computation against conventional one-bit-per-cell DRAM-based memory. We use \tool{} to minimize the hypervector size while delivering a target query accuracy of \codein{99\%} for 2 BPC ReRAM, 3-BPC ReRAM, and conventional DRAM.

\proseheading{MLC ReRAMs.}  ReRAM is an emerging resistive memory technology that is prone to bit corruption but delivers fast access times, non-volatility, and improved density. We investigate the benefits of 2BPC and 3BPC ReRAM, which have raw bit error rates of \codein{0.0215} and \codein{0.1273}, respectively. All ReRAM error measurements were collected by characterizing a ReRAM storage array fabricated 130nm logic CMOS process in the BEOL with ECC disabled.~\cite{wei2023pba,hsieh2019high,le2021radar}

\proseheading{Analysis.}  Figure~\ref{fig:storage} presents the number of memory cells required to store each hypervector as a function of the benchmark size for conventional, 2BPC ReRAM (\codein{2bpc}), and 3BPC ReRAM (\codein{3bpc}) hardware platforms. \tool{} produces hypervectors that require 76-8400 binary memory cells for conventional memory. \tool{} produces 94-10064 bit hypervectors that use 47-5032 memory cells for 2 BPC ReRAM, netting an additional 1.614x-1.677x cell reduction over conventional binary memory. Though the overall hypervector size increases for 2 BPC memory, the 2x improvement in data density for this memory technology subsumes this size increase. \tool{} produces 270-27361 bit hypervectors that use 90-9121 memory cells for 3 BPC ReRAM. Though 3 BPC ReRAM is denser than 2 BPC ReRAM, it does not net density improvements because the benchmark HD computations require substantially larger hypervectors to execute accurately in the presence of hardware error.

\subsection{\tool{} Performance Analysis with Analog CAMs}\label{sec:results:cam}

\begin{table}[t]
\footnotesize
\centering
\setlength{\tabcolsep}{2pt}
\begin{tabular}{|l|lll|llll|ll|}
\hline
\textbf{} & \multicolumn{3}{c|}{\codecap{processor}} & \multicolumn{4}{c|}{\codecap{caches}} & \multicolumn{2}{c|}{\codecap{memory}}\\
\textbf{platform} & \textbf{cores} & \textbf{threads/core} & \textbf{frequency} & \textbf{L1D} & \textbf{L1I} & \textbf{L2} & \textbf{L3} & \textbf{main} & \textbf{specialized}\\
\hline
\codecap{micro} & 2 & 1 & 1GHz & 32KB, 8w & 64KB/8w  & 2MB/16w & 16MB/16w & 3GB & - \\
\codecap{multi} & 10 & 2 & 2.8GHz & 32KB/8w & 32KB/8w & 256KB/4w& 20MB/16w & 3GB & - \\
\codecap{cam} & 2 & 1 & 1GHz & 32KB/8w & 64KB/8w  & 2MB/16w & 16MB/16w & 3GB &  Analog CAM\\
\hline
\end{tabular}
\caption{Benchmark hardware platforms. All caches are $k$-way (\codecap{kw}) associative, L1, L2 are private. }
\vspace{-0.16in}
\label{tbl:hwbaselines}
\end{table}

\begin{figure}
\begin{subfigure}[b]{0.28\textwidth}
\includegraphics[width=\linewidth]{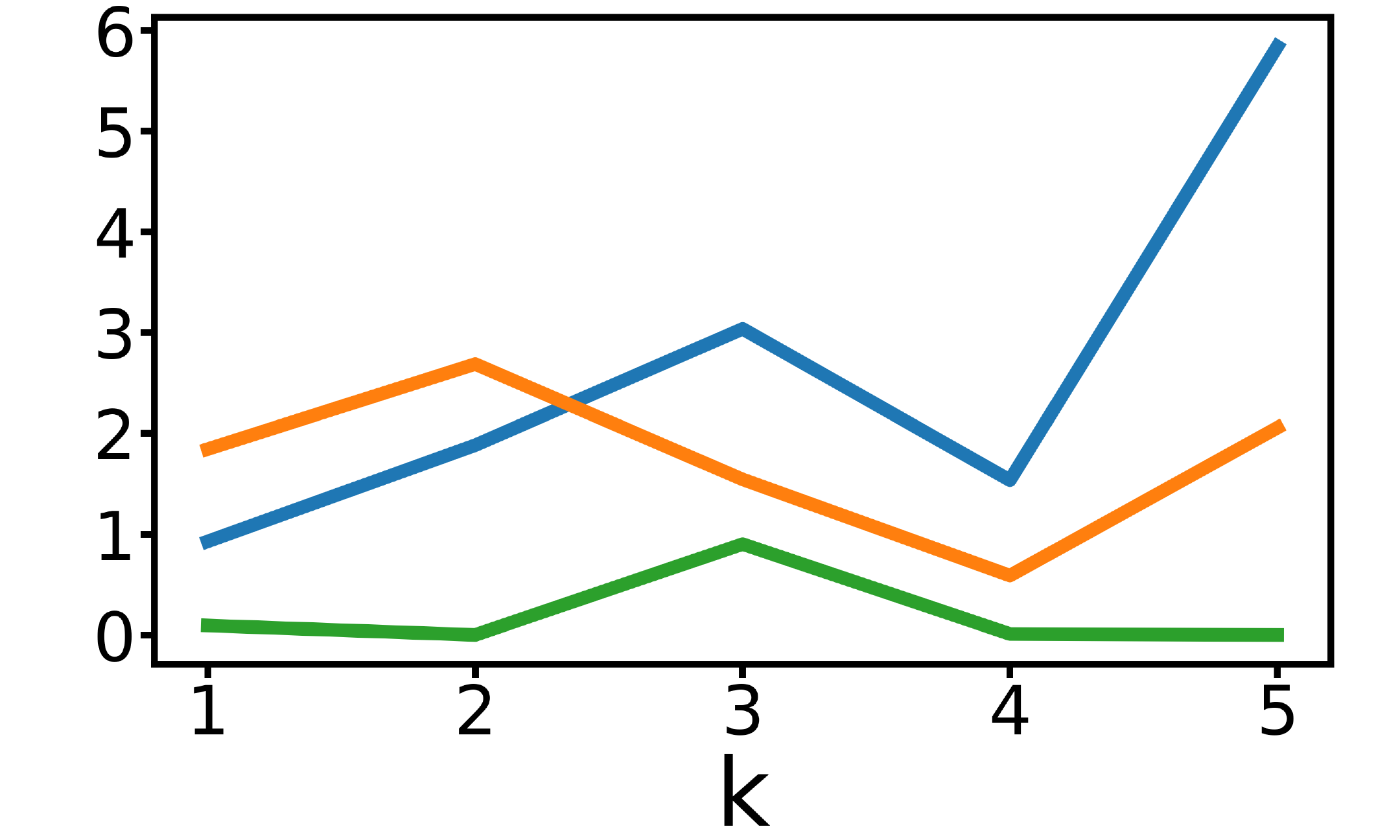}
\vspace{-0.25in}
\caption{\codein{set}}
\label{fig:gem5:set}
\end{subfigure}
\begin{subfigure}[b]{0.28\textwidth}
\includegraphics[width=\linewidth]{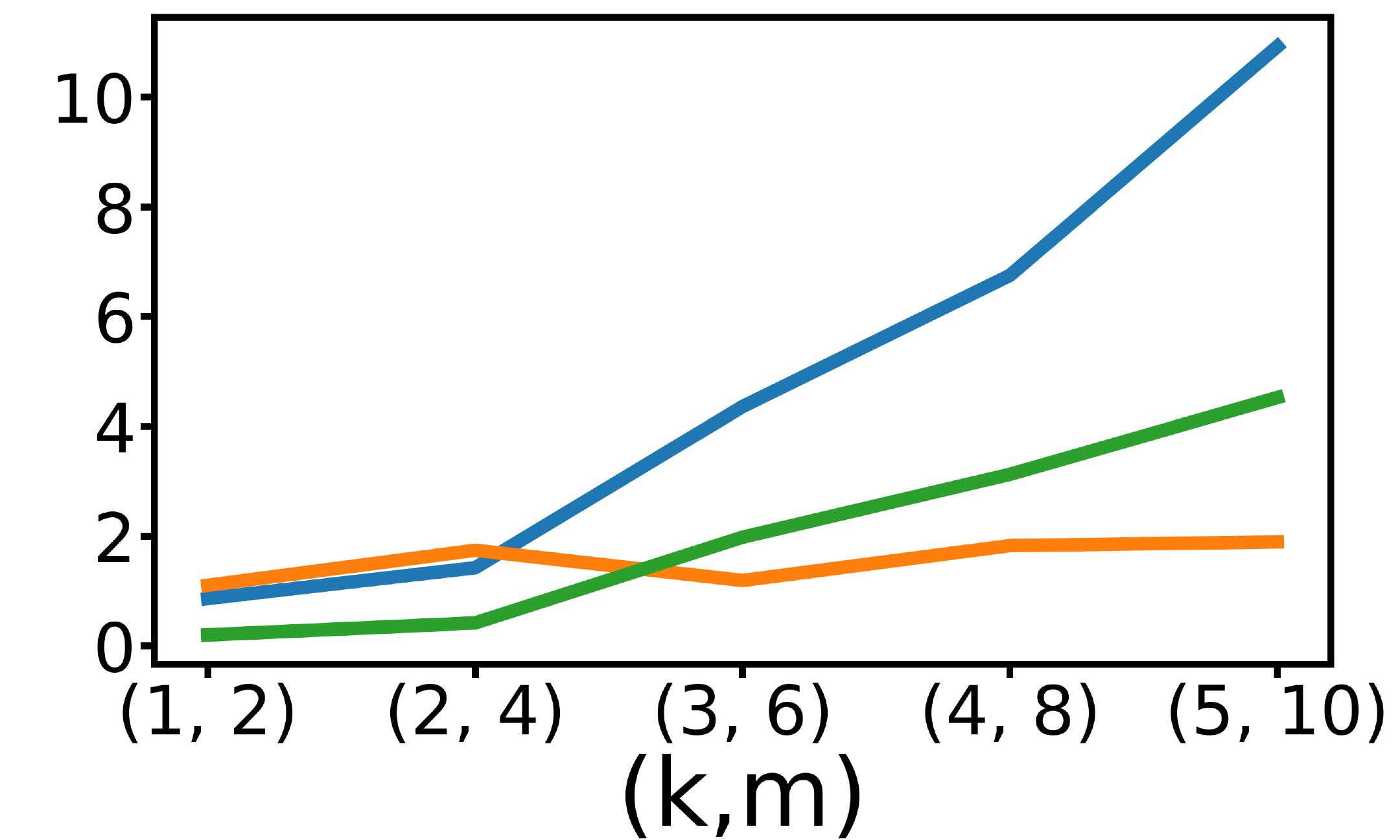}
\vspace{-0.25in}
\caption{\codein{db-match}}
\label{fig:gem5:match}
\end{subfigure}
\begin{subfigure}[b]{0.28\textwidth}
\includegraphics[width=\linewidth]{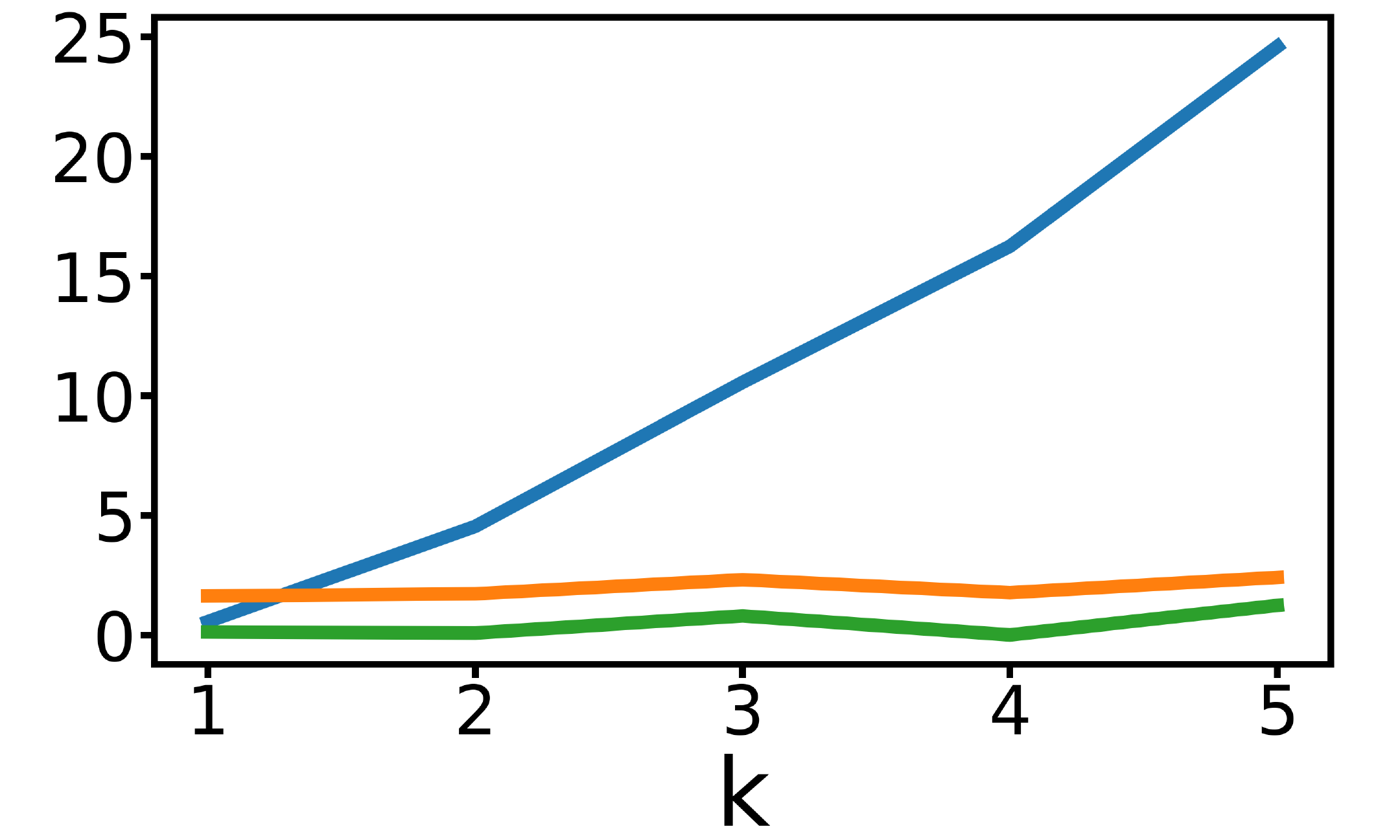}
\vspace{-0.25in}
\caption{\codein{kgraph}}
\label{fig:gem5:kg}
\end{subfigure}
\begin{subfigure}[b]{0.28\textwidth}
\includegraphics[width=\linewidth]{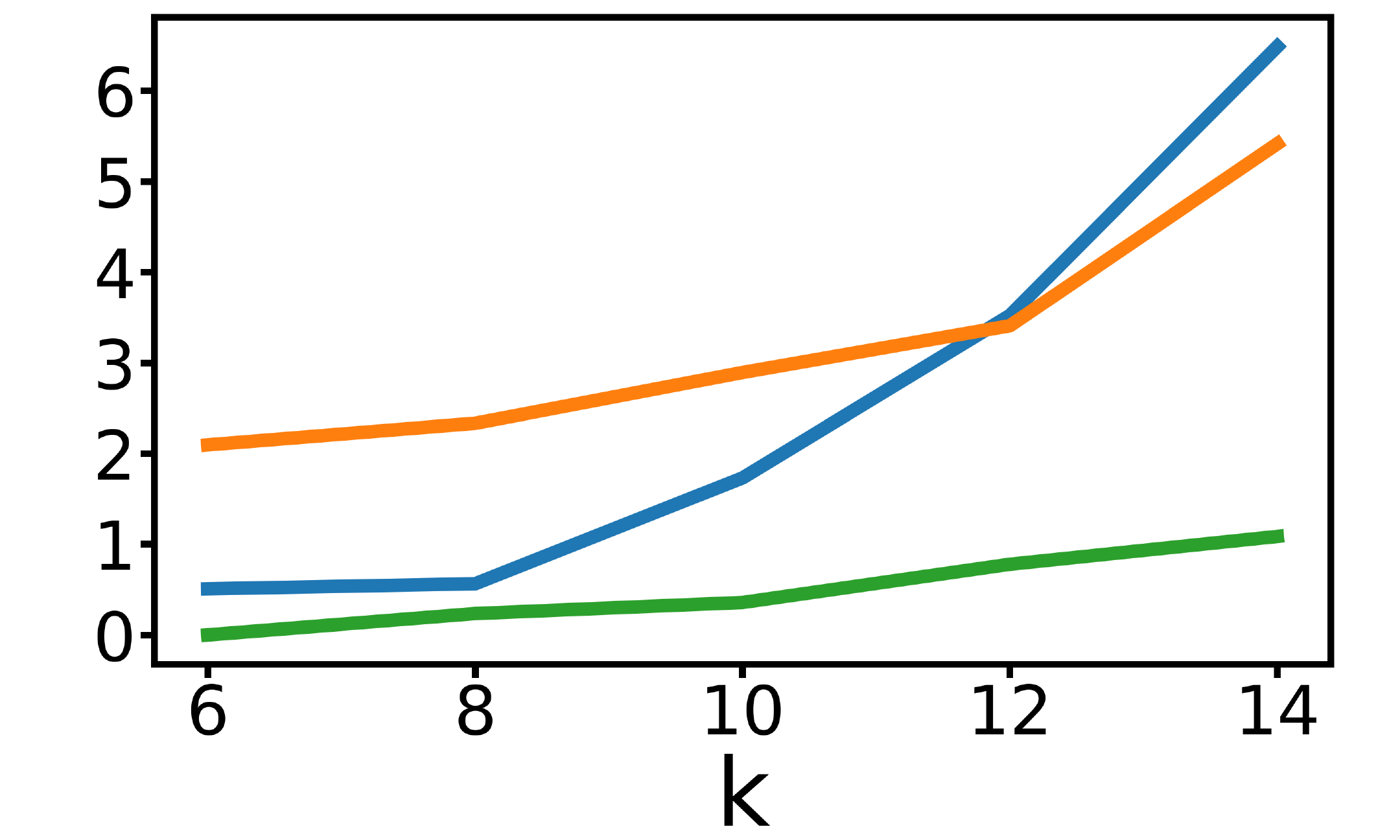}
\vspace{-0.25in}
\caption{\codein{nfa}}
\label{fig:gem5:nfa}
\end{subfigure}
\begin{subfigure}[b]{0.28\textwidth}
\includegraphics[width=\linewidth]{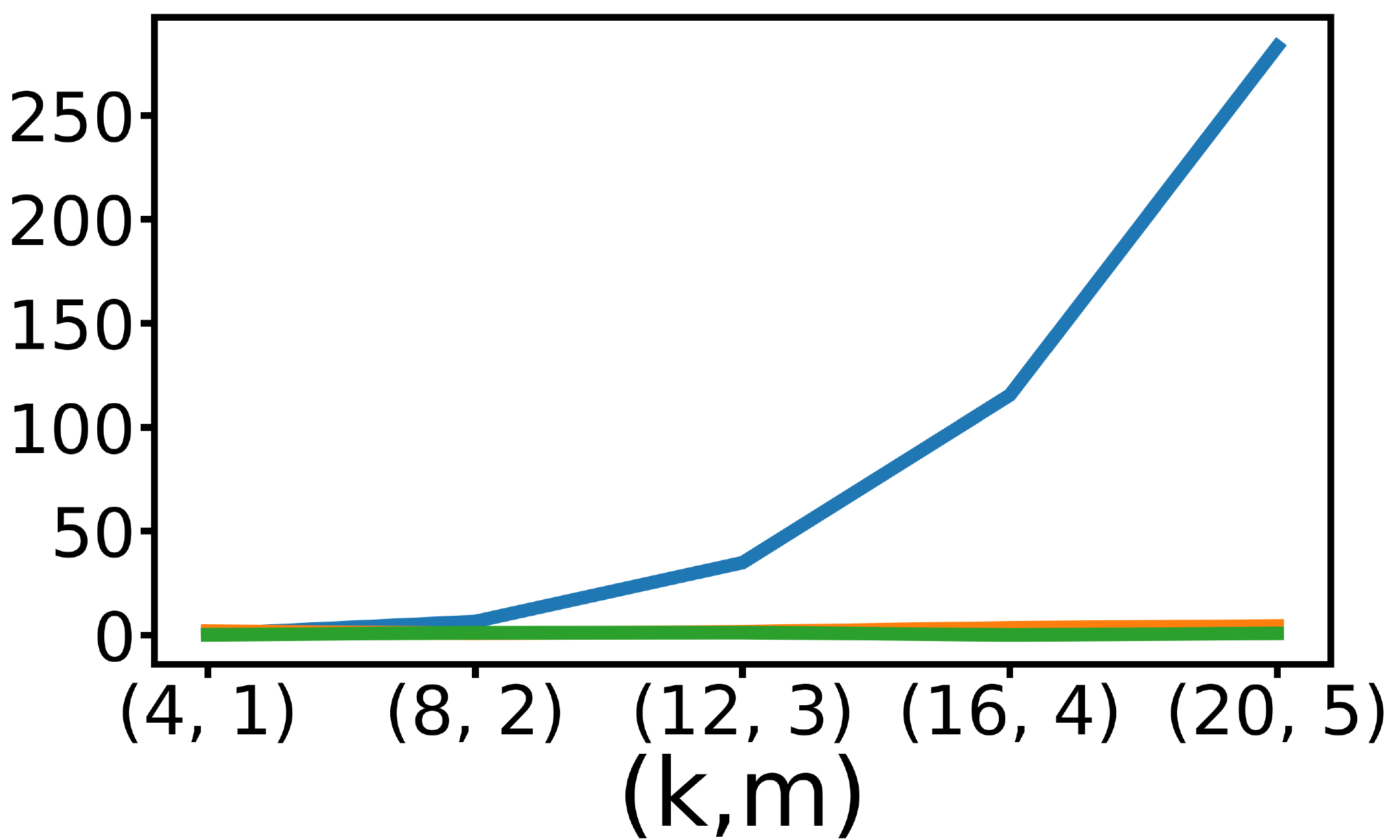}
\vspace{-0.25in}
\caption{\codein{db-analogy}}
\label{fig:gem5:wta:analogy}
\end{subfigure}
\caption{Y-axis is simulated runtime (ms) of a single query with gem5 averaged over 20 runs. The \swatch{skyblue} is \codein{micro}, \swatch{tangerine} is 
 \codein{multi}, \swatch{pixelgreen} is \codein{cam}.} 
 \label{fig:gem5}
\end{figure}

We evaluate the benchmark applications on embedded (\codein{micro}), multicore (\codein{multi}), and emerging hardware (\codein{cam}) platforms. Each architecture is simulated with the \codein{x86} \codein{gem5} simulator using the processor and cache hierarchy presented in Table~\ref{tbl:hwbaselines}.~\cite{binkert2011gem5} The embedded and multicore baselines are based on the Intel Atom x7211E and Intel Core i9-10900E x86 architectures, respectively, and the CAM hardware platform uses the embedded architecture coupled with an analog CAM that efficiently performs item memory lookups to implement HD computations, but introduces error into the computation. To ensure an iso-accuracy comparison, we use \tool{} to optimize each execution to achieve a 99\% accuracy on the respective hardware platform.~\footnote{for \codecap{db-analogy} simulation, we set the item memory and codebook sizes to be $\frac14$ of the actual size to avoid memory issues with the \codecap{gem5-x86} simulator.}  The benchmarks are parallelized and implemented in C, where the query hypervector is built in parallel, and the item memory/query distances are computed in parallel. For the CAM hardware platform, the software instead computes the item memory/query distances by dispatching a single query to the CAM. 

\proseheading{The Analog CAM.} The \codein{cam} hardware platform has the same baseline characteristics as the embedded hardware platform but also offers a ReRAM-based Analog CAM that uses Ohm's and Kirchhoff's laws to perform in-memory, parallel hamming distance calculations against a query bit vector.~\cite{imani2017exploring} Because the analog CAM both performs analog computation and uses an emerging device technology, the associated hamming distance calculations are unreliable and complete with a bit error rate of ~$0.14\%$. The analog CAM is extremely fast and completes the entire item memory query in 2.74-11.90 nanoseconds, provided 6-100 item memory rows are in use.  The associated latency increases with row width, and the bit error rate increases with row width and device density. We parametrize the CAM to use 10k-bit rows, where > 10k-bit hypervectors are split across rows; these row distances are summed in the analog domain with Kirchhoff's law. The communication costs between the embedded system and the CAM are modeled as a DRAM access, and the threshold query latency is conservatively approximated with the WTA lookup latency. We build performance and hardware error models by fitting regressions to the figures presented in the associated paper.~\cite{imani2017exploring}

\proseheading{Analysis} Figure~\ref{fig:gem5} presents the simulated runtime of a single query as a function of the benchmark size, averaged over 20 executions. On average, the \codein{multi}, \codein{micro}, and \codein{cam} executions take 0.589-5.418, 0.385-283.226, and 0.000-4.526 milliseconds respectively, depending on the benchmark. We observe that, unsurprisingly,  the \codein{multi} executions complete much faster than the \codein{micro} executions for large benchmarks and only execute slightly slower on small benchmarks. The multicore platform has 5x more cores at its disposal and operates at 2.8x the frequency of the embedded system, so it, therefore, can complete HD computations much faster, provided the synchronization overheads amortize.

However, once the embedded platform is paired with a CAM, the embedded platform's performance becomes competitive and, in some cases, better than the multicore system. The \codein{cam} executions are 2.16x-389.51x faster than \codein{micro} executions for all benchmarks and 1.35x-62.55x faster than \codein{multi} executions for 21 out of 25 benchmarks. The \codein{cam} delivers substantial performance improvements because it significantly accelerates the item-memory search, which is usually the bottleneck when the query encoding is simple. We observe that these performance benefits hold, even though \tool{} optimizes the CAM executions to use 1.011x-1.012x larger hypervectors to ensure iso-accuracy in the presence of hardware error. The \codein{multi} executions are faster than \codein{cam} in the largest 3 \codein{db-match} benchmarks because the query encoding uses bundling operation and, therefore, becomes the bottleneck.

\proseheading{Discussion} While HD computing is, at a glance, more resource-inefficient than classical computation for this class of applications, the HD computing paradigm enables the use of emerging hardware technologies that drastically accelerate computation and enable dense storage, such as CAMs and MLC ReRAM. These emerging technologies only implement highly restrictive subsets of computational operators and are error-prone, which can often lead to unpredictable effects on classical computations. Moreover, because the HDC model uses a distributed information representation, it is also amenable to several program optimizations that do not typically apply to classical programs. For example, the \tool{} per-query hypervector sizes and thresholds can be used to compute hypervector distances over smaller sets of bits soundly or to terminate distance calculations early when a match or not-match is guaranteed. Moreover, the binding, bundling, and permutation operation implementations and hypervector sizing can be altered to improve performance, provided the distance relationships between input and output vectors hold. Second, because the encoded information is evenly distributed, algorithms can use highly unusual program transformations. For example, HD data structures can be combined by splicing hypervectors together, distances can be computed over any segment of hypervector bits in any order, queries over data structure hypervectors can be fielded in the middle of a data structure update, and HD queries can be prematurely interrupted to receive a computational result.

\section{Related Work}\label{sec:relatedwork}

\proseheading{HDC/VSA.} Hyperdimensional computing (HDC), or vector symbolic architecture (VSA), is a highly general cognitive computing paradigm that operates on binary, integer, real-valued, or complex-valued hypervectors~\cite{plate1994distributed,plate2003holographic,yu2022understanding,kleyko2021survey1,kleyko2021survey2,kleyko2022vector}. Researchers have implemented VSA computations with emerging hardware platforms~\cite{halawani2021fused,imani2017exploring,imani2019sparsehd,karunaratne2020memory,poduval2021stochd,langenegger2023memory}. We focus on HD computing with binary spatter codes~\cite{kanerva1997fully} because this computational model is easy to implement in hardware, and amenable to theoretical analysis.

%\noindent\textbf{Bloom Filters.} Bloom filters are a type of approximate data structure that stores set of of elements %s\luke{Bloom Filter~\cite{bloom1970space}}

\proseheading{Theoretical Analysis of HD Computation.} Kanerva derived the distance distributions for the set-recall ~\cite{kanerva1997fully}, and Kleyko derived distance distributions for subset-recall~\cite{kleyko2016holographic}. We use Kanerva's and Kleyko's theoretical results to develop the Type I and Type II QDS analysis employed by \tool{}. Researchers have also studied theoretical capacity and recall accuracy of winner-take-all queries over VSA item memories~\cite{frady2018theory,gallant2013representing,thomas2021theoretical,kleyko2020perceptron,plate1994distributed,kleyko2023efficient}. We extend the perception theory~\cite{frady2018theory} to develop \tool{}'s WTA accuracy analysis.

\proseheading{Horizontal Thresholding.} Kleyko et al. derived an approach for horizontally thresholding distance distributions for subset-set recall queries.~\cite{kleyko2016holographic} This thresholding does not apply to Type III QDS queries, discards distance sub-ranges, and does not estimate query accuracy or solve for the optimal distance threshold. In contrast, \tool{} analyzes a broader set of data structure queries (Type I-Type 3 QDS), and derives an optimal distance threshold to use. Our approach also uses the entire range of distance values when evaluating a query.

\proseheading{Hypervector Size Minimization for Classification.} Prior work has primarily focused on dynamic approaches toward hypervector parameter optimization. Imani et al. dynamically tuned hypervector size to explore the trade-off between computational efficiency and accuracy.~\cite{imani2018hierarchical} Morris et al. decomposed computational hypervectors into lower-dimensional vectors of a dynamically selected size.~\cite{morris2019comphd} Basaklar et al. reduced the hypervector size by tuning the level hypervector construction for classification tasks.~\cite{basaklar2021hypervector} Other works have tuned application-specific hypervector parameters, such as level hypervector chunk sizes, to reduce resource usage while maintaining classification accuracy.~\cite{imani2019sparsehd} All approaches mentioned above are heuristic techniques that leverage dynamic tuning over representative inputs and therefore do not offer static guarantees. In contrast, \tool{} employs statically sound analytical methods that deliver accuracy guarantees, even in the presence of hardware error.

\section{Conclusion}

We presented \tool{}, a framework for \textit{statically} optimizing HD computation parameters to minimize resource usage in the presence of hardware error. \tool{} produced parametrizations that generalize across queries and data structures and satisfy a target accuracy on expectation. \tool{} improved on dynamic parameter tuning-based approaches that potentially overfit to test data, provide no guarantees, and take orders of magnitude more time to find parametrizations. We demonstrated that \tool{}'s analysis results could be leveraged to perform aggressive space-saving optimizations without compromising result fidelity and to systematically analyze emerging technologies' benefits and drawbacks while maintaining iso-accuracy. With analysis and programming systems such as \tool{}, we can enable the development of principled program optimizations that effectively reduce the resource requirements of HD computations without compromising accuracy.

\begin{acks}
This research was supported by the Stanford SystemX Alliance and ACCESS – AI Chip Center for Emerging Smart Systems,
sponsored by InnoHK funding, Hong Kong SAR.
\end{acks}

\section*{Data-Availability Statement}
The software that supports Sections~\ref{sec:results} and \ref{sec:emerging:results}
is available on Zenodo~\cite{zenodo}.

\bibliography{references}


\begin{thebibliography}{62}

%%% the Web address.

\ifx \showCODEN    \undefined \def \showCODEN     #1{\unskip}     \fi
\ifx \showDOI      \undefined \def \showDOI       #1{#1}\fi
\ifx \showISBNx    \undefined \def \showISBNx     #1{\unskip}     \fi
\ifx \showISBNxiii \undefined \def \showISBNxiii  #1{\unskip}     \fi
\ifx \showISSN     \undefined \def \showISSN      #1{\unskip}     \fi
\ifx \showLCCN     \undefined \def \showLCCN      #1{\unskip}     \fi
\ifx \shownote     \undefined \def \shownote      #1{#1}          \fi
\ifx \showarticletitle \undefined \def \showarticletitle #1{#1}   \fi
\ifx \showURL      \undefined \def \showURL       {\relax}        \fi
% invisible to TeX
\providecommand\bibfield[2]{#2}
\providecommand\bibinfo[2]{#2}
\providecommand\natexlab[1]{#1}
\providecommand\showeprint[2][]{arXiv:#2}

\bibitem[Achour and Rinard(2015)]%
        {achour2015approximate}
\bibfield{author}{\bibinfo{person}{Sara Achour} {and} \bibinfo{person}{Martin~C
  Rinard}.} \bibinfo{year}{2015}\natexlab{}.
\newblock \showarticletitle{Approximate computation with outlier detection in
  topaz}.
\newblock \bibinfo{journal}{\emph{Acm Sigplan Notices}} \bibinfo{volume}{50},
  \bibinfo{number}{10} (\bibinfo{year}{2015}), \bibinfo{pages}{711--730}.
\newblock
\urldef\tempurl%
\url{https://doi.org/10.1145/2858965.2814314}
\showDOI{\tempurl}


\bibitem[Basaklar et~al\mbox{.}(2021)]%
        {basaklar2021hypervector}
\bibfield{author}{\bibinfo{person}{Toygun Basaklar}, \bibinfo{person}{Yigit
  Tuncel}, \bibinfo{person}{Shruti~Yadav Narayana}, \bibinfo{person}{Suat
  Gumussoy}, {and} \bibinfo{person}{Umit~Y Ogras}.}
  \bibinfo{year}{2021}\natexlab{}.
\newblock \showarticletitle{Hypervector design for efficient hyperdimensional
  computing on edge devices}.
\newblock \bibinfo{journal}{\emph{arXiv preprint arXiv:2103.06709}}
  (\bibinfo{year}{2021}).
\newblock
\urldef\tempurl%
\url{https://doi.org/10.48550/arXiv.2103.06709}
\showDOI{\tempurl}


\bibitem[Binkert et~al\mbox{.}(2011)]%
        {binkert2011gem5}
\bibfield{author}{\bibinfo{person}{Nathan Binkert}, \bibinfo{person}{Bradford
  Beckmann}, \bibinfo{person}{Gabriel Black}, \bibinfo{person}{Steven~K
  Reinhardt}, \bibinfo{person}{Ali Saidi}, \bibinfo{person}{Arkaprava Basu},
  \bibinfo{person}{Joel Hestness}, \bibinfo{person}{Derek~R Hower},
  \bibinfo{person}{Tushar Krishna}, \bibinfo{person}{Somayeh Sardashti},
  {et~al\mbox{.}}} \bibinfo{year}{2011}\natexlab{}.
\newblock \showarticletitle{The gem5 simulator}.
\newblock \bibinfo{journal}{\emph{ACM SIGARCH computer architecture news}}
  \bibinfo{volume}{39}, \bibinfo{number}{2} (\bibinfo{year}{2011}),
  \bibinfo{pages}{1--7}.
\newblock
\urldef\tempurl%
\url{https://doi.org/10.1145/2024716.2024718}
\showDOI{\tempurl}


\bibitem[Clarkson et~al\mbox{.}(2023)]%
        {clarkson2023capacity}
\bibfield{author}{\bibinfo{person}{Kenneth~L Clarkson},
  \bibinfo{person}{Shashanka Ubaru}, {and} \bibinfo{person}{Elizabeth Yang}.}
  \bibinfo{year}{2023}\natexlab{}.
\newblock \showarticletitle{Capacity Analysis of Vector Symbolic
  Architectures}.
\newblock \bibinfo{journal}{\emph{arXiv preprint arXiv:2301.10352}}
  (\bibinfo{year}{2023}).
\newblock
\urldef\tempurl%
\url{https://doi.org/10.48550/arXiv.2301.10352}
\showDOI{\tempurl}


\bibitem[Eggimann et~al\mbox{.}(2021)]%
        {eggimann20215}
\bibfield{author}{\bibinfo{person}{Manuel Eggimann}, \bibinfo{person}{Abbas
  Rahimi}, {and} \bibinfo{person}{Luca Benini}.}
  \bibinfo{year}{2021}\natexlab{}.
\newblock \showarticletitle{A 5 $\mu$w standard cell memory-based configurable
  hyperdimensional computing accelerator for always-on smart sensing}.
\newblock \bibinfo{journal}{\emph{IEEE Transactions on Circuits and Systems I:
  Regular Papers}} \bibinfo{volume}{68}, \bibinfo{number}{10}
  (\bibinfo{year}{2021}), \bibinfo{pages}{4116--4128}.
\newblock
\urldef\tempurl%
\url{https://doi.org/10.1109/TCSI.2021.3100266}
\showDOI{\tempurl}


\bibitem[Frady et~al\mbox{.}(2018)]%
        {frady2018theory}
\bibfield{author}{\bibinfo{person}{E~Paxon Frady}, \bibinfo{person}{Denis
  Kleyko}, {and} \bibinfo{person}{Friedrich~T Sommer}.}
  \bibinfo{year}{2018}\natexlab{}.
\newblock \showarticletitle{A theory of sequence indexing and working memory in
  recurrent neural networks}.
\newblock \bibinfo{journal}{\emph{Neural Computation}} \bibinfo{volume}{30},
  \bibinfo{number}{6} (\bibinfo{year}{2018}), \bibinfo{pages}{1449--1513}.
\newblock
\urldef\tempurl%
\url{https://doi.org/10.1162/neco_a_01084}
\showDOI{\tempurl}


\bibitem[Gallant and Okaywe(2013)]%
        {gallant2013representing}
\bibfield{author}{\bibinfo{person}{Stephen~I Gallant} {and}
  \bibinfo{person}{T~Wendy Okaywe}.} \bibinfo{year}{2013}\natexlab{}.
\newblock \showarticletitle{Representing objects, relations, and sequences}.
\newblock \bibinfo{journal}{\emph{Neural computation}} \bibinfo{volume}{25},
  \bibinfo{number}{8} (\bibinfo{year}{2013}), \bibinfo{pages}{2038--2078}.
\newblock
\urldef\tempurl%
\url{https://doi.org/10.1162/NECO_a_00467}
\showDOI{\tempurl}


\bibitem[Gayler and Levy(2009)]%
        {gayler2009distributed}
\bibfield{author}{\bibinfo{person}{Ross~W Gayler} {and}
  \bibinfo{person}{Simon~D Levy}.} \bibinfo{year}{2009}\natexlab{}.
\newblock \showarticletitle{A distributed basis for analogical mapping}. In
  \bibinfo{booktitle}{\emph{New Frontiers in Analogy Research; Proc. of 2nd
  Intern. Analogy Conf}}, Vol.~\bibinfo{volume}{9}.
\newblock


\bibitem[Grossi et~al\mbox{.}(2019)]%
        {grossi2019resistive}
\bibfield{author}{\bibinfo{person}{Alessandro Grossi}, \bibinfo{person}{Elisa
  Vianello}, \bibinfo{person}{Mohamed~M Sabry}, \bibinfo{person}{Marios
  Barlas}, \bibinfo{person}{Laurent Grenouillet}, \bibinfo{person}{Jean
  Coignus}, \bibinfo{person}{Edith Beigne}, \bibinfo{person}{Tony Wu},
  \bibinfo{person}{Binh~Q Le}, \bibinfo{person}{Mary~K Wootters},
  {et~al\mbox{.}}} \bibinfo{year}{2019}\natexlab{}.
\newblock \showarticletitle{Resistive RAM endurance: Array-level
  characterization and correction techniques targeting deep learning
  applications}.
\newblock \bibinfo{journal}{\emph{IEEE Transactions on Electron Devices}}
  \bibinfo{volume}{66}, \bibinfo{number}{3} (\bibinfo{year}{2019}),
  \bibinfo{pages}{1281--1288}.
\newblock
\urldef\tempurl%
\url{https://doi.org/10.1109/TED.2019.2894387}
\showDOI{\tempurl}


\bibitem[Halawani et~al\mbox{.}(2021)]%
        {halawani2021fused}
\bibfield{author}{\bibinfo{person}{Yasmin Halawani}, \bibinfo{person}{Eman
  Hassan}, \bibinfo{person}{Baker Mohammad}, {and} \bibinfo{person}{Hani
  Saleh}.} \bibinfo{year}{2021}\natexlab{}.
\newblock \showarticletitle{Fused RRAM-based shift-add architecture for
  efficient hyperdimensional computing paradigm}. In
  \bibinfo{booktitle}{\emph{2021 IEEE International Midwest Symposium on
  Circuits and Systems (MWSCAS)}}. IEEE, \bibinfo{pages}{179--182}.
\newblock
\urldef\tempurl%
\url{https://doi.org/10.1109/MWSCAS47672.2021.9531748}
\showDOI{\tempurl}


\bibitem[Heddes et~al\mbox{.}(2022)]%
        {heddes2022hyperdimensional}
\bibfield{author}{\bibinfo{person}{Mike Heddes}, \bibinfo{person}{Igor Nunes},
  \bibinfo{person}{Tony Givargis}, \bibinfo{person}{Alexandru Nicolau}, {and}
  \bibinfo{person}{Alex Veidenbaum}.} \bibinfo{year}{2022}\natexlab{}.
\newblock \showarticletitle{Hyperdimensional hashing: a robust and efficient
  dynamic hash table}. In \bibinfo{booktitle}{\emph{Proceedings of the 59th
  ACM/IEEE Design Automation Conference}}. \bibinfo{pages}{907--912}.
\newblock
\urldef\tempurl%
\url{https://doi.org/10.1145/3489517.3530553}
\showDOI{\tempurl}


\bibitem[Hsieh et~al\mbox{.}(2019)]%
        {hsieh2019high}
\bibfield{author}{\bibinfo{person}{ER Hsieh}, \bibinfo{person}{M Giordano},
  \bibinfo{person}{B Hodson}, \bibinfo{person}{A Levy}, \bibinfo{person}{SK
  Osekowsky}, \bibinfo{person}{RM Radway}, \bibinfo{person}{YC Shih},
  \bibinfo{person}{W Wan}, \bibinfo{person}{TF Wu}, \bibinfo{person}{X Zheng},
  {et~al\mbox{.}}} \bibinfo{year}{2019}\natexlab{}.
\newblock \showarticletitle{High-density multiple bits-per-cell 1T4R RRAM array
  with gradual SET/RESET and its effectiveness for deep learning}. In
  \bibinfo{booktitle}{\emph{2019 IEEE International Electron Devices Meeting
  (IEDM)}}. IEEE, \bibinfo{pages}{35--6}.
\newblock
\urldef\tempurl%
\url{https://doi.org/10.1109/IEDM19573.2019.8993514}
\showDOI{\tempurl}


\bibitem[Imani et~al\mbox{.}(2019a)]%
        {imani2019quanthd}
\bibfield{author}{\bibinfo{person}{Mohsen Imani}, \bibinfo{person}{Samuel
  Bosch}, \bibinfo{person}{Sohum Datta}, \bibinfo{person}{Sharadhi
  Ramakrishna}, \bibinfo{person}{Sahand Salamat}, \bibinfo{person}{Jan~M
  Rabaey}, {and} \bibinfo{person}{Tajana Rosing}.}
  \bibinfo{year}{2019}\natexlab{a}.
\newblock \showarticletitle{Quanthd: A quantization framework for
  hyperdimensional computing}.
\newblock \bibinfo{journal}{\emph{IEEE Transactions on Computer-Aided Design of
  Integrated Circuits and Systems}} \bibinfo{volume}{39}, \bibinfo{number}{10}
  (\bibinfo{year}{2019}), \bibinfo{pages}{2268--2278}.
\newblock
\urldef\tempurl%
\url{https://doi.org/10.1109/TCAD.2019.2954472}
\showDOI{\tempurl}


\bibitem[Imani et~al\mbox{.}(2018)]%
        {imani2018hierarchical}
\bibfield{author}{\bibinfo{person}{Mohsen Imani}, \bibinfo{person}{Chenyu
  Huang}, \bibinfo{person}{Deqian Kong}, {and} \bibinfo{person}{Tajana
  Rosing}.} \bibinfo{year}{2018}\natexlab{}.
\newblock \showarticletitle{Hierarchical hyperdimensional computing for energy
  efficient classification}. In \bibinfo{booktitle}{\emph{Proceedings of the
  55th Annual Design Automation Conference}}. \bibinfo{pages}{1--6}.
\newblock
\urldef\tempurl%
\url{https://doi.org/10.1145/3195970.3196060}
\showDOI{\tempurl}


\bibitem[Imani et~al\mbox{.}(2017a)]%
        {imani2017voicehd}
\bibfield{author}{\bibinfo{person}{Mohsen Imani}, \bibinfo{person}{Deqian
  Kong}, \bibinfo{person}{Abbas Rahimi}, {and} \bibinfo{person}{Tajana
  Rosing}.} \bibinfo{year}{2017}\natexlab{a}.
\newblock \showarticletitle{Voicehd: Hyperdimensional computing for efficient
  speech recognition}. In \bibinfo{booktitle}{\emph{2017 IEEE international
  conference on rebooting computing (ICRC)}}. IEEE, \bibinfo{pages}{1--8}.
\newblock
\urldef\tempurl%
\url{https://doi.org/10.1109/ICRC.2017.8123650}
\showDOI{\tempurl}


\bibitem[Imani et~al\mbox{.}(2017b)]%
        {imani2017exploring}
\bibfield{author}{\bibinfo{person}{Mohsen Imani}, \bibinfo{person}{Abbas
  Rahimi}, \bibinfo{person}{Deqian Kong}, \bibinfo{person}{Tajana Rosing},
  {and} \bibinfo{person}{Jan~M Rabaey}.} \bibinfo{year}{2017}\natexlab{b}.
\newblock \showarticletitle{Exploring hyperdimensional associative memory}. In
  \bibinfo{booktitle}{\emph{2017 IEEE International Symposium on High
  Performance Computer Architecture (HPCA)}}. IEEE, \bibinfo{pages}{445--456}.
\newblock
\urldef\tempurl%
\url{https://doi.org/10.1109/HPCA.2017.28}
\showDOI{\tempurl}


\bibitem[Imani et~al\mbox{.}(2019b)]%
        {imani2019fach}
\bibfield{author}{\bibinfo{person}{Mohsen Imani}, \bibinfo{person}{Sahand
  Salamat}, \bibinfo{person}{Saransh Gupta}, \bibinfo{person}{Jiani Huang},
  {and} \bibinfo{person}{Tajana Rosing}.} \bibinfo{year}{2019}\natexlab{b}.
\newblock \showarticletitle{Fach: Fpga-based acceleration of hyperdimensional
  computing by reducing computational complexity}. In
  \bibinfo{booktitle}{\emph{Proceedings of the 24th Asia and South Pacific
  Design Automation Conference}}. \bibinfo{pages}{493--498}.
\newblock
\urldef\tempurl%
\url{https://doi.org/10.1145/3287624.3287667}
\showDOI{\tempurl}


\bibitem[Imani et~al\mbox{.}(2019c)]%
        {imani2019sparsehd}
\bibfield{author}{\bibinfo{person}{Mohsen Imani}, \bibinfo{person}{Sahand
  Salamat}, \bibinfo{person}{Behnam Khaleghi}, \bibinfo{person}{Mohammad
  Samragh}, \bibinfo{person}{Farinaz Koushanfar}, {and} \bibinfo{person}{Tajana
  Rosing}.} \bibinfo{year}{2019}\natexlab{c}.
\newblock \showarticletitle{Sparsehd: Algorithm-hardware co-optimization for
  efficient high-dimensional computing}. In \bibinfo{booktitle}{\emph{2019 IEEE
  27th Annual International Symposium on Field-Programmable Custom Computing
  Machines (FCCM)}}. IEEE, \bibinfo{pages}{190--198}.
\newblock
\urldef\tempurl%
\url{https://doi.org/10.1109/FCCM.2019.00034}
\showDOI{\tempurl}


\bibitem[Jones and Mewhort(2007)]%
        {jones2007representing}
\bibfield{author}{\bibinfo{person}{Michael~N Jones} {and}
  \bibinfo{person}{Douglas~JK Mewhort}.} \bibinfo{year}{2007}\natexlab{}.
\newblock \showarticletitle{Representing word meaning and order information in
  a composite holographic lexicon.}
\newblock \bibinfo{journal}{\emph{Psychological review}} \bibinfo{volume}{114},
  \bibinfo{number}{1} (\bibinfo{year}{2007}), \bibinfo{pages}{1}.
\newblock
\urldef\tempurl%
\url{https://doi.org/10.1037/0033-295X.114.1.1}
\showDOI{\tempurl}


\bibitem[Kanerva(2009)]%
        {kanerva2009hyperdimensional}
\bibfield{author}{\bibinfo{person}{Pentti Kanerva}.}
  \bibinfo{year}{2009}\natexlab{}.
\newblock \showarticletitle{Hyperdimensional computing: An introduction to
  computing in distributed representation with high-dimensional random
  vectors}.
\newblock \bibinfo{journal}{\emph{Cognitive computation}} \bibinfo{volume}{1},
  \bibinfo{number}{2} (\bibinfo{year}{2009}), \bibinfo{pages}{139--159}.
\newblock


\bibitem[Kanerva(2010)]%
        {kanerva2010we}
\bibfield{author}{\bibinfo{person}{Pentti Kanerva}.}
  \bibinfo{year}{2010}\natexlab{}.
\newblock \showarticletitle{What we mean when we say" What's the dollar of
  Mexico?": Prototypes and mapping in concept space}. In
  \bibinfo{booktitle}{\emph{2010 AAAI fall symposium series}}.
\newblock


\bibitem[Kanerva(2014)]%
        {kanerva2014computing}
\bibfield{author}{\bibinfo{person}{Pentti Kanerva}.}
  \bibinfo{year}{2014}\natexlab{}.
\newblock \showarticletitle{Computing with 10,000-bit words}. In
  \bibinfo{booktitle}{\emph{2014 52nd annual Allerton conference on
  communication, control, and computing (Allerton)}}. IEEE,
  \bibinfo{pages}{304--310}.
\newblock
\urldef\tempurl%
\url{https://doi.org/10.1109/ALLERTON.2014.7028470}
\showDOI{\tempurl}


\bibitem[Kanerva(2018)]%
        {kanerva2018computing}
\bibfield{author}{\bibinfo{person}{Pentti Kanerva}.}
  \bibinfo{year}{2018}\natexlab{}.
\newblock \showarticletitle{Computing with high-dimensional vectors}.
\newblock \bibinfo{journal}{\emph{IEEE Design \& Test}} \bibinfo{volume}{36},
  \bibinfo{number}{3} (\bibinfo{year}{2018}), \bibinfo{pages}{7--14}.
\newblock
\urldef\tempurl%
\url{https://doi.org/10.1109/MDAT.2018.2890221}
\showDOI{\tempurl}


\bibitem[Kanerva et~al\mbox{.}(1997)]%
        {kanerva1997fully}
\bibfield{author}{\bibinfo{person}{Pentti Kanerva} {et~al\mbox{.}}}
  \bibinfo{year}{1997}\natexlab{}.
\newblock \showarticletitle{Fully distributed representation}.
\newblock \bibinfo{journal}{\emph{PAT}} \bibinfo{volume}{1},
  \bibinfo{number}{5} (\bibinfo{year}{1997}), \bibinfo{pages}{10000}.
\newblock


\bibitem[Karunaratne et~al\mbox{.}(2020)]%
        {karunaratne2020memory}
\bibfield{author}{\bibinfo{person}{Geethan Karunaratne},
  \bibinfo{person}{Manuel Le~Gallo}, \bibinfo{person}{Giovanni Cherubini},
  \bibinfo{person}{Luca Benini}, \bibinfo{person}{Abbas Rahimi}, {and}
  \bibinfo{person}{Abu Sebastian}.} \bibinfo{year}{2020}\natexlab{}.
\newblock \showarticletitle{In-memory hyperdimensional computing}.
\newblock \bibinfo{journal}{\emph{Nature Electronics}} \bibinfo{volume}{3},
  \bibinfo{number}{6} (\bibinfo{year}{2020}), \bibinfo{pages}{327--337}.
\newblock
\urldef\tempurl%
\url{https://doi.org/10.1038/s41565-023-01357-8}
\showDOI{\tempurl}


\bibitem[Kim et~al\mbox{.}(2020)]%
        {kim2020geniehd}
\bibfield{author}{\bibinfo{person}{Yeseong Kim}, \bibinfo{person}{Mohsen
  Imani}, \bibinfo{person}{Niema Moshiri}, {and} \bibinfo{person}{Tajana
  Rosing}.} \bibinfo{year}{2020}\natexlab{}.
\newblock \showarticletitle{Geniehd: Efficient dna pattern matching accelerator
  using hyperdimensional computing}. In \bibinfo{booktitle}{\emph{2020 Design,
  Automation \& Test in Europe Conference \& Exhibition (DATE)}}. IEEE,
  \bibinfo{pages}{115--120}.
\newblock
\urldef\tempurl%
\url{https://doi.org/10.23919/DATE48585.2020.9116397}
\showDOI{\tempurl}


\bibitem[Kleyko et~al\mbox{.}(2023a)]%
        {kleyko2023efficient}
\bibfield{author}{\bibinfo{person}{Denis Kleyko}, \bibinfo{person}{Connor
  Bybee}, \bibinfo{person}{Ping-Chen Huang}, \bibinfo{person}{Christopher~J
  Kymn}, \bibinfo{person}{Bruno~A Olshausen}, \bibinfo{person}{E~Paxon Frady},
  {and} \bibinfo{person}{Friedrich~T Sommer}.}
  \bibinfo{year}{2023}\natexlab{a}.
\newblock \showarticletitle{Efficient decoding of compositional structure in
  holistic representations}.
\newblock \bibinfo{journal}{\emph{Neural Computation}} \bibinfo{volume}{35},
  \bibinfo{number}{7} (\bibinfo{year}{2023}), \bibinfo{pages}{1159--1186}.
\newblock
\urldef\tempurl%
\url{https://doi.org/10.1162/neco_a_01590}
\showDOI{\tempurl}


\bibitem[Kleyko et~al\mbox{.}(2022)]%
        {kleyko2022vector}
\bibfield{author}{\bibinfo{person}{Denis Kleyko}, \bibinfo{person}{Mike
  Davies}, \bibinfo{person}{Edward~Paxon Frady}, \bibinfo{person}{Pentti
  Kanerva}, \bibinfo{person}{Spencer~J Kent}, \bibinfo{person}{Bruno~A
  Olshausen}, \bibinfo{person}{Evgeny Osipov}, \bibinfo{person}{Jan~M Rabaey},
  \bibinfo{person}{Dmitri~A Rachkovskij}, \bibinfo{person}{Abbas Rahimi},
  {et~al\mbox{.}}} \bibinfo{year}{2022}\natexlab{}.
\newblock \showarticletitle{Vector Symbolic Architectures as a Computing
  Framework for Emerging Hardware}.
\newblock \bibinfo{journal}{\emph{Proc. IEEE}} \bibinfo{volume}{110},
  \bibinfo{number}{10} (\bibinfo{year}{2022}), \bibinfo{pages}{1538--1571}.
\newblock
\urldef\tempurl%
\url{https://doi.org/10.1109/JPROC.2022.3209104}
\showDOI{\tempurl}


\bibitem[Kleyko et~al\mbox{.}(2016)]%
        {kleyko2016holographic}
\bibfield{author}{\bibinfo{person}{Denis Kleyko}, \bibinfo{person}{Evgeny
  Osipov}, \bibinfo{person}{Alexander Senior}, \bibinfo{person}{Asad~I Khan},
  {and} \bibinfo{person}{Ya{\c{s}}ar~Ahmet {\c{S}}ekerciog{\u{g}}lu}.}
  \bibinfo{year}{2016}\natexlab{}.
\newblock \showarticletitle{Holographic graph neuron: A bioinspired
  architecture for pattern processing}.
\newblock \bibinfo{journal}{\emph{IEEE transactions on neural networks and
  learning systems}} \bibinfo{volume}{28}, \bibinfo{number}{6}
  (\bibinfo{year}{2016}), \bibinfo{pages}{1250--1262}.
\newblock
\urldef\tempurl%
\url{https://doi.org/10.1109/TNNLS.2016.2535338}
\showDOI{\tempurl}


\bibitem[Kleyko et~al\mbox{.}(2023b)]%
        {kleyko2021survey2}
\bibfield{author}{\bibinfo{person}{Denis Kleyko}, \bibinfo{person}{Dmitri
  Rachkovskij}, \bibinfo{person}{Evgeny Osipov}, {and} \bibinfo{person}{Abbas
  Rahimi}.} \bibinfo{year}{2023}\natexlab{b}.
\newblock \showarticletitle{A survey on hyperdimensional computing aka vector
  symbolic architectures, part ii: Applications, cognitive models, and
  challenges}.
\newblock \bibinfo{journal}{\emph{Comput. Surveys}} \bibinfo{volume}{55},
  \bibinfo{number}{9} (\bibinfo{year}{2023}), \bibinfo{pages}{1--52}.
\newblock
\urldef\tempurl%
\url{https://doi.org/10.1145/3558000}
\showDOI{\tempurl}


\bibitem[Kleyko et~al\mbox{.}(2021)]%
        {kleyko2021survey1}
\bibfield{author}{\bibinfo{person}{Denis Kleyko}, \bibinfo{person}{Dmitri~A
  Rachkovskij}, \bibinfo{person}{Evgeny Osipov}, {and} \bibinfo{person}{Abbas
  Rahimi}.} \bibinfo{year}{2021}\natexlab{}.
\newblock \showarticletitle{A Survey on Hyperdimensional Computing aka Vector
  Symbolic Architectures, Part I: Models and Data Transformations}.
\newblock \bibinfo{journal}{\emph{ACM Computing Surveys (CSUR)}}
  (\bibinfo{year}{2021}).
\newblock
\urldef\tempurl%
\url{https://doi.org/A Survey on Hyperdimensional Computing aka Vector Symbolic
  Architectures}
\showDOI{\tempurl}


\bibitem[Kleyko et~al\mbox{.}(2020)]%
        {kleyko2020autoscaling}
\bibfield{author}{\bibinfo{person}{Denis Kleyko}, \bibinfo{person}{Abbas
  Rahimi}, \bibinfo{person}{Ross~W Gayler}, {and} \bibinfo{person}{Evgeny
  Osipov}.} \bibinfo{year}{2020}\natexlab{}.
\newblock \showarticletitle{Autoscaling bloom filter: controlling trade-off
  between true and false positives}.
\newblock \bibinfo{journal}{\emph{Neural Computing and Applications}}
  \bibinfo{volume}{32} (\bibinfo{year}{2020}), \bibinfo{pages}{3675--3684}.
\newblock
\urldef\tempurl%
\url{https://doi.org/10.1007/s00521-019-04397-1}
\showDOI{\tempurl}


\bibitem[Kleyko et~al\mbox{.}(2023c)]%
        {kleyko2020perceptron}
\bibfield{author}{\bibinfo{person}{Denis Kleyko}, \bibinfo{person}{Antonello
  Rosato}, \bibinfo{person}{Edward~Paxon Frady}, \bibinfo{person}{Massimo
  Panella}, {and} \bibinfo{person}{Friedrich~T. Sommer}.}
  \bibinfo{year}{2023}\natexlab{c}.
\newblock \showarticletitle{Perceptron Theory Can Predict the Accuracy of
  Neural Networks}.
\newblock \bibinfo{journal}{\emph{IEEE Transactions on Neural Networks and
  Learning Systems}} (\bibinfo{year}{2023}), \bibinfo{pages}{1--15}.
\newblock
\urldef\tempurl%
\url{https://doi.org/10.1109/TNNLS.2023.3237381}
\showDOI{\tempurl}


\bibitem[Langenegger et~al\mbox{.}(2023)]%
        {langenegger2023memory}
\bibfield{author}{\bibinfo{person}{Jovin Langenegger}, \bibinfo{person}{Geethan
  Karunaratne}, \bibinfo{person}{Michael Hersche}, \bibinfo{person}{Luca
  Benini}, \bibinfo{person}{Abu Sebastian}, {and} \bibinfo{person}{Abbas
  Rahimi}.} \bibinfo{year}{2023}\natexlab{}.
\newblock \showarticletitle{In-memory factorization of holographic perceptual
  representations}.
\newblock \bibinfo{journal}{\emph{Nature Nanotechnology}}
  (\bibinfo{year}{2023}), \bibinfo{pages}{1--7}.
\newblock
\urldef\tempurl%
\url{https://doi.org/10.1038/s41565-023-01357-8}
\showDOI{\tempurl}


\bibitem[Le et~al\mbox{.}(2021)]%
        {le2021radar}
\bibfield{author}{\bibinfo{person}{Binh~Q Le}, \bibinfo{person}{Akash Levy},
  \bibinfo{person}{Tony~F Wu}, \bibinfo{person}{Robert~M Radway},
  \bibinfo{person}{E~Ray Hsieh}, \bibinfo{person}{Xin Zheng},
  \bibinfo{person}{Mark Nelson}, \bibinfo{person}{Priyanka Raina},
  \bibinfo{person}{H-S~Philip Wong}, \bibinfo{person}{Simon Wong},
  {et~al\mbox{.}}} \bibinfo{year}{2021}\natexlab{}.
\newblock \showarticletitle{RADAR: A fast and energy-efficient programming
  technique for multiple bits-per-cell RRAM arrays}.
\newblock \bibinfo{journal}{\emph{IEEE Transactions on Electron Devices}}
  \bibinfo{volume}{68}, \bibinfo{number}{9} (\bibinfo{year}{2021}),
  \bibinfo{pages}{4397--4403}.
\newblock
\urldef\tempurl%
\url{https://doi.org/10.1109/TED.2021.3097975}
\showDOI{\tempurl}


\bibitem[Li et~al\mbox{.}(2016)]%
        {li2016hyperdimensional}
\bibfield{author}{\bibinfo{person}{Haitong Li}, \bibinfo{person}{Tony~F Wu},
  \bibinfo{person}{Abbas Rahimi}, \bibinfo{person}{Kai-Shin Li},
  \bibinfo{person}{Miles Rusch}, \bibinfo{person}{Chang-Hsien Lin},
  \bibinfo{person}{Juo-Luen Hsu}, \bibinfo{person}{Mohamed~M Sabry},
  \bibinfo{person}{S~Burc Eryilmaz}, \bibinfo{person}{Joon Sohn},
  {et~al\mbox{.}}} \bibinfo{year}{2016}\natexlab{}.
\newblock \showarticletitle{Hyperdimensional computing with 3D VRRAM in-memory
  kernels: Device-architecture co-design for energy-efficient, error-resilient
  language recognition}. In \bibinfo{booktitle}{\emph{2016 IEEE International
  Electron Devices Meeting (IEDM)}}. IEEE, \bibinfo{pages}{16--1}.
\newblock
\urldef\tempurl%
\url{https://doi.org/10.1109/IEDM.2016.7838428}
\showDOI{\tempurl}


\bibitem[Misailovic et~al\mbox{.}(2014)]%
        {misailovic2014chisel}
\bibfield{author}{\bibinfo{person}{Sasa Misailovic}, \bibinfo{person}{Michael
  Carbin}, \bibinfo{person}{Sara Achour}, \bibinfo{person}{Zichao Qi}, {and}
  \bibinfo{person}{Martin~C Rinard}.} \bibinfo{year}{2014}\natexlab{}.
\newblock \showarticletitle{Chisel: Reliability-and accuracy-aware optimization
  of approximate computational kernels}.
\newblock \bibinfo{journal}{\emph{ACM Sigplan Notices}} \bibinfo{volume}{49},
  \bibinfo{number}{10} (\bibinfo{year}{2014}), \bibinfo{pages}{309--328}.
\newblock
\urldef\tempurl%
\url{https://doi.org/10.1145/2714064.2660231}
\showDOI{\tempurl}


\bibitem[Montagna et~al\mbox{.}(2018)]%
        {montagna2018pulp}
\bibfield{author}{\bibinfo{person}{Fabio Montagna}, \bibinfo{person}{Abbas
  Rahimi}, \bibinfo{person}{Simone Benatti}, \bibinfo{person}{Davide Rossi},
  {and} \bibinfo{person}{Luca Benini}.} \bibinfo{year}{2018}\natexlab{}.
\newblock \showarticletitle{PULP-HD: Accelerating brain-inspired
  high-dimensional computing on a parallel ultra-low power platform}. In
  \bibinfo{booktitle}{\emph{2018 55th ACM/ESDA/IEEE Design Automation
  Conference (DAC)}}. IEEE, \bibinfo{pages}{1--6}.
\newblock
\urldef\tempurl%
\url{https://doi.org/10.1145/3195970.3196096}
\showDOI{\tempurl}


\bibitem[Morris et~al\mbox{.}(2019)]%
        {morris2019comphd}
\bibfield{author}{\bibinfo{person}{Justin Morris}, \bibinfo{person}{Mohsen
  Imani}, \bibinfo{person}{Samuel Bosch}, \bibinfo{person}{Anthony Thomas},
  \bibinfo{person}{Helen Shu}, {and} \bibinfo{person}{Tajana Rosing}.}
  \bibinfo{year}{2019}\natexlab{}.
\newblock \showarticletitle{CompHD: Efficient hyperdimensional computing using
  model compression}. In \bibinfo{booktitle}{\emph{2019 IEEE/ACM International
  Symposium on Low Power Electronics and Design (ISLPED)}}. IEEE,
  \bibinfo{pages}{1--6}.
\newblock
\urldef\tempurl%
\url{https://doi.org/10.1109/ISLPED.2019.8824908}
\showDOI{\tempurl}


\bibitem[Nagaev and Chebotarev(2011)]%
        {nagaev2011bound}
\bibfield{author}{\bibinfo{person}{SV Nagaev} {and} \bibinfo{person}{VI
  Chebotarev}.} \bibinfo{year}{2011}\natexlab{}.
\newblock \showarticletitle{On the bound of proximity of the binomial
  distribution to the normal one}. In \bibinfo{booktitle}{\emph{Doklady
  Mathematics}}, Vol.~\bibinfo{volume}{83}. Springer, \bibinfo{pages}{19--21}.
\newblock
\urldef\tempurl%
\url{https://doi.org/10.1134/S1064562411010030}
\showDOI{\tempurl}


\bibitem[Osipov et~al\mbox{.}(2017)]%
        {osipov2017associative}
\bibfield{author}{\bibinfo{person}{Evgeny Osipov}, \bibinfo{person}{Denis
  Kleyko}, {and} \bibinfo{person}{Alexander Legalov}.}
  \bibinfo{year}{2017}\natexlab{}.
\newblock \showarticletitle{Associative synthesis of finite state automata
  model of a controlled object with hyperdimensional computing}. In
  \bibinfo{booktitle}{\emph{IECON 2017-43rd Annual Conference of the IEEE
  Industrial Electronics Society}}. IEEE, \bibinfo{pages}{3276--3281}.
\newblock
\urldef\tempurl%
\url{https://doi.org/10.1109/IECON.2017.8216554}
\showDOI{\tempurl}


\bibitem[Pashchenko et~al\mbox{.}(2020)]%
        {pashchenko2020search}
\bibfield{author}{\bibinfo{person}{Dmitry~V Pashchenko},
  \bibinfo{person}{Dmitry~A Trokoz}, \bibinfo{person}{Alexey~I Martyshkin},
  \bibinfo{person}{Mihail~P Sinev}, {and} \bibinfo{person}{Boris~L Svistunov}.}
  \bibinfo{year}{2020}\natexlab{}.
\newblock \showarticletitle{Search for a substring of characters using the
  theory of non-deterministic finite automata and vector-character
  architecture}.
\newblock \bibinfo{journal}{\emph{Bulletin of Electrical Engineering and
  Informatics}} \bibinfo{volume}{9}, \bibinfo{number}{3}
  (\bibinfo{year}{2020}), \bibinfo{pages}{1238--1250}.
\newblock
\urldef\tempurl%
\url{https://doi.org/10.11591/eei.v9i3.1720}
\showDOI{\tempurl}


\bibitem[Plate(1994)]%
        {plate1994distributed}
\bibfield{author}{\bibinfo{person}{Tony~A Plate}.}
  \bibinfo{year}{1994}\natexlab{}.
\newblock \bibinfo{booktitle}{\emph{Distributed representations and nested
  compositional structure}}.
\newblock \bibinfo{publisher}{Citeseer}.
\newblock


\bibitem[Plate(2000)]%
        {plate2000analogy}
\bibfield{author}{\bibinfo{person}{Tony~A Plate}.}
  \bibinfo{year}{2000}\natexlab{}.
\newblock \showarticletitle{Analogy retrieval and processing with distributed
  vector representations}.
\newblock \bibinfo{journal}{\emph{Expert systems}} \bibinfo{volume}{17},
  \bibinfo{number}{1} (\bibinfo{year}{2000}), \bibinfo{pages}{29--40}.
\newblock
\urldef\tempurl%
\url{https://doi.org/10.1111/1468-0394.00125}
\showDOI{\tempurl}


\bibitem[Plate(2003)]%
        {plate2003holographic}
\bibfield{author}{\bibinfo{person}{Tony~A Plate}.}
  \bibinfo{year}{2003}\natexlab{}.
\newblock \showarticletitle{Holographic Reduced Representation: Distributed
  representation for cognitive structures}.
\newblock  (\bibinfo{year}{2003}).
\newblock


\bibitem[Poduval et~al\mbox{.}(2021)]%
        {poduval2021stochd}
\bibfield{author}{\bibinfo{person}{Prathyush Poduval}, \bibinfo{person}{Zhuowen
  Zou}, \bibinfo{person}{Hassan Najafi}, \bibinfo{person}{Houman Homayoun},
  {and} \bibinfo{person}{Mohsen Imani}.} \bibinfo{year}{2021}\natexlab{}.
\newblock \showarticletitle{Stochd: Stochastic hyperdimensional system for
  efficient and robust learning from raw data}. In
  \bibinfo{booktitle}{\emph{2021 58th ACM/IEEE Design Automation Conference
  (DAC)}}. IEEE, \bibinfo{pages}{1195--1200}.
\newblock
\urldef\tempurl%
\url{https://doi.org/10.1109/DAC18074.2021.9586166}
\showDOI{\tempurl}


\bibitem[Rachkovskij and Slipchenko(2012)]%
        {rachkovskij2012similarity}
\bibfield{author}{\bibinfo{person}{Dmitri~A Rachkovskij} {and}
  \bibinfo{person}{Serge~V Slipchenko}.} \bibinfo{year}{2012}\natexlab{}.
\newblock \showarticletitle{Similarity-based retrieval with structure-sensitive
  sparse binary distributed representations}.
\newblock \bibinfo{journal}{\emph{Computational Intelligence}}
  \bibinfo{volume}{28}, \bibinfo{number}{1} (\bibinfo{year}{2012}),
  \bibinfo{pages}{106--129}.
\newblock
\urldef\tempurl%
\url{https://doi.org/10.1111/j.1467-8640.2011.00423.x}
\showDOI{\tempurl}


\bibitem[Rahimi et~al\mbox{.}(2016)]%
        {rahimi2016hyperdimensional}
\bibfield{author}{\bibinfo{person}{Abbas Rahimi}, \bibinfo{person}{Simone
  Benatti}, \bibinfo{person}{Pentti Kanerva}, \bibinfo{person}{Luca Benini},
  {and} \bibinfo{person}{Jan~M Rabaey}.} \bibinfo{year}{2016}\natexlab{}.
\newblock \showarticletitle{Hyperdimensional biosignal processing: A case study
  for EMG-based hand gesture recognition}. In \bibinfo{booktitle}{\emph{2016
  IEEE International Conference on Rebooting Computing (ICRC)}}. IEEE,
  \bibinfo{pages}{1--8}.
\newblock
\urldef\tempurl%
\url{https://doi.org/10.1109/ICRC.2016.7738683}
\showDOI{\tempurl}


\bibitem[Rahimi et~al\mbox{.}(2017)]%
        {rahimi2017high}
\bibfield{author}{\bibinfo{person}{Abbas Rahimi}, \bibinfo{person}{Sohum
  Datta}, \bibinfo{person}{Denis Kleyko}, \bibinfo{person}{Edward~Paxon Frady},
  \bibinfo{person}{Bruno Olshausen}, \bibinfo{person}{Pentti Kanerva}, {and}
  \bibinfo{person}{Jan~M Rabaey}.} \bibinfo{year}{2017}\natexlab{}.
\newblock \showarticletitle{High-dimensional computing as a nanoscalable
  paradigm}.
\newblock \bibinfo{journal}{\emph{IEEE Transactions on Circuits and Systems I:
  Regular Papers}} \bibinfo{volume}{64}, \bibinfo{number}{9}
  (\bibinfo{year}{2017}), \bibinfo{pages}{2508--2521}.
\newblock
\urldef\tempurl%
\url{https://doi.org/10.1109/TCSI.2017.2705051}
\showDOI{\tempurl}


\bibitem[Rahimi et~al\mbox{.}(2018)]%
        {rahimi2018efficient}
\bibfield{author}{\bibinfo{person}{Abbas Rahimi}, \bibinfo{person}{Pentti
  Kanerva}, \bibinfo{person}{Luca Benini}, {and} \bibinfo{person}{Jan~M
  Rabaey}.} \bibinfo{year}{2018}\natexlab{}.
\newblock \showarticletitle{Efficient biosignal processing using
  hyperdimensional computing: Network templates for combined learning and
  classification of exg signals}.
\newblock \bibinfo{journal}{\emph{Proc. IEEE}} \bibinfo{volume}{107},
  \bibinfo{number}{1} (\bibinfo{year}{2018}), \bibinfo{pages}{123--143}.
\newblock
\urldef\tempurl%
\url{https://doi.org/10.1109/JPROC.2018.2871163}
\showDOI{\tempurl}


\bibitem[Schlegel et~al\mbox{.}(2021)]%
        {schlegel2021multivariate}
\bibfield{author}{\bibinfo{person}{Kenny Schlegel}, \bibinfo{person}{Florian
  Mirus}, \bibinfo{person}{Peer Neubert}, {and} \bibinfo{person}{Peter
  Protzel}.} \bibinfo{year}{2021}\natexlab{}.
\newblock \showarticletitle{Multivariate time series analysis for driving style
  classification using neural networks and hyperdimensional computing}. In
  \bibinfo{booktitle}{\emph{2021 IEEE Intelligent Vehicles Symposium (IV)}}.
  IEEE, \bibinfo{pages}{602--609}.
\newblock
\urldef\tempurl%
\url{https://doi.org/10.1109/IV48863.2021.9576028}
\showDOI{\tempurl}


\bibitem[Schlegel et~al\mbox{.}(2022)]%
        {schlegel2022hdc}
\bibfield{author}{\bibinfo{person}{Kenny Schlegel}, \bibinfo{person}{Peer
  Neubert}, {and} \bibinfo{person}{Peter Protzel}.}
  \bibinfo{year}{2022}\natexlab{}.
\newblock \showarticletitle{HDC-MiniROCKET: Explicit time encoding in time
  series classification with hyperdimensional computing}. In
  \bibinfo{booktitle}{\emph{2022 International Joint Conference on Neural
  Networks (IJCNN)}}. IEEE, \bibinfo{pages}{1--8}.
\newblock
\urldef\tempurl%
\url{https://doi.org/10.1109/IJCNN55064.2022.9892158}
\showDOI{\tempurl}


\bibitem[Sharif et~al\mbox{.}(2021)]%
        {sharif2021approxtuner}
\bibfield{author}{\bibinfo{person}{Hashim Sharif}, \bibinfo{person}{Yifan
  Zhao}, \bibinfo{person}{Maria Kotsifakou}, \bibinfo{person}{Akash Kothari},
  \bibinfo{person}{Ben Schreiber}, \bibinfo{person}{Elizabeth Wang},
  \bibinfo{person}{Yasmin Sarita}, \bibinfo{person}{Nathan Zhao},
  \bibinfo{person}{Keyur Joshi}, \bibinfo{person}{Vikram~S Adve},
  {et~al\mbox{.}}} \bibinfo{year}{2021}\natexlab{}.
\newblock \showarticletitle{ApproxTuner: a compiler and runtime system for
  adaptive approximations}. In \bibinfo{booktitle}{\emph{Proceedings of the
  26th ACM SIGPLAN Symposium on Principles and Practice of Parallel
  Programming}}. \bibinfo{pages}{262--277}.
\newblock
\urldef\tempurl%
\url{https://doi.org/10.1145/3437801.3446108}
\showDOI{\tempurl}


\bibitem[Shulaker et~al\mbox{.}(2014)]%
        {shulaker2014monolithic}
\bibfield{author}{\bibinfo{person}{Max~M Shulaker}, \bibinfo{person}{Tony~F
  Wu}, \bibinfo{person}{Asish Pal}, \bibinfo{person}{Liang Zhao},
  \bibinfo{person}{Yoshio Nishi}, \bibinfo{person}{Krishna Saraswat},
  \bibinfo{person}{H-S~Philip Wong}, {and} \bibinfo{person}{Subhasish Mitra}.}
  \bibinfo{year}{2014}\natexlab{}.
\newblock \showarticletitle{Monolithic 3D integration of logic and memory:
  Carbon nanotube FETs, resistive RAM, and silicon FETs}. In
  \bibinfo{booktitle}{\emph{2014 IEEE International Electron Devices Meeting}}.
  IEEE, \bibinfo{pages}{27--4}.
\newblock
\urldef\tempurl%
\url{https://doi.org/10.1109/IEDM.2014.7047120}
\showDOI{\tempurl}


\bibitem[Simpkin et~al\mbox{.}(2019)]%
        {simpkin2019constructing}
\bibfield{author}{\bibinfo{person}{Chris Simpkin}, \bibinfo{person}{Ian
  Taylor}, \bibinfo{person}{Graham~A Bent}, \bibinfo{person}{Geeth de Mel},
  \bibinfo{person}{Swati Rallapalli}, \bibinfo{person}{Liang Ma}, {and}
  \bibinfo{person}{Mudhakar Srivatsa}.} \bibinfo{year}{2019}\natexlab{}.
\newblock \showarticletitle{Constructing distributed time-critical applications
  using cognitive enabled services}.
\newblock \bibinfo{journal}{\emph{Future Generation Computer Systems}}
  \bibinfo{volume}{100} (\bibinfo{year}{2019}), \bibinfo{pages}{70--85}.
\newblock
\urldef\tempurl%
\url{https://doi.org/10.1016/j.future.2019.04.010}
\showDOI{\tempurl}


\bibitem[Theiss et~al\mbox{.}(2022)]%
        {theiss2022unpaired}
\bibfield{author}{\bibinfo{person}{Justin Theiss}, \bibinfo{person}{Jay
  Leverett}, \bibinfo{person}{Daeil Kim}, {and} \bibinfo{person}{Aayush
  Prakash}.} \bibinfo{year}{2022}\natexlab{}.
\newblock \showarticletitle{Unpaired Image Translation via Vector Symbolic
  Architectures}. In \bibinfo{booktitle}{\emph{Computer Vision--ECCV 2022: 17th
  European Conference, Tel Aviv, Israel, October 23--27, 2022, Proceedings,
  Part XXI}}. Springer, \bibinfo{pages}{17--32}.
\newblock
\urldef\tempurl%
\url{https://doi.org/10.1007/978-3-031-19803-8_2}
\showDOI{\tempurl}


\bibitem[Thomas et~al\mbox{.}(2021)]%
        {thomas2021theoretical}
\bibfield{author}{\bibinfo{person}{Anthony Thomas}, \bibinfo{person}{Sanjoy
  Dasgupta}, {and} \bibinfo{person}{Tajana Rosing}.}
  \bibinfo{year}{2021}\natexlab{}.
\newblock \showarticletitle{Theoretical Foundations of Hyperdimensional
  Computing}.
\newblock \bibinfo{journal}{\emph{Journal of Artificial Intelligence Research}}
   \bibinfo{volume}{72} (\bibinfo{year}{2021}), \bibinfo{pages}{215--249}.
\newblock
\urldef\tempurl%
\url{https://doi.org/10.48550/arXiv.2010.07426}
\showDOI{\tempurl}


\bibitem[Wei et~al\mbox{.}(2023)]%
        {wei2023pba}
\bibfield{author}{\bibinfo{person}{Anjiang Wei}, \bibinfo{person}{Akash Levy},
  \bibinfo{person}{Pu~(Luke) Yi}, \bibinfo{person}{Robert Radway},
  \bibinfo{person}{Priyanka Raina}, \bibinfo{person}{Subhasish Mitra}, {and}
  \bibinfo{person}{Sara Achour}.} \bibinfo{year}{2023}\natexlab{}.
\newblock \showarticletitle{PBA: Percentile-Based Level Allocation for
  Multiple-Bits-Per-Cell RRAM}. In \bibinfo{booktitle}{\emph{ICCAD}}.
\newblock


\bibitem[Wu et~al\mbox{.}(2018)]%
        {wu2018hyperdimensional}
\bibfield{author}{\bibinfo{person}{Tony~F Wu}, \bibinfo{person}{Haitong Li},
  \bibinfo{person}{Ping-Chen Huang}, \bibinfo{person}{Abbas Rahimi},
  \bibinfo{person}{Gage Hills}, \bibinfo{person}{Bryce Hodson},
  \bibinfo{person}{William Hwang}, \bibinfo{person}{Jan~M Rabaey},
  \bibinfo{person}{H-S~Philip Wong}, \bibinfo{person}{Max~M Shulaker},
  {et~al\mbox{.}}} \bibinfo{year}{2018}\natexlab{}.
\newblock \showarticletitle{Hyperdimensional computing exploiting carbon
  nanotube FETs, resistive RAM, and their monolithic 3D integration}.
\newblock \bibinfo{journal}{\emph{IEEE Journal of Solid-State Circuits}}
  \bibinfo{volume}{53}, \bibinfo{number}{11} (\bibinfo{year}{2018}),
  \bibinfo{pages}{3183--3196}.
\newblock
\urldef\tempurl%
\url{https://doi.org/10.1109/JSSC.2018.2870560}
\showDOI{\tempurl}


\bibitem[Yerxa et~al\mbox{.}(2018)]%
        {yerxa2018hyperdimensional}
\bibfield{author}{\bibinfo{person}{Thomas Yerxa}, \bibinfo{person}{Alexander
  Anderson}, {and} \bibinfo{person}{Eric Weiss}.}
  \bibinfo{year}{2018}\natexlab{}.
\newblock \showarticletitle{The hyperdimensional stack machine}.
\newblock \bibinfo{journal}{\emph{Cognitive Computing}} (\bibinfo{year}{2018}),
  \bibinfo{pages}{1--2}.
\newblock


\bibitem[Yi and Achour(2023)]%
        {zenodo}
\bibfield{author}{\bibinfo{person}{Pu~(Luke) Yi} {and} \bibinfo{person}{Sara
  Achour}.} \bibinfo{year}{2023}\natexlab{}.
\newblock \bibinfo{title}{{Artifact for the OOPSLA 2023 Article "Hardware-Aware
  Static Optimization of Hyperdimensional Computations"}}.
\newblock
\newblock
\urldef\tempurl%
\url{https://doi.org/10.5281/zenodo.8329813}
\showDOI{\tempurl}


\bibitem[Yu et~al\mbox{.}(2022)]%
        {yu2022understanding}
\bibfield{author}{\bibinfo{person}{Tao Yu}, \bibinfo{person}{Yichi Zhang},
  \bibinfo{person}{Zhiru Zhang}, {and} \bibinfo{person}{Christopher~M De~Sa}.}
  \bibinfo{year}{2022}\natexlab{}.
\newblock \showarticletitle{Understanding hyperdimensional computing for
  parallel single-pass learning}.
\newblock \bibinfo{journal}{\emph{Advances in Neural Information Processing
  Systems}}  \bibinfo{volume}{35} (\bibinfo{year}{2022}),
  \bibinfo{pages}{1157--1169}.
\newblock


\end{thebibliography}

% \newpage
% \input{supplementary}

\end{document}